\documentclass[11pt]{article}
\usepackage{latexsym,epsfig,subfigure,graphicx,caption,amsmath,amssymb,amsfonts,dsfont,mathrsfs,mathabx,accents}

\newcommand{\scrL}{{\mathscr{L}}}
\newcommand{\scrP}{{\mathscr{P}}}

\newcommand{\calA}{{\cal{A}}}

\newcommand{\calC}{{\cal{C}}}

\newcommand{\calE}{{\cal{E}}}
\newcommand{\calF}{{\cal{F}}}

\newcommand{\calH}{{\cal{H}}}
\newcommand{\calJ}{{\cal{J}}}

\newcommand{\calI}{{\cal{I}}}

\newcommand{\calM}{{\cal{M}}}
\newcommand{\calN}{{\cal{N}}}

\newcommand{\calQ}{{\cal{Q}}}

\newcommand{\calS}{{\cal{S}}}
\newcommand{\calT}{{\cal{T}}}

\newcommand{\calV}{{\cal{V}}}
\newcommand{\calX}{{\cal{X}}}
\newcommand{\calY}{{\cal{Y}}}
\newcommand{\calZ}{{\cal{Z}}}

\def\cC{{\Bbb C}}
\def\eE{{\Bbb E}}

\def\nN{{\Bbb N}}
\def\pP{{\Bbb P}}
\def\qQ{{\Bbb Q}}
\def\rR{{\Bbb R}}
\def\zZ{{\Bbb Z}}

\newcommand{\sfJ}{{\mathsf J}}
\newcommand{\sfL}{{\mathsf L}}

\newcommand{\bl}{{\mathbf l}}

\newcommand{\bx}{{\mathbf x}}
\newcommand{\bX}{{\mathbf X}}
\newcommand{\by}{{\mathbf y}}
\newcommand{\bY}{{\mathbf Y}}

\newcommand{\Su}{S_{\mathrm u}}
\newcommand{\Tu}{T_{\mathrm u}}
\newcommand{\Vu}{V_{\mathrm u}}
\newcommand{\deltaNP}{\delta_{\textrm{NP}}}

\newcommand{\Var}{{\mathrm{Var}}}
\newcommand{\Cov}{{\mathrm{Cov}}}
\newcommand{\supp}{\mathrm{supp}}
\newcommand{\avg}{\mathrm{avg}}

\newcommand{\Ans}{A_\mathrm{ns}}
\newcommand{\Abns}{\breve{A}_\mathrm{ns}}
\newcommand{\Cin}{\mathcal{C}^\mathrm{in}}
\newcommand{\Cout}{\mathcal{C}^\mathrm{out}}
\newcommand{\Jcap}{{\mathsf{J}_\mathrm{cap}}}
\newcommand{\Xcap}{{\mathcal{X}_\mathrm{cap}}}
\newcommand{\CTin}{{\calC_{\calT_0}^\mathrm{in}}}
\newcommand{\CTout}{{\calC_\calT^\mathrm{out}}}

\newtheorem{definition}{Definition}[section] 
\newtheorem{remark}{Remark}[section] 
\newtheorem{theorem}{Theorem}[section] 
\newtheorem{lemma}[theorem]{Lemma} 
\newtheorem{proposition}[theorem]{Proposition}

\renewcommand{\theequation}{\arabic{section}.\arabic{equation}}

\begin{document} 
 
\bibliographystyle{ieeetr} 

\title{The Log-Volume of Optimal Codes for Memoryless Channels, Asymptotically Within A Few Nats}

\author{Pierre~Moulin 
\thanks{The author is with the ECE Department, the Coordinated Science Laboratory,
and the Beckman Institute at the University of Illinois at Urbana-Champaign,
Urbana, IL 61801, USA. This research was initiated during the author's
sabbatical stay at the Chinese University of Hong Kong (January-March 2010) and
supported by DARPA under the ITMANET program and by NSF under grant CCF 12-19145.
This work was presented in part at the DARPA ITMANET meeting in Austin, TX, May 2010,
at the ITA workshop in San Diego, CA, Feb. 2012,
at ISIT in Cambridge, MA, July 2012, and at ISIT in Istanbul, Turkey, July 2013.}}

\date{October 31, 2013. \\ Revised May 25, 2015, and June 11, 2016}

\maketitle

\begin{abstract}
Shannon's analysis of the fundamental capacity limits for memoryless communication
channels has been refined over time. In this paper,
the maximum volume $M_\avg^*(n,\epsilon)$ of length-$n$ codes subject to an average decoding error
probability $\epsilon$ is shown to satisfy the following tight asymptotic lower and upper bounds as $n \to \infty$:
\[ \underline{A}_\epsilon + o(1) \le \log M_\avg^*(n,\epsilon) 
	- [nC - \sqrt{nV_\epsilon} \,Q^{-1}(\epsilon) + \frac{1}{2} \log n] \le \overline{A}_\epsilon + o(1) \]
where $C$ is the Shannon capacity, $V_\epsilon$ the $\epsilon$-channel dispersion, 
or second-order coding rate, $Q$ the tail probability of the normal distribution,
and the constants $\underline{A}_\epsilon$ and $\overline{A}_\epsilon$ are explicitly identified. 
This expression holds under mild regularity assumptions on the channel, including nonsingularity.
The gap $\overline{A}_\epsilon - \underline{A}_\epsilon$ is one nat for weakly symmetric channels
in the Cover-Thomas sense, and typically a few nats for other symmetric channels, for the binary symmetric channel,
and for the $Z$ channel.
The derivation is based on strong large-deviations analysis and refined central limit asymptotics. 
A random coding scheme that achieves the lower bound is presented. The codewords are drawn from
a capacity-achieving input distribution modified by an $O(1/\sqrt{n})$ correction term.
\end{abstract}


\newpage

\section{Introduction}
\label{sec:intro}

Shannon's seminal paper \cite{Shannon48} introduced the fundamental capacity limits
for memoryless communication channels. For any channel code of length $n$ and tolerable
decoding error probability $\epsilon$, the maximum volume of the code
is given by $M^*(n,\epsilon) = e^{nC + o(n)}$. The $o(n)$ term is significant
for practical values of $n$, hence much effort went into characterizing it
in the early 1960's \cite{Weiss60,Wolfowitz61,Dobrushin61,Strassen62,Kemperman62}.
It was discovered that under regularity conditions, the $o(n)$ term is of the form
$- \sqrt{nV_\epsilon} \,Q^{-1}(\epsilon) + o(\sqrt{n})$
where $V_\epsilon$ is the $\epsilon$-channel dispersion, or second-order coding rate.
The $o(\sqrt{n})$ term was found to be $O(\log n)$ for discrete memoryless channels (DMCs), in a remarkable
paper by Strassen \cite{Strassen62} which pioneered the Neyman-Pearson (NP) hypothesis testing approach
to the source and channel coding converse as well as the use of refined asymptotics
for source and channel coding. This line of research seemed forgotten until its 
revival by Polyanskiy {\em et al.} \cite{Polyanskiy10,Polyanskiy10b} and Hayashi \cite{Hayashi09}.
Thus
\begin{equation}
   \log M^*(n,\epsilon) = nC - \sqrt{nV_\epsilon} \,Q^{-1}(\epsilon) + O(\log n)
\label{eq:expansion}
\end{equation}
subject to some regularity conditions on the channel law.
This holds under both the maximum and the average error probability criteria,
and the corresponding maximum code volumes are denoted by $M_{\max}^*(n,\epsilon)$ 
and $M_\avg^*(n,\epsilon)$, respectively. 

The appeal of asymptotic expansions such as (\ref{eq:expansion}) is that they convey
significant insights into the essence of the problem and they are practically useful.
Indeed the remainder $O(\cdot)$ term can be bounded and sometimes neglected for moderately
large values of $n$, as shown numerically in \cite{Polyanskiy10}.

The third-order term in (\ref{eq:expansion}) has been characterized in several studies. It is equal to 
$\frac{1}{2} \log n + O(1)$ for symmetric DMCs (see \cite[footnote~p.~692]{Strassen62}, which
discusses a result by Dobrushin \cite[Eqn. (75)]{Dobrushin61}) and to $O(1)$ for the binary erasure channel (BEC). 
For DMCs with finite input alphabet $\calX$ and output alphabet $\calY$, under regularity
assumptions on the channel law, the third-order term is sandwiched between $O(1)$ and
$[1+o(1)] (|\calX|-\frac{1}{2}) \log n$ \cite{Polyanskiy10}. If the DMC is nonsingular (has positive reverse
dispersion), the third-order term is lower-bounded by $\frac{1}{2} \log n + O(1)$ \cite[Sec.~3.4.5]{Polyanskiy10b}.
For the additive white Gaussian channel (AWGN) under an average (resp. maximum) power input constraint, 
the third-order term is sandwiched between $O(1)$ and $\frac{3}{2} \log n + O(1)$ 
(resp. $\frac{1}{2} \log n + O(1)$). More recent results on the third-order term, concurrent with the 2013 version of
this manuscript \cite{Moulin13b}, include a $\frac{1}{2} \log n + O(1)$ upper bound by Tomamichel and Tan \cite{Tomamichel13}
(using an $\epsilon$ hypothesis testing divergence approach and an auxiliary distribution that is a mixture
of product distributions over $\calY^n$) and an $O(1)$ upper bound by Altug and Wagner \cite{Altug13} for singular DMCs.
Haim {\em et al.} \cite{Haim13} recognized the importance of tie-breaking for ML decoding
on the BSC, which presumably affects asymptotics beyond the third-order term.
While the $\log n$ term is generally significant for moderate values of $n$,
the $O(1)$ term might be comparable to $\log n$ and is therefore of great interest as well
in finite-blocklength analyses.

The state of the art in 2010 motivated us to undertake a more refined analysis of the problem,
in which asymptotic {\em equalities} for the relevant error probabilities
are obtained using strong large-deviations analysis, which are
closely related to Laplace's method for asymptotic expansion of integrals \cite{Moulin10}---\cite{Moulin13}.
A strong large-deviations analysis provides an asymptotic expansion for
the probability of rare events such as $\{\sum_{i=1}^n U_i \ge na\}$ where
the random variables $U_i, \,1 \le i \le n$ are independent and identically distributed
(iid), and $a$ is strictly larger than the mean of $U_1$ \cite{Blackwell59,Bahadur60}. 
Under regularity conditions, the expansion is of the form
\[ \exp\left\{ - n\Lambda(a) - \frac{1}{2} \log n + \gamma(a) + o(1)\right\} \]
where $\Lambda(\cdot)$ denotes the large-deviations function for $U_1$ and $\gamma(\cdot)$ is a real-valued function.
In contrast, the ordinary (``weak'') large-deviations analysis
merely states that the aforementioned probability vanishes as $\exp\{ - n\Lambda(a) + o(n)\}$. \footnote{
	The view expressed in \cite[p.~2337]{Polyanskiy10}:
	``Inherently, the large deviations approach does not capture the subexponential behavior''
	applies only to the ``weak'' large-deviations analysis.}

Using this approach, we derive the following sharp asymptotic bounds (Theorem~\ref{thm:main}). 
For channels with finite $\calX$ and $\calY$ and positive dispersion, under regularity conditions, \footnote{
	In the original version of this paper \cite{Moulin13b}, stronger assumptions were given, including
	a unique capacity-achieving input distribution.}
\begin{eqnarray}
   \underline{A}_\epsilon + o(1) 
	& \le & \log M_\avg^*(n,\epsilon) - \left[ nC - \sqrt{nV_\epsilon} \,Q^{-1}(\epsilon) 
		+ \frac{1}{2} \log n \right] \nonumber \\
	& \le & \overline{A}_\epsilon + o(1) .
\label{eq:main}
\end{eqnarray}
Hence the third-order term in the asymptotic expansion is $\frac{1}{2} \log n$, and the fourth-order term 
can be lower- and upper-bounded by two constants $\underline{A}_\epsilon$ and $\overline{A}_\epsilon$ 
that are easily computable and are independent of $n$.
The constant gap $\overline{A}_\epsilon - \underline{A}_\epsilon$ between the lower and upper bounds 
of \eqref{eq:main} is equal to 1 nat for a class of symmetric channels. Moreover the gap is typically in the range 
1---3 nats for other symmetric channels and for the nonsymmetric Z channel. 
This gap represents the ``few nats'' mentioned in the title of this paper. 

The derivation of both the lower and upper bounds requires a refined analysis of the behavior of averages
of independent random variables in the central regime. This analysis is rooted in two-term
Edgeworth expansions \cite{Wallace58,Feller71,Barndorff91,DasGupta08} and more specifically
in work by Cram\'{e}r \cite{Cramer38} and Ess\'{e}en \cite{Esseen45} during the 1930's and 1940's.
The Berry-Ess\'{e}en theorem that was used by Strassen \cite{Strassen62},
Polyanskiy {\em et al.} \cite{Polyanskiy10}, and Hayashi \cite{Hayashi09} provides a bound for deviation from
Gaussianity but is not sharp enough for our purpose.

To obtain the lower bound in \eqref{eq:main}, we apply strong large deviations to analyze 
the error probability of random coding schemes with maximum-likelihood (ML) decoding. 
The only inequality used in this analysis is the classical union-of-events bound, 
which turns out to be remarkably tight.

To obtain the upper bound in \eqref{eq:main}, we use the technique
introduced by Polyanskiy {\em et al.} \cite{Polyanskiy10} for proving converse theorems:
it provides upper bounds for $M_{\max}^*(n,\epsilon)$ and $M_{\avg}^*(n,\epsilon)$ in terms of maxmin
optimization problems whose payoff function is the type-II error probability of a NP test
at significance level $1-\epsilon$. This is a powerful idea which extends
the NP-testing paradigm of Strassen \cite[pp.~711, 712]{Strassen62}
and Kemperman \cite[Theorem~7.1]{Kemperman62}. A key step in deriving our converse is to solve
an asymptotic Neyman-Pearson game $\inf_P \sup_Q \beta_n(P,Q)$ where $P$ ranges over an
$o(n^{-1/4})$ enlargement of the set of $\epsilon$-dispersion-achieving input distributions,
and $Q$ is a distribution on $\calY$. 
We exploit this result to derive asymptotic upper bounds on $M_{\max}^*(n,\epsilon)$ and $M_\avg^*(n,\epsilon)$; 
we show that these asymptotic upper bounds are in fact {\em identical}. Prior work gave
the inequality $M_{\avg}^*(n,\epsilon) \le \sqrt{n} M_{\max}^*(n,\epsilon(1+n^{-1/2}))$
\cite[p.~2332]{Polyanskiy10}.

The paper is divided into nine sections and nine appendices and is organized as follows. 
Sec.~\ref{sec:prelim} introduces the notation, basic definitions,
and variational properties of information functions.
Sec.~\ref{sec:main} states the main result and applies it to a class of symmetric channels
and to the (asymmetric) Z channel. Sec.~\ref{sec:optimization}
presents several optimization lemmas that are used throughout the paper.
Sec.~\ref{sec:asymptotics} reviews exact asymptotics in the central limit regime,
in particular the Cram\'{e}r-Ess{\'{e}en} theorem, and strong large deviations
asymptotics, with an emphasis on the general approach by Chaganty and Sethuramanan \cite{Chaganty93}.
Sec.~\ref{sec:strongLD-HT} presents new results on strong large deviations
for likelihood ratios, including the case of varying component distributions (Sec.~\ref{sec:LR-vary}),
and exact asymptotics of NP tests (Sec.~\ref{sec:NP}).
The lower bound on the log-volume of optimal codes is obtained via a random-coding
argument in Sec.~\ref{sec:achieve}. The upper bound is obtained
via a strong converse in Sec.~\ref{sec:converse}. The paper concludes with a discussion
in Sec.~\ref{sec:discuss}. For ease of reading, most proofs are presented in the appendices.
The technical material relevant to the achievability results may be found in Sec.~\ref{sec:intro}, 
Lemma~\ref{lem:zetan-P}, and Secs.~\ref{sec:CLT}, and \ref{sec:achieve}. 
The material relevant to the converse includes everything but Sec.~\ref{sec:achieve}.

\section{Preliminaries}
\label{sec:prelim}
\setcounter{equation}{0}

\subsection{Notation}

We use uppercase letters for random variables, lowercase letters for their individual values, 
calligraphic letters for alphabets, boldface letters for sequences, and sans serif fonts for matrices. 
The set of all probability distributions on $\calX$ is denoted by $\scrP(\calX)$.
The support of a distribution $P$ is denoted by $\supp\{P\}$.
If $\calX$ is discrete, $P$ is a probability mass function (pmf).
Mathematical expectation, variance, and covariance with respect to $P$
are denoted by the symbols $\eE_P$, $\Var_P$, and $\Cov_P$, respectively.
The probability density function (pdf)
of the normal random variable is denoted by $\phi(x),\,x\in\rR$ and the cumulative
distribution function (cdf) by $\Phi(x) = 1-Q(x)$.

The indicator function of a set $\calA$ is denoted by $\mathds1\{x \in \calA\}$.
All logarithms are natural logarithms. The symbol $\mathrm{Proj}_\calV (\cdot)$ denotes
orthogonal projection onto a vector space $\calV$.

The symbol $f(n) \sim g(n)$ denotes asymptotic equality:
$\lim_{n \rightarrow \infty} \frac{f(n)}{g(n)} = 1$.
The notation $f(n) = o(g(n))$ (small oh, also denoted by $f(n) \ll g(n)$) indicates that
$\lim_{n \rightarrow \infty} \frac{f(n)}{g(n)} = 0$. The notation $f(n) = O(g(n))$ indicates that 
$\limsup_{n \rightarrow \infty} |\frac{f(n)}{g(n)}| < \infty$.
In case the sequences $f(n)$ and $g(n)$ depend on a variable $P$ lying in a set $\scrP_n$ (possibly 
dependent on $n$), the same symbols are used to denote uniform convergence over $\{\scrP_n\}_{n \ge 1}$, 
e.g., $f(P,n) = o(g(P,n))$ indicates that 
$\lim_{n \rightarrow \infty} \sup_{P\in\scrP_n} \frac{f(P,n)}{g(P,n)} = 0$.

Consider a DMC $(W,\calX,\calY)$ with input and output alphabets
$\calX$ and $\calY$ respectively and conditional pmf 
$W(\cdot|x) \in \scrP(\calY)$, also denoted by $W_x(\cdot)$, for each $x\in\calX$.
Given an input pmf $P$ on $\calX$, denote by $P \times W$
the joint pmf on $\calX \times \calY$ and by $(PW)$ the marginal output pmf, i.e., 
\[
  (PW)(y) \triangleq \sum_{x\in\calX} P(x) W(y|x) \quad \forall y\in\calY .
\]

Kullback-Leibler (KL) divergence between two distributions $P$ and $Q$ on a common alphabet
is denoted by $D(P\|Q) \triangleq \eE_P [ \log \frac{dP}{dQ} ]$, divergence variance by
$V(P\|Q) \triangleq \eE_P [ \log \frac{dP}{dQ} ]^2 - D^2(P\|Q)$, and divergence centralized third moment by
$T(P\|Q) \triangleq \eE_P [ \log \frac{dP}{dQ} - D(P\|Q)]^3$. R\'{e}nyi divergence of order $\alpha \ne 1$ 
is denoted by $D_\alpha(P\|Q) = \frac{1}{\alpha-1} \log \eE_Q [\frac{dP}{dQ}]^\alpha$;
recall that $\lim_{\alpha \to 1} D_\alpha(P\|Q) = D(P\|Q)$. 

Given two alphabets $\calX$ and $\calY$, a $\calX$-valued random variable $X$ distributed as $P$,
and two conditional distributions $W$ and $Q$ on a $\calY$-valued random variable $Y$ given $X$,
denote by $D(W\|Q|P) = \eE_{P \times W} \left[ \log \frac{W(Y|X)}{Q(Y|X)} \right]$ the conditional KL divergence
between $W$ and $Q$ given $P$, and likewise by $V(W\|Q|P) = \eE_P [V(W_X\|Q)]$ and $T(W\|Q|P) = \eE_P [T(W_X\|Q)]$
the conditional divergence variance and the conditional divergence third central moment. 
Conditional R\'{e}nyi divergence of order $\alpha \ne 1$ is defined as
$D_\alpha(W\|Q|P) \triangleq \eE_P [D_\alpha(W_X\|Q)]$ \cite{Csiszar95}.

\subsection{Definitions}

We review some standard definitions.
A real random variable $L$ is said to be of the {\em lattice type} if there exist
numbers $d$ and $l_0$ such that $L$ belongs to the lattice $\{ l_0 + kd, \,k\in\zZ \}$
with probability 1. The largest $d$ for which this holds is called the {\em span}
of the lattice, and $l_0$ is the {\em offset}. The sum of a nonlattice random variable and an independent
lattice random variable is a nonlattice random variable. The sum of two independent lattice random variables 
is a lattice random variable if and only if the ratio of their spans is a rational number.
For each $d > 0$, define the discrete set
\begin{equation}
   \scrP_\textrm{lat}(d) \triangleq \left\{ Q \in \scrP(\calY) ~:~ \log \frac{W(Y|X)}{Q(Y)} 
	\mathrm{~is~a~lattice~rv~with~span~} d \right\} .
\label{eq:P-lat}
\end{equation}
The union of these sets over all $d > 0$ forms a sieve for $\scrP(\calY)$, where $d$ is the mesh size.

The information density is defined as
\begin{equation}
   i(x,y) = \log \frac{W(y|x)}{(PW)(y)} , \quad x\in\calX, y\in\calY .
\label{eq:i}
\end{equation}
For some DMCs with capacity-achieving input distribution, the random variable $i(X,Y)$
is of the lattice type. For instance $i(X,Y) \in \{ \log (2\theta), \,\log (2-2\theta) \}$
for the BSC with crossover probability $\theta \ne \{0, \frac{1}{2}, 1\}$ and uniform
input distribution; the span of the lattice is $\log | \frac{\theta}{1-\theta}|$. 
However for almost every {\em asymmetric} binary channel,
as well as for almost every {\em nonbinary} channel (symmetric or not), $i(X;Y)$ is not of the lattice type. 
We refer to capacity problems where $i(X,Y)$ is of the lattice type as the {\em lattice case}.

The following moments of the random variable $i(X,Y)$ with respect to
the joint distribution $P \times W$ are used throughout this paper:
the {\em unconditional} mean (= mutual information)
\begin{eqnarray}
   I(P;W) = \eE_{P \times W} [i(X,Y)] 
	& = & D(P \times W \| P \times (PW)) , 
\label{eq:mi}
\end{eqnarray}
the {\em conditional} mean (given $X=x$)
\[ D(W_x\|(PW)) = \eE_{W_x}[i(x,Y)] , \quad x\in \calX , \]
the {\em unconditional} information variance
\begin{eqnarray}
   \Vu(P;W) = \Var_{P \times W} [i(X,Y)] 
	& = & V(P \times W \| P \times (PW)) ,
\label{eq:U}
\end{eqnarray}
the {\em conditional} information variance (given $X$)
\begin{eqnarray}
   V(P;W) & = & \eE\{ \Var[i(X,Y)|X]\} = \sum_{x\in\calX} P(x) V(W_x \| PW) ,
\label{eq:V}
\end{eqnarray}
the {\em reverse dispersion} \cite[p.~89]{Polyanskiy10}
\begin{equation}
    V^r(P;W) =  \eE\{\Var[i(X,Y)|Y]\}  ,
\label{eq:Vr}
\end{equation}
the {\em unconditional} third central moment
\begin{eqnarray}
   \Tu(P;W) & = & T(P \times W \| P \times (PW)) ,
\label{eq:thirdVu}
\end{eqnarray}
the {\em conditional} third central moment (given $X$)
\begin{eqnarray}
   T(P;W) & = & \sum_{x\in\calX} P(x) T(W_x \| PW) ,
\label{eq:thirdV}
\end{eqnarray}
the {\em unconditional} skewness
\begin{equation}
   \Su(P;W) = \frac{\Tu(P;W)}{[\Vu(P;W)]^{3/2}} ,
\label{eq:u-skew}
\end{equation}
and the {\em conditional} skewness
\begin{equation}
   S(P;W) = \frac{T(P;W)}{[V(P;W)]^{3/2}} .
\label{eq:skew}
\end{equation}
It follows from these definitions that
\begin{equation}
   V(P;W) = V(W\|(PW)|P) \quad \mathrm{and} \quad T(P;W) = T(W\|(PW)|P) .
\label{eq:VPW-PQ}
\end{equation}
Define the shorthand $t_\epsilon = Q^{-1}(\epsilon)$.
The following functions will also be used throughout:
\begin{equation}
   F_\epsilon(W\|Q|P) \triangleq \frac{1}{2} t_\epsilon^2 - \frac{1}{6} \frac{T(W\|Q|P)}{V(W\|Q|P)} (t_\epsilon^2 - 1) 
		+ \frac{1}{2} \log (2\pi V(W\|Q|P))
\label{eq:FQ}
\end{equation}
and
\begin{equation}
  \zeta_n(P,Q) \triangleq n D(W\|Q|P) - \sqrt{n V(W\|Q|P)} \,t_\epsilon + F_\epsilon(W\|Q|P)
\label{eq:zeta-PQ}
\end{equation}
where $P \in \scrP(\calX)$ and $Q \in \calQ \triangleq \{ Q=(PW):~P \in \scrP(\calX)\} \subset \scrP(\calY)$.
Evaluating \eqref{eq:FQ} and \eqref{eq:zeta-PQ} at $Q = (PW)$ and using \eqref{eq:VPW-PQ}, we obtain the functions
\begin{equation}
   F_\epsilon(P;W) \triangleq \frac{1}{2} t_\epsilon^2 - \frac{1}{6} \frac{T(P;W)}{V(P;W)} (t_\epsilon^2 - 1) 
		+ \frac{1}{2} \log (2\pi V(P;W))
\label{eq:F}
\end{equation}
and
\begin{eqnarray}
  \zeta_n(P;W) & \triangleq & nI(P;W) - \sqrt{nV(P;W)} \,t_\epsilon + F_\epsilon(P,W) \nonumber \\
	& = & \zeta_n(P,(PW)) . \label{eq:zetaPQ-PW} 
\label{eq:zeta-PW}
\end{eqnarray}

Shannon capacity is given by $C = \sup_{P \in \scrP(\calX)} I(P;W)$.
Denote by $Q^* \in \scrP(\calY)$ the capacity-achieving output distribution,
which is unique. Define the set of capacity-achieving input distributions
\begin{equation}
   \Pi \triangleq  \{ P\in\scrP(\calX) :~(PW) = Q^* \}
\label{eq:Pi}
\end{equation}
and the vector $\Delta$ with components
\begin{equation}
   \Delta(x) = D(W_x\|Q^*) - C , \quad x\in\calX .
\label{eq:Delta}
\end{equation}
Let $\Xcap \triangleq \{x\in\calX :~\Delta(x)=0\}$.
It is well known that $\Delta(x) \le 0$ for all $x\in\calX$ and that $\mathrm{supp}(P) \subseteq \Xcap$
is a necessary condition for $P\in\Pi$.
With a slight abuse of notation, we define $\scrP(\Xcap) \triangleq \{P\in\scrP(\calX) : \mathrm{supp}(P) \subseteq \Xcap\}$.

For $P\in\Pi$, we have $V(P;W) = \sum_x P(x) \tilde{v}(x)$ where $\tilde{v}(x) \triangleq V(W_x\|Q^*)$. Moreover
the gradient of $V(P;W)$, for $P\in\Pi$, is given by $\frac{\partial V(P;W)}{\partial P(x)} = \tilde{v}(x)$, $x\in\calX$.
Define
\begin{equation}
   V_{\min} \triangleq \min_{P\in\Pi} V(P;W) , \quad V_{\max} \triangleq \max_{P\in\Pi} V(P;W) ,
\label{eq:VminVmax}
\end{equation}
the $\epsilon$ channel dispersion
\begin{equation}
   V_\epsilon = \begin{cases}  V_{\min} & : \,\epsilon \le \frac{1}{2} \\  V_{\max} & :\,\mathrm{else,} \end{cases}
\label{eq:Veps}
\end{equation}
and the subset of $\epsilon$-dispersion-achieving distributions
\begin{equation}
   \Pi^* \triangleq  \left\{ P\in\Pi :~\sum_{x\in\calX} P(x) \tilde{v}(x) = V_\epsilon \right\} 
	: \quad \epsilon \ne \frac{1}{2}
\label{eq:Pi*} 
\end{equation}
and $\Pi^* = \Pi$ if $\epsilon = \frac{1}{2}$.
Both $\Pi$ and $\Pi^*$ are convex polytopes. For $\epsilon \ne \frac{1}{2}$, the set $\Pi^*$ is generally a singleton, 
except on a set of channels of measure zero.
Any dispersion-achieving input distribution has support in the following subset of $\Xcap$:
\begin{equation}
   \calX^* \triangleq \cup_{P\in\Pi^*} \mathrm{supp}(P) .
\label{eq:X*}
\end{equation}
We will extensively use the following $\delta$-enlargements of the sets $\Pi$ and $\Pi^*$:
\begin{eqnarray}
   \Pi(\delta) & \triangleq & \{P\in\scrP(\calX) : \min_{P'\in\Pi} \|P-P'\|_\infty \le \delta \} , \label{eq:Pi-delta} \\
   \Pi^*(\delta) & \triangleq & \{P\in\scrP(\calX) : \min_{P'\in\Pi^*} \|P-P'\|_\infty \le \delta \} , \quad \delta > 0 . \label{eq:Pi*-delta}
\end{eqnarray}

Define the {\em reverse channel} with input distribution $(PW)$ and output distribution $P$ as
\begin{equation}
  \widecheck{W}(x|y) \triangleq \frac{W(y|x) \,P(x)}{(PW)(y)} , \quad x\in\calX, \,y\in\calY .
\label{eq:Wreverse}
\end{equation}
Also define
\begin{equation}
   \rho(P;W) \triangleq 1 - \frac{V^r(P;W)}{\Vu(P;W)} ,
\label{eq:rho}
\end{equation}
which will be interpreted in Sec.~\ref{sec:achieve} as a normalized correlation coefficient. 
For now, note that $\rho(P;W) \in [0,1]$ as a consequence of the law of total variance:
\begin{eqnarray}
    \Var[i(X;Y)] & = & \eE\{ \Var[i(X;Y)|Y]\} + \Var[\eE(i(X;Y)|Y)] \nonumber \\
    \Rightarrow \quad \Vu(P;W) & = & V^r(P;W) + \Var_{(PW)} [D(\widecheck{W}_Y \| P)] .
\label{eq:rho2}
\end{eqnarray}

A DMC $(W,\calX,\calY)$ is said to be weakly symmetric in the Cover-Thomas
(CT) sense if the rows $W_x(\cdot)$ of the channel transition probability matrix are permutations
of each other and all the column sums are equal.
See \cite{Xie08} for a detailed discussion of symmetric channels.

The empirical distribution on $\calX$ ($n$-type) of a sequence $\bx \in \calX^n$ is defined by
\begin{equation}
  \hat{P}_\bx(x) \triangleq \frac{1}{n} \sum_{i=1}^n \mathds1\{x_i = x\}, \quad x \in \calX .
\label{eq:P-emp}
\end{equation}
We denote by $\scrP_n(\calX) \triangleq \{ \hat{P}_\bx , \,\bx\in\calX^n\}$ the set of $n$-types.

\subsection{Encoder, Decoder, and Error Probabilities}

The message $m$ to be transmitted is drawn uniformly from the message set $\calM_n = \{1,2,\cdots,M_n\}$.
A deterministic code is a pair of encoder mapping $f_n ~:~ \calM_n \to {\mathsf F} \subset \calX^n$, $\bx(m)=f_n(m)$,
and decoder mapping $g_n ~:~ \calY^n \to \calM_n$, $\hat{m} = g_n(\by)$. 
The code has volume (or size) $M_n$ and rate $R_n \triangleq \frac{1}{n} \log M_n$.
More generally, we assume a stochastic encoder where the codeword assigned to message $m$ is drawn from
a conditional probability distribution $P_{\bX|M=m}$, and a stochastic decoder whose output
$\widehat{M} \in \calM_n$ is drawn from a conditional distribution $P_{\widehat{M}|\bY}$.

The error probability for message $m \in \calM_n$ is given by
\[ e(m) \triangleq \sum_{\bx\in\calX^n} P_{\bX|M}(\bx|m) \sum_{\by\in\calY^n} W^n(\by|\bx) [1- P_{\widehat{M}|\bY}(m|\by)] . \]
Both the {\em average error probability} 
\[ P_{e,\avg}(f_n,g_n,W) \triangleq \frac{1}{M_n} \sum_{m\in\calM_n} e(m) \]
and the {\em maximum error probability}
\[ P_{e,\max}(f_n,g_n,W) \triangleq \max_{m\in\calM_n} e(m) \]
are considered in this paper. For $n \ge 1$ and $\epsilon \in (0,1)$, denote by
\[ M_{\avg}^*(n,\epsilon) \triangleq \max \{ M_n ~:~ \exists (f_n,g_n) ~:~ P_{e,\avg}(f_n,g_n,W) \le \epsilon \} \]
and
\[ M_{\max}^*(n,\epsilon) \triangleq \max \{ M_n ~:~ \exists (f_n,g_n) ~:~ P_{e,\max}(f_n,g_n,W) \le \epsilon \} \]
the maximum possible value of $M_n$ for arbitrary $(n,\epsilon)$ codes under the average and maximum
error probability criteria, respectively. We will also consider constant-composition codes, 
which are particularly useful in problems with maximum-cost constraints on the codewords \cite{Moulin12b}.

\subsection{Fisher Information Matrix}

Since $I(P;W)$ is a concave function of $P$, its Hessian is nonnegative definite.
Moreover, if $P^*$ is the unique maximizer of $I(\cdot;W)$, the Hessian at $P^*$ is positive definite.
The gradient of $I(P;W)$ is given by
\begin{equation}
   \frac{\partial I(P;W)}{\partial P(x)}  = D(W_x\|(PW)) - 1 , \quad x\in\calX
\label{eq:del-I}
\end{equation}
and the negative Hessian by
\begin{eqnarray}
   J_{xx'}^\calX (P,W) \triangleq - \frac{\partial^2 I(P;W)}{\partial P(x) \partial P(x')}  
	& = & - \frac{\partial D(W_x\|(PW))}{\partial P(x')}  \nonumber \\
    & = & \sum_{y\in\calY} \frac{W(y|x) \,W(y|x')}{(PW)(y)} , \quad \forall x,x' \in\calX .
\label{eq:fisher}
\end{eqnarray}
This expression was already derived by Strassen \cite[below (4.43)]{Strassen62} and can be interpreted as the Fisher information matrix 
for the probability vector $P$ given $Y$ \cite{Moulin13b}.
The matrix $J^\calX (P;W)$ is symmetric and nonnegative definite, and factorizes as $J^\calX (P;W) = \sfL \sfL^\top$ where 
\begin{equation}
   \sfL(x,y) = \frac{W(y|x)}{\sqrt{(PW)(y)}}, \quad x\in\calX, y\in\calY .
\label{eq:Ltheta}
\end{equation}
Moreover
\begin{equation}
   \sum_{x\in\calX} P(x') J_{xx'}^\calX (P,W) = 1 , \quad \forall x \in \calX .
\label{eq:JP}
\end{equation}

The matrix $J^\calX (P;W)$ is the same for all $P\in\Pi$, and we denote it by $\sfJ^\calX$ for short.
However, only its entries indexed by $x,x'\in\calX^*$ of \eqref{eq:X*} will affect the fourth-order term. \footnote{
	In Sec.~\ref{sec:optimization}, we also use the entries indexed by $x,x'\in\Xcap$.}
We therefore define a $|\calX| \times |\calX|$ matrix $\sfJ \succeq 0$ as follows. If $\calX=\calX^*$, then $\sfJ = J^\calX$.
Else $\sfJ$ coincides with $J^\calX (P;W)$ for $x,x'\in\calX^*$ and is zero elsewhere. 
Therefore $\mathrm{rank}(\sfJ) \le |\calX^*|$. Now define the following subspaces of $\rR^\calX$:
\begin{eqnarray}
   \scrL(\calX) & \triangleq & \{ h \in \rR^{\calX} : \sum_{x\in\calX} h(x) = 0\} , \label{eq:LX} \\
   \scrL(\calX^*) & \triangleq & \{ h \in \scrL(\calX) : \mathrm{supp}(h) \subseteq \calX^*\} . \label{eq:LX*}
\end{eqnarray}
We have
\begin{equation}
  h^\top \sfJ^\calX h = \eE_{Q^*} \left( \frac{(hW)(Y)}{Q^*(Y)} \right)^2  \ge 0 , \quad \forall h \in \scrL(\calX)
\label{eq:hJ2}
\end{equation}
where $(hW)(y) \triangleq \sum_{x\in\calX} h(x) W(y|x)$.
We refer to $\|h\|_\sfJ \triangleq \sqrt{h^\top \sfJ h}$, $h \in \scrL(\calX)$
as the {\em Fisher information metric} associated with $\{W(y|x)\}_{x\in\calX^*,y\in\calY}$, 
and to $h^\top \sfJ \tilde{h}$, $h, \tilde{h} \in \scrL(\calX)$, as the Fisher-Rao inner product.

It also follows from the factorization (\ref{eq:Ltheta}) that $\mathrm{rank}(\sfJ) \le \min(|\calX^*|, \,|\calY|)$. 
For instance, the noisy typewriter channel of \cite{Cover06} has $|\calX| = |\Xcap| = |\calX^*| = |\calY| = 26$
and rank($\sfJ$) = 25.

\subsection{Pseudo-Inverse} 

We define a pseudo-inverse matrix $\sfJ^+$ as follows. Assume $\sfJ$ has rank $r \ge 2$ 
and define the $r-1$ dimensional subspace
\begin{equation}
   \calH = \scrL(\calX^*) \cap \mathrm{ker}(\sfJ)^\perp .
\label{eq:H}
\end{equation}
Define the function
\begin{equation}
  \calE(g) \triangleq \sup_{h\in\calH} \left[ g^\top  h - \frac{1}{2} \|h\|_\sfJ^2 \right] , \quad g \in \rR^\calX
\label{eq:J+}
\end{equation}
which is the convex conjugate transform of the quadratic function $\frac{1}{2} \|h\|_\sfJ^2$.
The function depends on $g$ only via its projection onto $\calH$,
and the supremum is achieved by a unique $h=\sfJ^+ g \in \calH$.
Some elementary properties of the matrix $\sfJ^+$ are given below.

\begin{lemma}
The pseudo-inverse matrix $\sfJ^+$ satisfies the following properties:
\begin{description}
\item[(i)] $\sfJ^+$ is symmetric and nonnegative definite.
\item[(ii)] $\sfJ \sfJ^+ g = \sfJ^+ \sfJ g = g$ for all $g\in\calH$.
\item[(iii)] The supremum over $h\in\calH$ is achieved at 
   \begin{equation}
     \calE(g) = \frac{1}{2} \|g\|_{\sfJ^+}^2 \triangleq \frac{1}{2}  g^\top \sfJ^+ g .
	 \label{eq:normJ+}
   \end{equation}
\item[(iv)] If $\sfJ$ is invertible, then
	\begin{equation}
	   \sfJ^+ = \sfJ^{-1} - \frac{1}{1^\top \sfJ^{-1} 1} (\sfJ^{-1} 1)(\sfJ^{-1} 1)^\top
	\label{eq:J+finiteX}
	\end{equation}
	where $1 \in \rR^{|\calX|}$ is the all-one vector.
\item[(v)] If $\sfJ$ has rank $r$ and admits the eigenvector decomposition $\sfJ = \sum_{i=1}^r \lambda_i u_i u_i^\top$
	with positive eigenvalues $\lambda_i$ and associated eigenvectors $u_i \in \rR^\calX$, then
	\[ \sfJ^+ = \sum_{i=1}^r \lambda_i^{-1} u_i u_i^\top - \frac{1}{\sum_{i=1}^r \lambda_i^{-1}} v v^\top \]
	where $v = \sum_{i=1}^r \lambda_i^{-1} u_i$.
\end{description}
\label{lem:J+}
\end{lemma}

\subsection{Derivatives of Information Variance Functionals}

Define the following three vectors, which will be of interest for $P\in\Pi$:
\begin{eqnarray}
  v^{(P)}(x) & \triangleq & \frac{\partial V(P;W)}{\partial P(x)} , \label{eq:v} \\
  \tilde{v}(x) & \triangleq & V(W_x\|Q^*) \label{eq:vtilde} \\
  \breve{v}^{(P)}(x) & \triangleq &  \eE_P \left[ \frac{\partial V(W_X\|PW)}{\partial P(x)} \right] ,
	\quad x\in\calX . \label{eq:vbreve}
\end{eqnarray}
We also define 
\begin{equation}
   \Ans^{(P)} \triangleq \frac{1}{V_\epsilon} \|v^{(P)}\|_{\sfJ^+}^2 \quad \mathrm{and} \quad
   \Abns^{(P)} \triangleq \frac{1}{V_\epsilon} \|\breve{v}^{(P)}\|_{\sfJ^+}^2 , \quad P \in \Pi^* . 
\label{eq:Ans}
\end{equation}
As we shall see in Prop.~\ref{prop:symmetric}, these quantities are zero for CT weakly symmetric channels, 
hence the subscript ``ns''. 

\begin{lemma}
{\bf (i)} The vector-valued functions $\breve{v}^{(P)}$ and $v^{(P)}$, $P\in\Pi$ are respectively linear and affine over $\Pi$:
\begin{eqnarray}
   \breve{v}^{(P)}(x) & = & - 2 \,\Cov_{P \times W} \left( \left. 
		\frac{W_x(Y)}{Q^*(Y)} , \log \frac{W(Y|X)}{Q^*(Y)} \right| X \right) , \label{eq:vbreve-lemma} \\
   v^{(P)}(x) & = & \tilde{v}(x) + \breve{v}^{(P)}(x) . \label{eq:v-sum}
\end{eqnarray}
Moreover 
\[ \eE_P [\breve{v}^{(P)}(X)] = 0, \quad \eE_P [\tilde{v}(X)] = \eE_P [v^{(P)}(X)] = V_\epsilon . \]
{\bf (ii)} $\mathrm{Proj}_{\mathrm{ker}(\sfJ)}(v^{(P)}) = \mathrm{Proj}_{\mathrm{ker}(\sfJ)}(\tilde{v})$ for all $P\in\Pi$.

\noindent
{\bf (iii)} For any $P\in\Pi$,
\begin{eqnarray}
   \Vu(P) = V(P) , \label{eq:Vu=V} \\
   \Tu(P) = T(P) , \label{eq:Tu=T} \\
   \Su(P) = S(P) , \label{eq:Su=S}
\end{eqnarray}
and
\begin{equation}
   \frac{\partial [\Vu(P;W)-V(P;W)]}{\partial P(x)} = 2 I^2(P;W) , \quad x\in\calX .
\label{eq:delU=delV}
\end{equation}
Moreover
\begin{eqnarray}
   \breve{v}^{(P)}(x) & = & - 2 \,\Cov_{P \times W} \left( \frac{W_x(Y)}{Q^*(Y)} , \log \frac{W(Y|X)}{Q^*(Y)} \right) \nonumber \\
	& = & -2 [\eE_{W_x} D(\widecheck{W}_Y \| P) - C] , \quad x \in \calX .
\label{eq:vbreve2-lemma}
\end{eqnarray}
\label{lem:UV}
\end{lemma}
The identity \eqref{eq:Vu=V} was already established in \cite[Eqn.~(456)]{Polyanskiy10}.
The other statements are proved in Appendix~\ref{app:UV}.

\section{Main Result}
\label{sec:main}
\setcounter{equation}{0}

This section states our basic assumptions (Sec.~\ref{sec:assumptions}) and presents the main result 
(Sec.~\ref{sec:MainResult}) and several applications (Secs.~\ref{sec:sym}, \ref{sec:Z}, and \ref{sec:BSC}).

\subsection{Assumptions}
\label{sec:assumptions}

Consider the following three regularity conditions: 
\begin{description}
\item[(A1)] $V_{\min} > 0$. 
\item[(A2)] $\exists P\in\Pi^*$ such that $\rho(P;W) < 1$.
\item[(A3)] $\log \frac{W(Y|X)}{Q^*(Y)}$ is not a lattice random variable.
\end{description}

Assumption {\bf (A1)} excludes {\em exotic} channels, which have zero information variance.
The assumption corresponds to Cases~I and III in \cite[Theorem~48]{Polyanskiy10}; the extension
of our proof to cases II and IV of \cite{Polyanskiy10} is straightforward but tedious, and is omitted here.
In view of \eqref{eq:rho}, Assumption {\bf (A2)} is equivalent to the existence of a distribution $P\in\Pi^*$
with positive reverse dispersion $V^r(P;W) $.
The regularity assumptions {\bf (A1)} and {\bf (A2)} are not merely technical, as different asymptotic
behaviors can be observed when either assumption not satisfied \cite{Polyanskiy10,Altug13}.
For instance,  binary erasure channels do not satisfy {\bf (A2)}, 
and indeed the $\frac{1}{2} \log n$ term does not appear in the asymptotics of the log volume
\cite[Theorem~53]{Polyanskiy10}. Finally, Assumption {\bf (A3)} is introduced solely to avoid some technicalities.
This assumption does not hold for the BSC as well as for some pathological channels (see Sec.~\ref{ssec:BITO}). 
In view of the importance of the BSC, we derive its fundamental performance limits separately, in Sec.~\ref{sec:BSC}.

\subsection{Main Result}
\label{sec:MainResult}

Recall the definition of $\Ans^{(P)}$ and $\Abns^{(P)}$ in \eqref{eq:Ans}.
Define the shorthand $t_\epsilon \triangleq Q^{-1}(\epsilon)$ and the constants
\begin{eqnarray}
  \overline{A}_\epsilon & \triangleq & \max_{P\in\Pi^*} \left\{ \frac{t_\epsilon^2}{8} (\Ans^{(P)} - \Abns^{(P)}) 
	- \frac{S(P) \sqrt{V_\epsilon}}{6} (t_\epsilon^2 - 1) 
	+ \frac{t_\epsilon^2}{2} + \frac{1}{2} \log (2\pi V_\epsilon) \right\} , \label{eq:Aeps+} \\
  \underline{A}_\epsilon & \triangleq & \max_{P\in\Pi^*} \left\{ \frac{t_\epsilon^2}{8} \Ans^{(P)} 
	- \frac{S(P) \sqrt{V_\epsilon}}{6} (t_\epsilon^2 - 1) 
	+ \frac{t_\epsilon^2}{2} \frac{1-\rho(P)}{1+\rho(P)} + \frac{1}{2} \log (2\pi V_\epsilon) + \log \sqrt{1-\rho(P)^2} - 1 \right\} \nonumber \\
\label{eq:Aeps-}
\end{eqnarray}
where both maxima are achieved because $\Pi^*$ is a closed set, and the maximands are bounded over $\Pi^*$.

\begin{theorem} (Achievability).
Assume the regularity conditions {\bf (A1)}, {\bf (A2)}, and {\bf (A3)} hold. Then
\[ \log M_\avg^*(n,\epsilon) \ge nC - \sqrt{nV_\epsilon} \,Q^{-1}(\epsilon) 
		+ \frac{1}{2} \log n + \underline{A}_\epsilon + o(1) . 
\]
\label{thm:direct}
\end{theorem}
\begin{theorem} (Converse).
Assume the regularity conditions {\bf (A1)} and {\bf (A3)} hold. Then
\[ \log M_\avg^*(n,\epsilon) \le nC - \sqrt{nV_\epsilon} \,Q^{-1}(\epsilon) 
		+ \frac{1}{2} \log n + \overline{A}_\epsilon + o(1) . \]
\label{thm:converse}
\end{theorem}

\begin{theorem} 
Assume {\bf (A1)}, {\bf (A2)}, and {\bf (A3)} hold.
Then \eqref{eq:main} holds.
\label{thm:main}
\end{theorem}
Theorem~\ref{thm:direct} is proved in Sec.~\ref{sec:achieve} using a random coding argument, and
Theorem~\ref{thm:converse} is proved in Sec.~\ref{sec:converse}. 
Theorem~\ref{thm:main} follows. In the remainder of this section, we present several applications of 
Theorem~\ref{thm:main}.

\subsection{Symmetric Channels}
\label{sec:sym}

We specialize Theorem~\ref{thm:main} to the case of CT weakly symmetric channels (Prop.~\ref{prop:symmetric})
and apply the result to additive-noise DMCs. An example of Gallager-symmetric channel is also given.

\subsubsection{CT Weakly Symmetric Channels}
\label{sec:CTWS}

The first two properties in Prop.~\ref{prop:symmetric} below are well known \cite{Xie08}, and the next five 
follow immediately from the definitions of the various information functions and the fact that 
the likelihood-ratio vectors $W_x(\cdot)/Q^*(\cdot)$, $x\in\calX$ are permutation-invariant 
for CT weakly symmetric channels.
\begin{proposition}
Assume the channel $W$ is CT weakly symmetric, $|\calX| = |\calY|$, and $\sfJ > 0$. Then the following hold:
\begin{description}
\item[(i)] The capacity-achieving input distribution $P^*$ is unique and uniform over $\calX$.
\item[(ii)] The output distribution $Q^*$ is uniform over $\calY$.
\item[(iii)] The divergence variance $V(W_x\|Q^*)$ and third central moment $T(W_x\|Q^*)$ are independent of $x$.
\item[(iv)] The partial derivatives $\left. \frac{\partial}{\partial P(x)} V(P) \right|_{P=P^*}$ are independent of $x$.
	Equivalently, the gradient vector $\nabla V(P^*)$ has identical components.
\item[(v)] The correlation coefficient $\rho(P^*;W)=0$.
\item[(vi)] $\Ans^{(P^*)} = \Abns^{(P^*)} = 0$ in (\ref{eq:Ans}).
\item[(vii)] $\underline{A}_\epsilon = \overline{A}_\epsilon -1$, i.e., the lower and upper bounds
	in (\ref{eq:main}) differ by only one nat.
\end{description}
\label{prop:symmetric}
\end{proposition}
A fundamental consequence of the last property is that for CT weakly symmetric channels satisfying 
{\bf (A1)}---{\bf (A3)}, the upper and lower bounds on $\log M_\avg^*(n,\epsilon)$ differ only by one nat.

{\bf Special case: Discrete Additive-Noise Channels}.
Assume $\calX = \calY = \{0,1,\cdots,k-1\}$ and the channel is given by $Y=X+E$,
where the noise $E$ is distributed as $P_E$, and the sum is modulo $k$.
This is a CT weakly symmetric channel.
Then $i(x,y) = \log k + \log P_E(y-x)$ and
\[ I(P;W) = \log k - H(P_E), \quad V(P;W) = \Var[\log P_E(\cdot)] ,
	\quad T(P;W) = \eE[\log P_E(\cdot) + H(P_E)]^3 \]
where $H(p) = - \sum_x p(x) \log p(x)$ is the entropy function.
If $\log P_E(E)$ is not a lattice random variable then Theorem~\ref{thm:main} applies.

\subsubsection{Binary-Input, Ternary-Output Channel}
\label{ssec:BITO}

Consider the channel with $\calX=\{0,1\}$, $\calY=\{0,1,2\}$,
and $W = \begin{bmatrix} 1-2\theta & \theta & \theta \\ \theta & \theta & 1-2\theta \end{bmatrix}$,
parameterized by $\theta \in (0, \frac{1}{3})$.  This is a Gallager-symmetric channel but not
a CT weakly symmetric channel. The channel is the cascade of a BEC($\theta$) and 
a BSC($\frac{\theta}{1-\theta}$) and we refer to it as the BITO($\theta$) channel.
The extreme values $\theta=0$ and $\theta=\frac{1}{3}$ correspond to a noiseless
and a completely noisy channel, respectively. For $\theta \in (0, \frac{1}{3})$,
the capacity-achieving distribution $P^*$ is unique and uniform, and the capacity-achieving
output distribution is given by $Q^* = [\frac{1-\theta}{2}, \theta, \frac{1-\theta}{2}]$.
We shall verify that {\bf (A1)} and {\bf (A2)} hold.
The information density takes values $i(x,y) \in
\begin{bmatrix} \log \frac{2(1-2\theta)}{1-\theta} & 0 & \log \frac{2\theta}{1-\theta} \\
\log \frac{2\theta}{1-\theta} & 0 & \log \frac{2(1-2\theta)}{1-\theta} \end{bmatrix}$
and is a nonlattice random variable (hence {\bf (A3)} holds) for all values 
$\theta \in (0, \frac{1}{3})$ except for the finite set satisfying 
$\frac{2(1-2\theta)}{1-\theta} = (\frac{1-\theta}{2\theta})^k$ for some $k\in\nN$.
Capacity, information variance, and third central moment are given below as functions of $\theta$:
\begin{eqnarray*}
   C(\theta) & = & I(P^*;W) = (1-2\theta) \log \frac{2(1-2\theta)}{1-\theta} + \theta \log \frac{2\theta}{1-\theta} , \\
   V(\theta) & = & V(P^*;W) = (1-2\theta) \log^2 \frac{2(1-2\theta)}{1-\theta} 
		+ \theta \log^2 \frac{2\theta}{1-\theta} - C^2(\theta) , \\
   T(\theta) & = & T(P^*;W) = (1-2\theta) \left( \log \frac{2(1-2\theta)}{1-\theta} - C(\theta) \right)^3
		+ \theta \left( \log\frac{2\theta}{1-\theta} - C(\theta) \right)^3 .
\end{eqnarray*}
The information skewness is given by $S(\theta) = S(P^*;W) = \frac{T(\theta)}{V^{3/2}(\theta)}$
and the reverse channel by
\[ \widecheck{W}_0 = \left[ \frac{1-2\theta}{1-\theta} , \frac{\theta}{1-\theta} \right] , \quad
	\widecheck{W}_1 = \left[ \frac{1}{2}, \frac{1}{2} \right] , \quad
	\widecheck{W}_2 = \left[ \frac{\theta}{1-\theta}, \frac{1-2\theta}{1-\theta} \right] .
\]
Hence $D(\widecheck{W}_0\|P^*) = D(\widecheck{W}_2\|P^*) = \frac{C(\theta)}{1-\theta}$ and $D(\widecheck{W}_1\|P^*) = 0$.
From \eqref{eq:rho}, \eqref{eq:rho2}, and the above formulas, we obtain the correlation coefficient
\[ \rho(\theta) = \rho(P^*;W) = \frac{\theta \,C^2(\theta)}{(1-\theta)V(\theta)} . \]

By symmetry, the vectors $v$ and $\tilde{v}$ of \eqref{eq:v} and \eqref{eq:vtilde} have uniform entries, 
hence they are orthogonal to $\scrL(\calX)$. It follows from \eqref{eq:J+} that $\sfJ^+ v = \sfJ^+ \tilde{v} = 0$.
Also $\Ans = \Abns = 0$, and from \eqref{eq:Aeps+} and \eqref{eq:Aeps-}, the gap
$\overline{A}_\epsilon - \underline{A}_\epsilon = t_\epsilon^2 \frac{\rho(\theta)}{1+\rho(\theta)}
- \log \sqrt{1-\rho^2(\theta)} + 1$.
Finally note from the discussion above that the channel satisfies properties (i)(iii)(iv)(vi) 
of Prop.~\ref{prop:symmetric} but not (ii)(v)(vii).

\begin{figure}[h!]
\centering
\begin{tabular}{cc}
\includegraphics[width=2.2in]{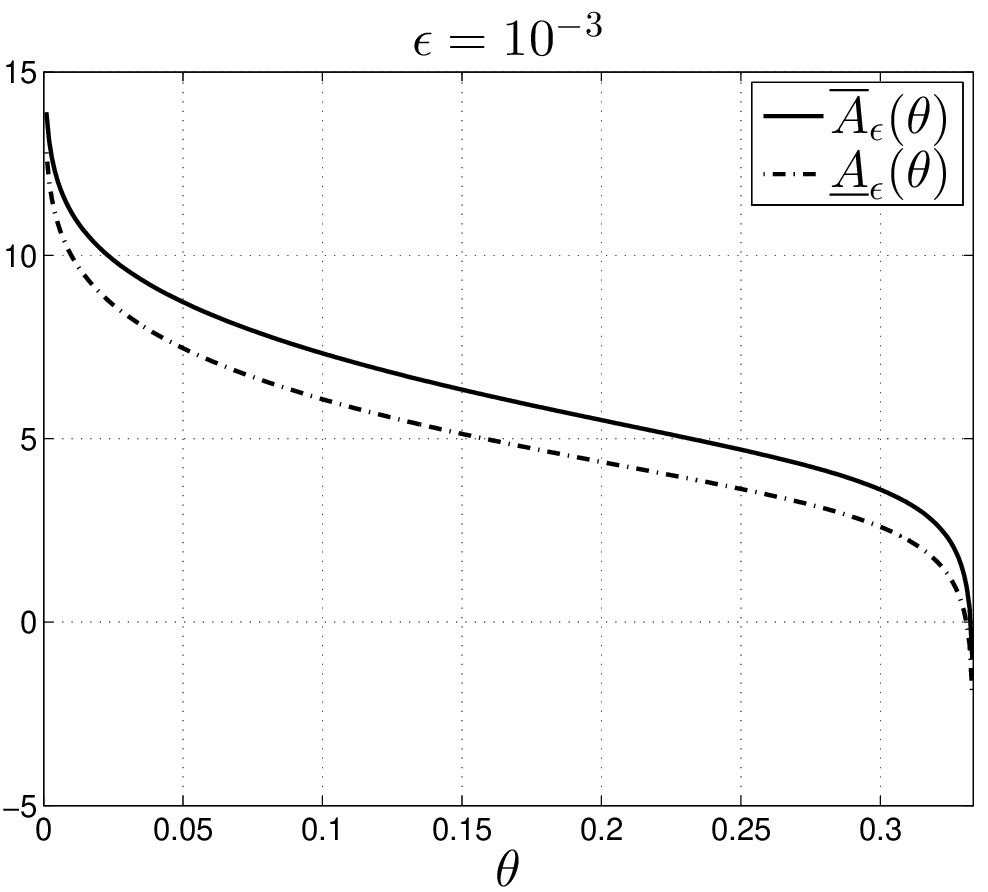} &
\includegraphics[width=2.2in]{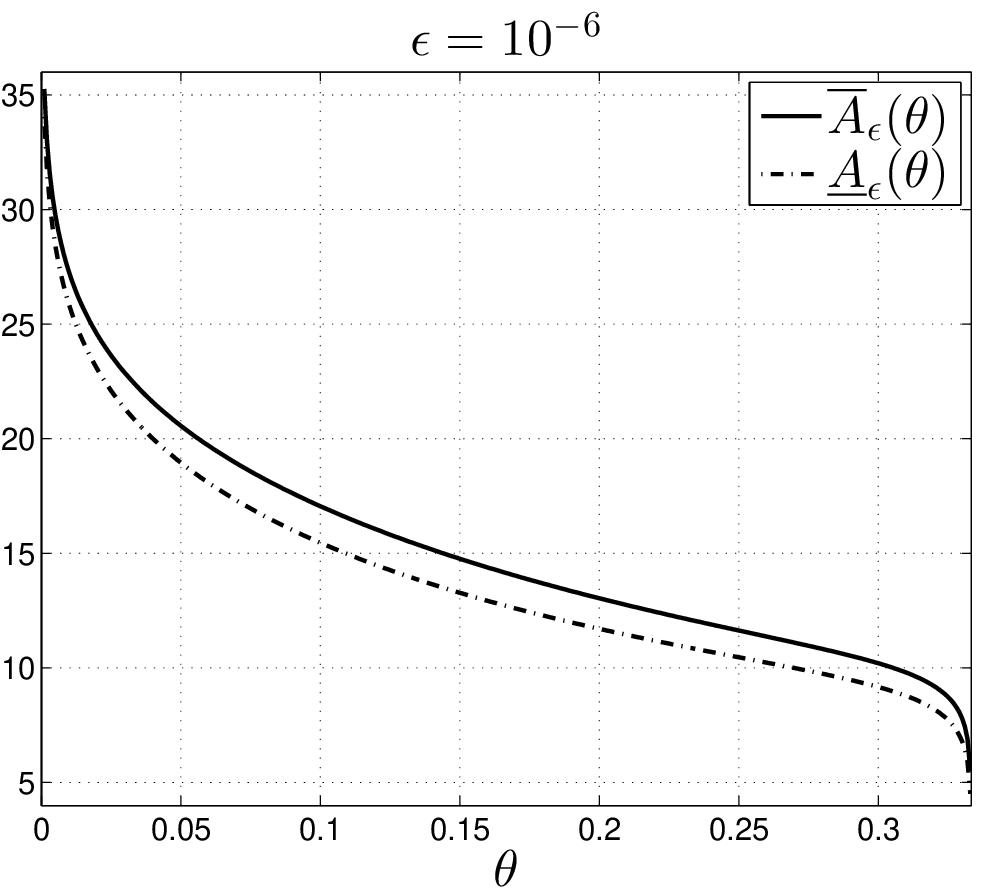} \\
(a) & (b) \\
\includegraphics[width=2.2in]{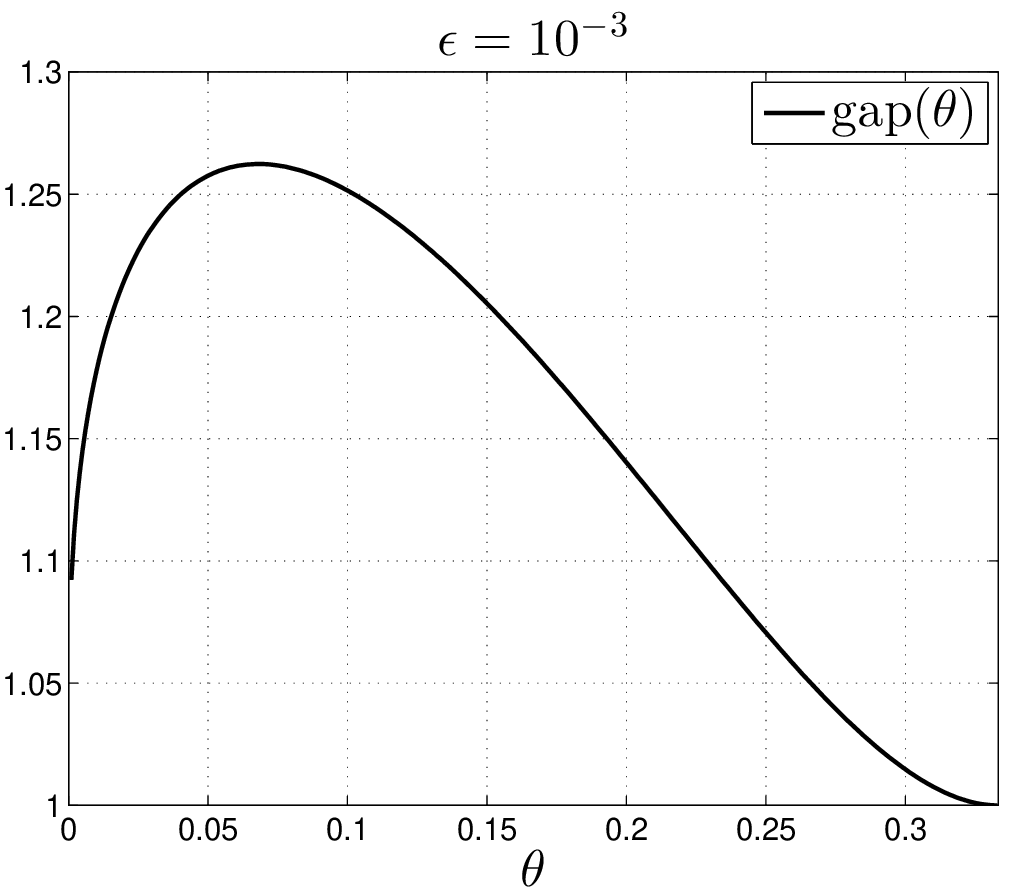} &
\includegraphics[width=2.2in]{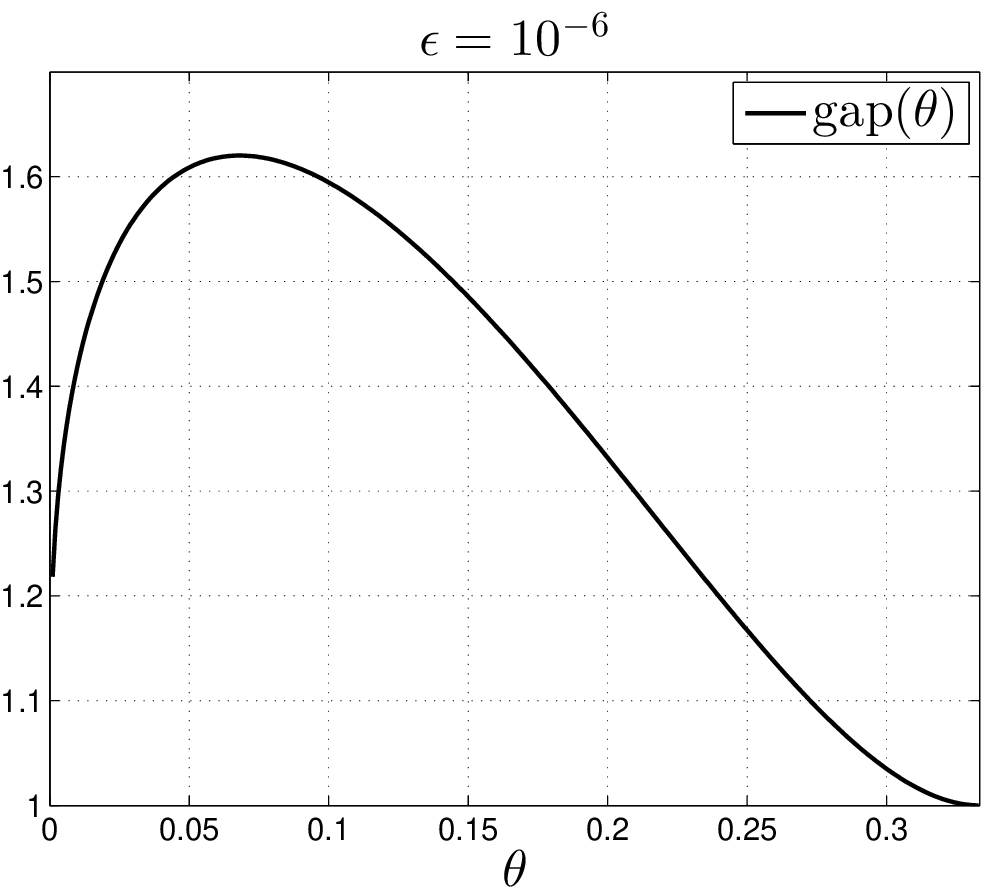} \\
(c) & (d)
\end{tabular}
\caption{\small{The constants $\overline{A}_\epsilon$ and $\underline{A}_\epsilon$ of \eqref{eq:Aeps-} \eqref{eq:Aeps+}
	for the BITO($\theta$) channel with $\theta \in (0, \frac{1}{3})$ and error probabilities 
	(a) $\epsilon = 10^{-3}$, and (b) $\epsilon = 10^{-6}$.
    The corresponding log-volume gaps $\overline{A}_\epsilon - \underline{A}_\epsilon$
	are shown in (c) and (d).}}
\label{fig:BITO-eps}
\end{figure}

\begin{figure}[h!]
\centering
\begin{minipage}{.5\textwidth}
  \centering
  \includegraphics[width=2.2in]{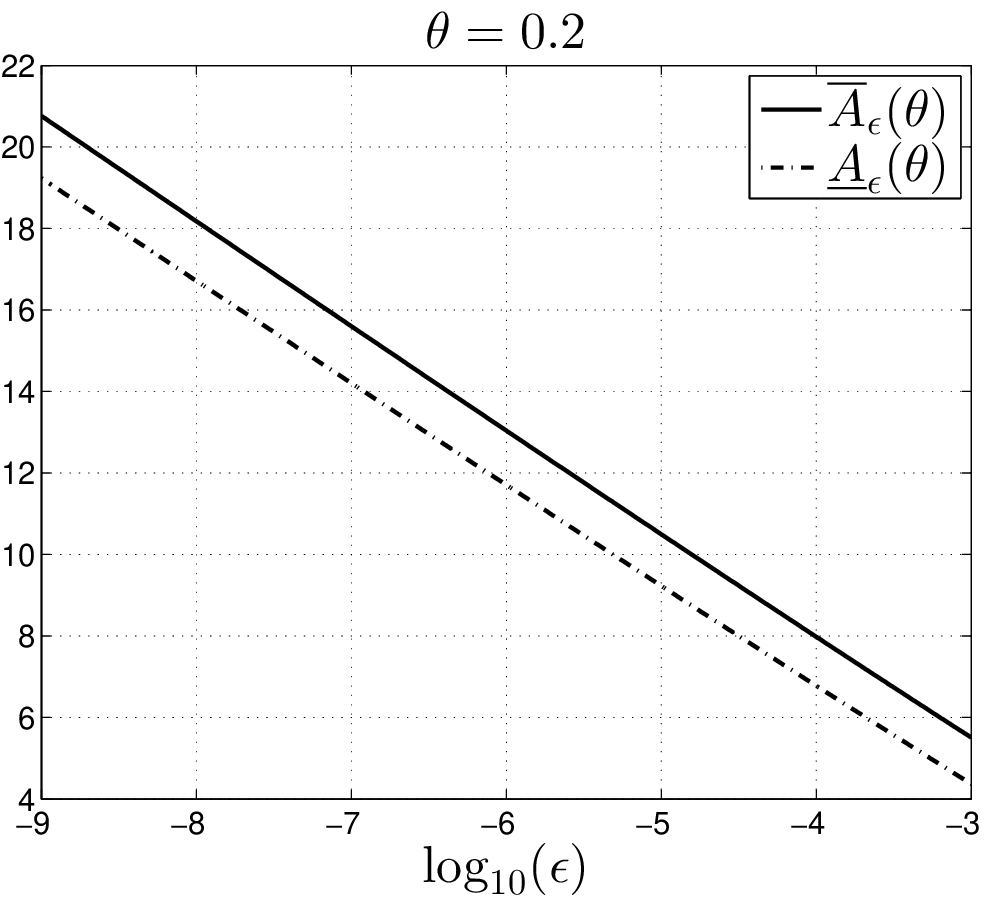}
  \caption{}{\small{The constants $\overline{A}_\epsilon$ and $\underline{A}_\epsilon$
	for the BITO channel with parameter $\theta=0.2$.
    The constants are plotted as a function of the log error probability}.}
  \label{fig:BITO-theta}
\end{minipage}%
\begin{minipage}{.5\textwidth}
  \centering
  \includegraphics[width=2.2in]{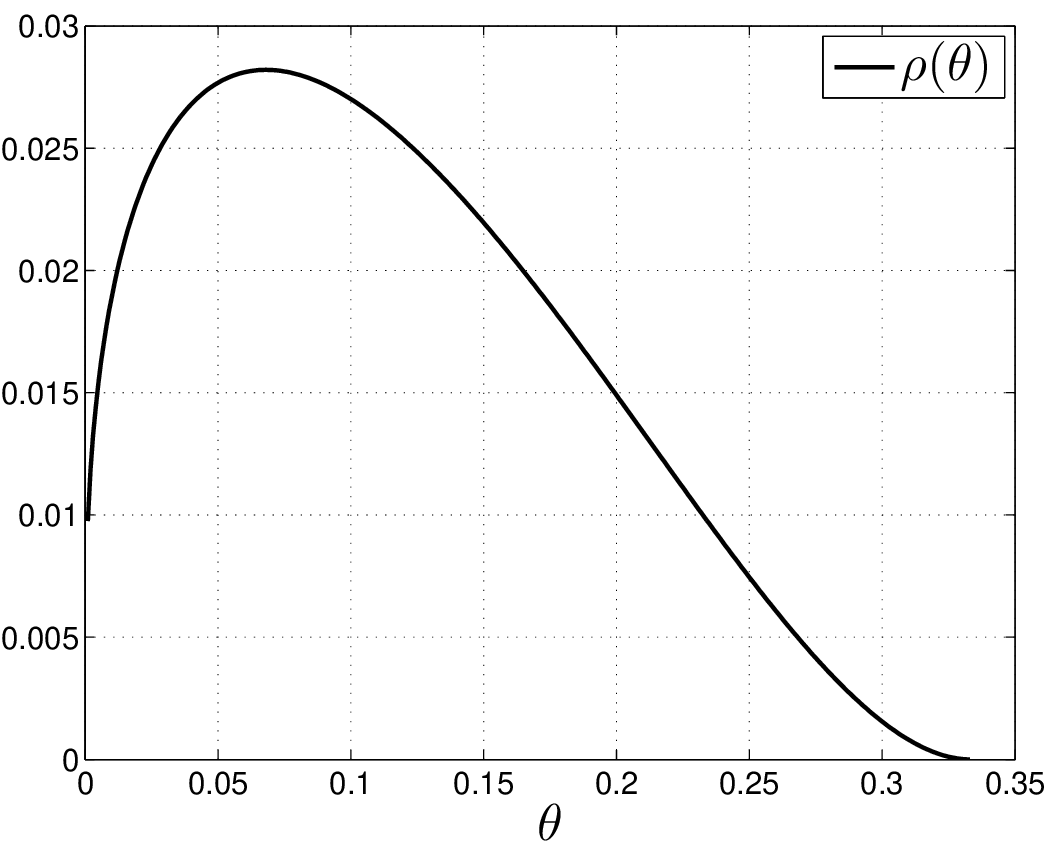}
  \caption{}{\small{Correlation coefficient $\rho$ for the BITO channel as a function of 
	the channel parameter $\theta$.}}
  \label{fig:BITO-rho}
\end{minipage}
\end{figure}

The expressions above have been evaluated numerically. 
It it seen from Figs.~\ref{fig:BITO-eps} and \ref{fig:BITO-theta} that the gap 
$\overline{A}_\epsilon - \underline{A}_\epsilon$ ranges from approximately 1 to 2 nats over
the range of $\epsilon$ and $\theta$ considered.

\subsection{Z Channel}
\label{sec:Z}

The Z channel with parameter $\theta \in [0,1]$ is a nonsymmetric channel with input and output alphabets
$\calX = \calY = \{0,1\}$ and transition probability matrix $W(y|x)$ specified by
$W(0|0)=1$ and $W(0|1) = \theta$. The capacity-achieving input distribution is given by
\[ P^*\{X=0\} = a \triangleq 1 - \frac{1}{(1-\theta)\left(1+\exp\{\frac{h_2(\theta)}{1-\theta}\}\right) } \ge \frac{1}{2} \]
where $h_2(p) = - p \log p - (1-p) \log(1-p)$ is the binary entropy function.
Capacity is given as a function of $\theta$ by
\begin{equation}
  C(\theta) = h_2((1-a)(1-\theta)) - (1-a) h_2(\theta) .
\label{eq:Clambda}
\end{equation} 
The capacity-achieving output distribution is given by $Q^*(0) = a + \theta(1-a)$. 
If $\theta \notin \{0,1\}$, Assumptions {\bf (A1)}---{\bf (A3)} hold, and Theorem~\ref{thm:main} applies. 

Denote by $d(p\|q)$, $v(p\|q)$ and $t(p\|q)$ the KL divergence, the divergence variance,
and the divergence third central moment between two Bernoulli random variables with parameters
$p$ and $q$, respectively. For a distribution $Q$ on $\calY$ with
$Q(0)=q$, we have $V(W_0\|Q) = T(W_0\|Q) = 0$, $V(W_1\|Q) = v(\theta\|q)$,
and $T(W_1\|Q) = t(\theta\|q)$. Thus
\[ V(W\|Q|P^*) = (1-a) \,v(\theta\|q) \quad \mathrm{and} \quad T(W\|Q|P^*) = (1-a) \,t(\theta\|q) . \]
In particular, putting $q=a+\theta(1-a)=Q^*(0)$ and thus $Q=Q^*$, we obtain respectively the conditional
information variance, the third conditional central moment, and the conditional skewness, 
all which may be viewed as functions of $\theta$:
\begin{eqnarray}
   V(\theta) & = & V(P^*;W) = (1-a) \,v(\theta\|a+\theta(1-a)) , \label{eq:Vtheta} \\
   T(\theta) & = & T(P^*;W) = (1-a) \,t(\theta\|a+\theta(1-a)) , \label{eq:Ttheta} \\
   S(\theta) & = & S(P^*;W) = \frac{T(\theta)}{V(\theta)^{3/2}} 
		= \frac{t(\theta\|a+\theta(1-a))}{\sqrt{1-a} \,v(\theta\|a+\theta(1-a))^{3/2}} . \label{eq:Stheta}
\end{eqnarray} 
The Fisher information matrix is obtained from \eqref{eq:fisher} (with $P=P^*$) as
\begin{equation}
   \sfJ(\theta) = \left( \begin{array}{ll} \frac{1}{a+\theta(1-a)} & \frac{\theta}{a+\theta(1-a)} \\ 
	\frac{\theta}{a+\theta(1-a)} & \frac{\theta^2}{a+\theta(1-a)} + \frac{1-\theta}{1-a} \end{array} \right)
\label{eq:Jtheta}
\end{equation}
and its determinant is $|\sfJ(\theta)| = \frac{1-\theta}{1-a}$. Then 
$\sfJ^{-1} 1 = (1, \frac{1-a}{a+\theta(1-a)})^\top$, and the rank-one matrix $\sfJ^+$ is 
obtained from \eqref{eq:J+finiteX}.

The reverse channel with input distribution $Q^*$ is given by 
$\widecheck{W}(0|1) = 0$ and $\widecheck{W}(0|0) = \frac{a}{a+\theta(1-a)}$.
Hence
\begin{eqnarray}
   D(\widecheck{W}_1\|P^*) & = & \log \frac{1}{1-a} \nonumber \\
   D(\widecheck{W}_0\|P^*) & = & \frac{\theta(1-a)}{a+\theta(1-a)} \log \theta - \log [a+\theta(1-a)] .
\label{eq:Dwtilde}
\end{eqnarray}
As $\theta \to 1$, we have $a \to 1 - 1/e$, $C(\theta) \to 0$, $D(\widecheck{W}_1\|P^*) \to 1$, 
and $D(\widecheck{W}_0\|P^*) \to 0$.

The vector $\breve{v}$ can be evaluated from \eqref{eq:vbreve2-lemma}:
\begin{eqnarray*}
   \breve{v}(0) & = & - 2 \eE_{W_0} [D(\widecheck{W}_Y\|P^*) - C(\theta)] 
						= - 2 [D(\widecheck{W}_0\|P^*) - C(\theta)] \\
   \breve{v}(1) & = & - 2 \eE_{W_1} [D(\widecheck{W}_Y\|P^*) - C(\theta)]
						= - 2 [\theta D(\widecheck{W}_0\|P^*) + (1-\theta) D(\widecheck{W}_1\|P^*) - C(\theta)] .
\end{eqnarray*}
The gradient vector $v$ of \eqref{eq:v} is obtained from \eqref{eq:v-sum} as
\begin{eqnarray*}
   v(0) & = & V(W_0\|Q^*) + \breve{v}(0) = - 2 [D(\widecheck{W}_0\|P^*) - C(\theta)] , \\
   v(1) & = & V(W_1\|Q^*) + \breve{v}(1) \\
		& = & v(\theta\|a+\theta(1-a)) - 2 [\theta D(\widecheck{W}_0\|P^*) 
		+ (1-\theta) D(\widecheck{W}_1\|P^*) - C(\theta)] .
\end{eqnarray*}
The expressions $\Ans$ and $\Abns$ in \eqref{eq:Ans} can be evaluated using the expressions above.

Finally, the correlation coefficient $\rho(\theta)$ is obtained from \eqref{eq:rho}, \eqref{eq:rho2},
\eqref{eq:Vtheta}, and \eqref{eq:Dwtilde} as
\begin{eqnarray}
   \rho(\theta) 
	& = & \frac{1}{V(\theta)} \left[ (a + \theta(1-a)) \left( \frac{\theta(1-a)}{a+\theta(1-a)} \log \theta 
		- \log [a+\theta(1-a)] \right)^2  \right. \nonumber \\
	&& \qquad \left. + (1-a)(1-\theta) \log^2 \frac{1}{1-a}  - C(\theta)^2 \right] .
\label{eq:rholambda}
\end{eqnarray}

\begin{figure}[h!]
\centering
\begin{tabular}{cc}
\includegraphics[width=2.2in]{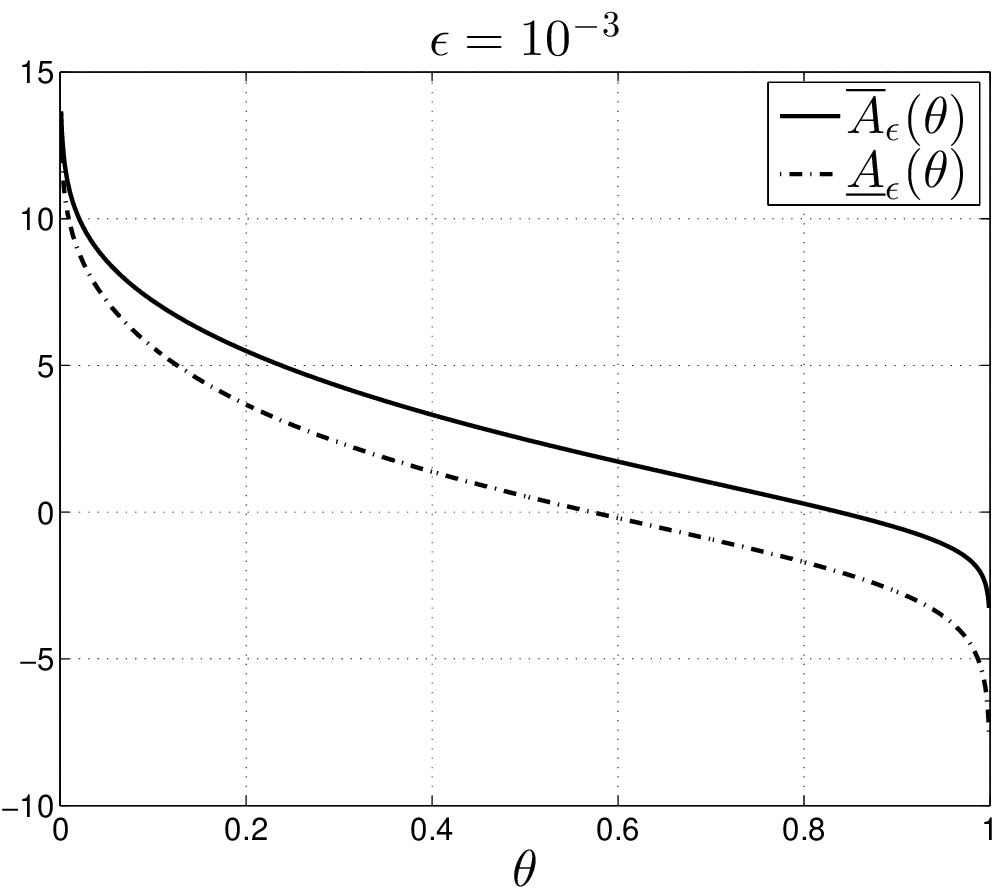} &
\includegraphics[width=2.2in]{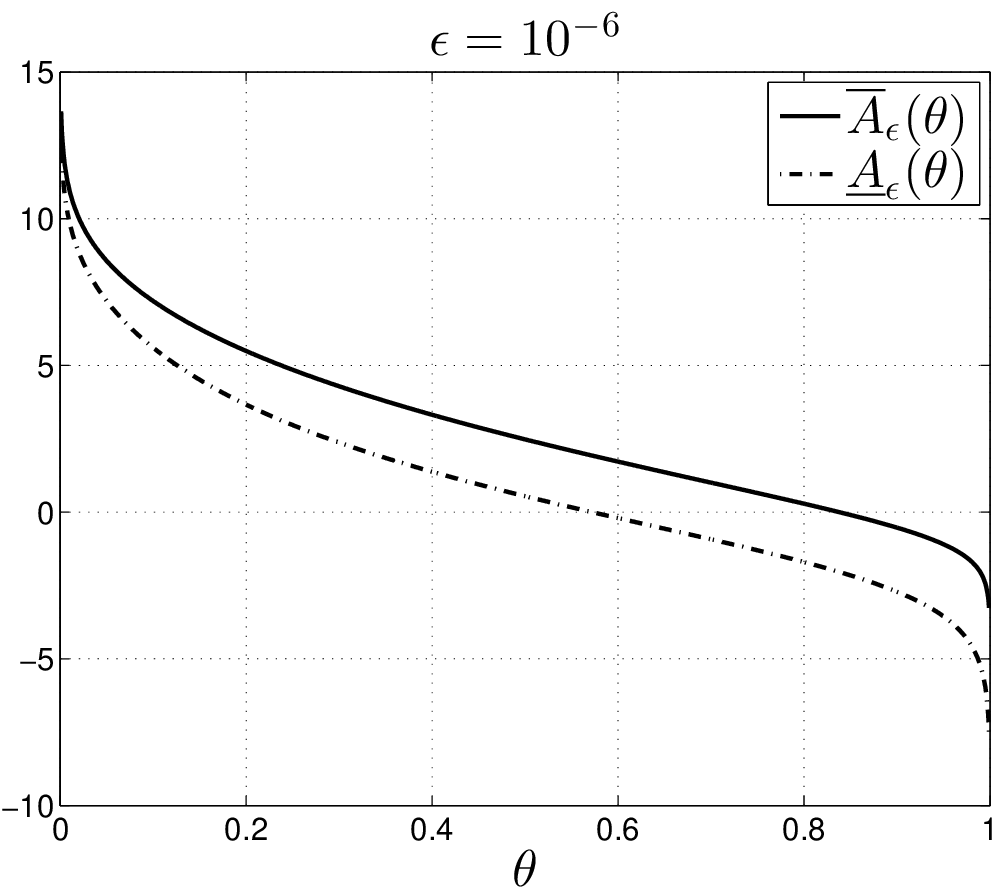} \\
(a) & (b) \\
\includegraphics[width=2.2in]{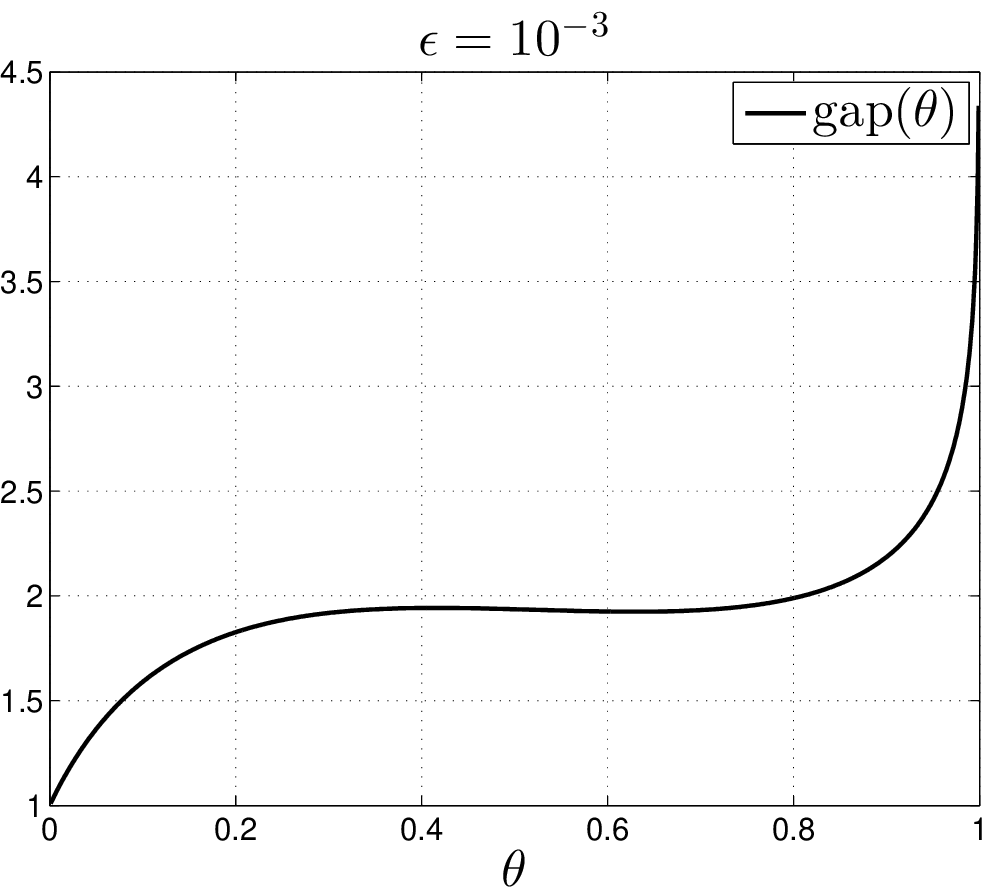} &
\includegraphics[width=2.2in]{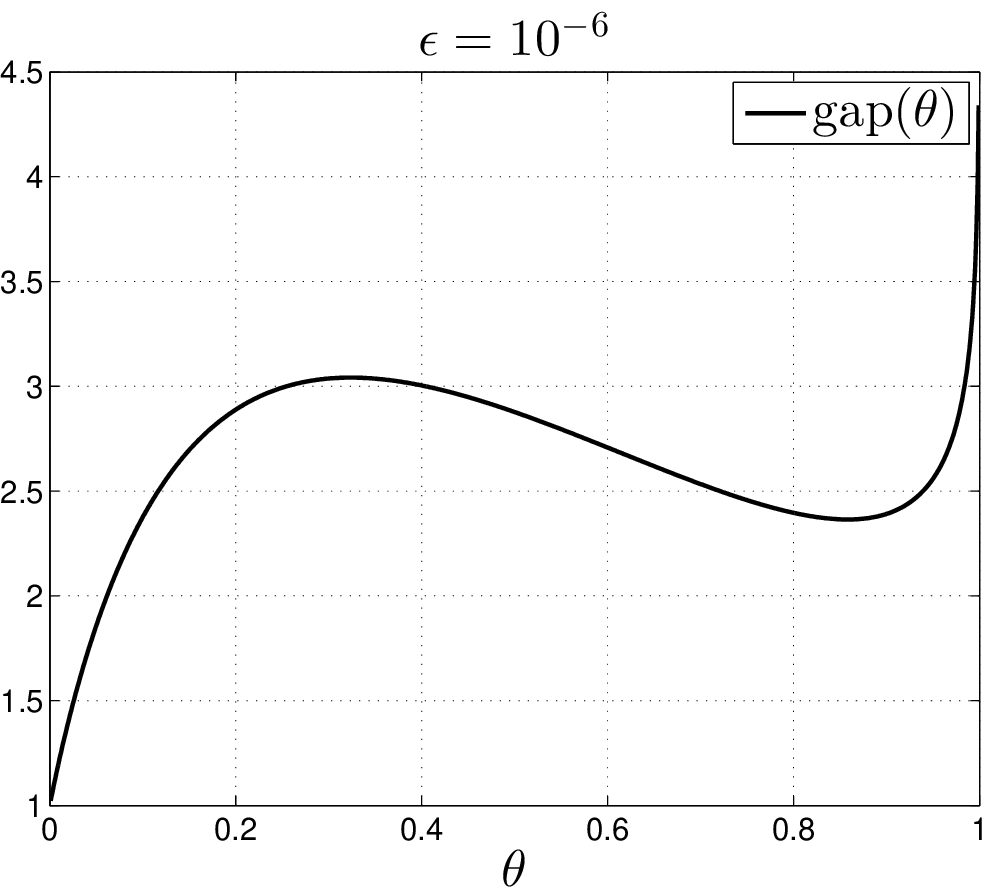} \\
(c) & (d)
\end{tabular}
\caption{\small{The constants $\overline{A}_\epsilon$ and $\underline{A}_\epsilon$ of \eqref{eq:Aeps-} \eqref{eq:Aeps+}
	for the Z channel with crossover probability $\theta \in (0,1)$ and error probabilities 
	(a) $\epsilon = 10^{-3}$, and (b) $\epsilon = 10^{-6}$.
    The corresponding log-volume gaps $\overline{A}_\epsilon - \underline{A}_\epsilon$
	are shown in (c) and (d).}}
\label{fig:Z-eps}
\end{figure}

\begin{figure}[h!]
\centering
\begin{minipage}{.5\textwidth}
  \centering
  \includegraphics[width=2.2in]{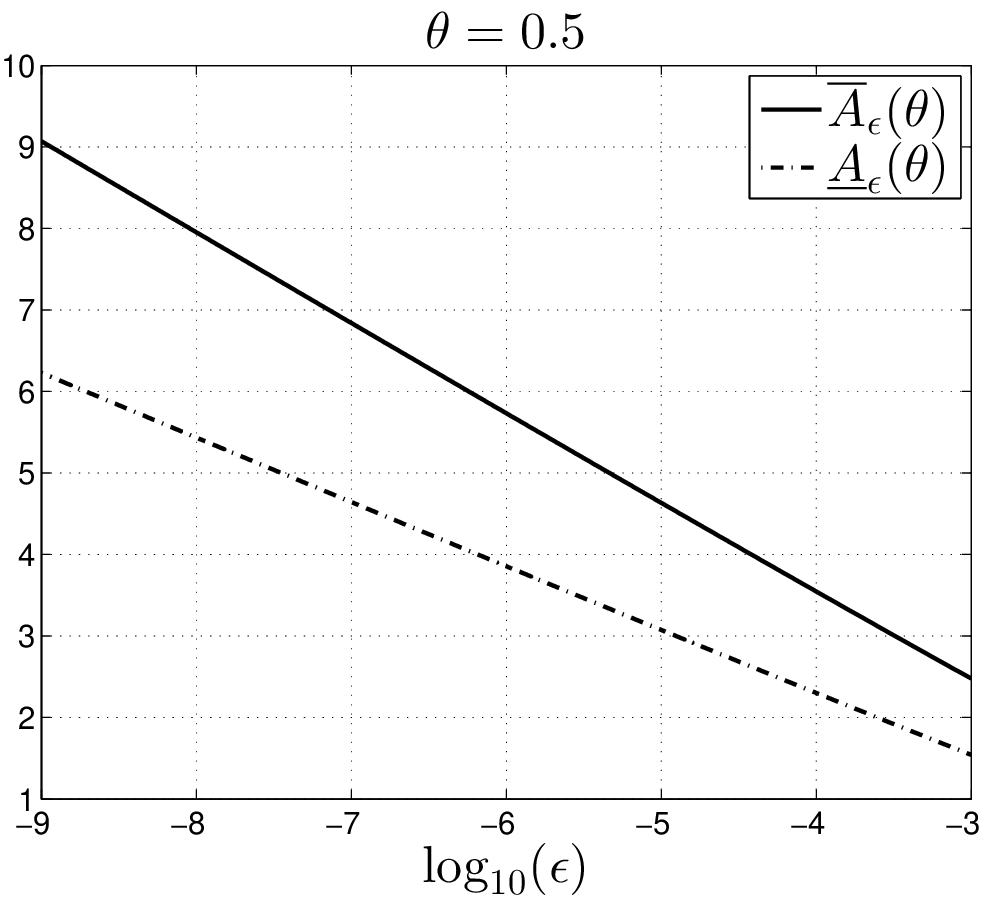}
  \caption{}{\small{The constants $\overline{A}_\epsilon$ and $\underline{A}_\epsilon$
	for the Z channel with crossover probability $\theta=\frac{1}{2}$.
    The constants are plotted as a function of the log error probability.}}
  \label{fig:Z-theta}
\end{minipage}%
\begin{minipage}{.5\textwidth}
  \centering
  \includegraphics[width=2.2in]{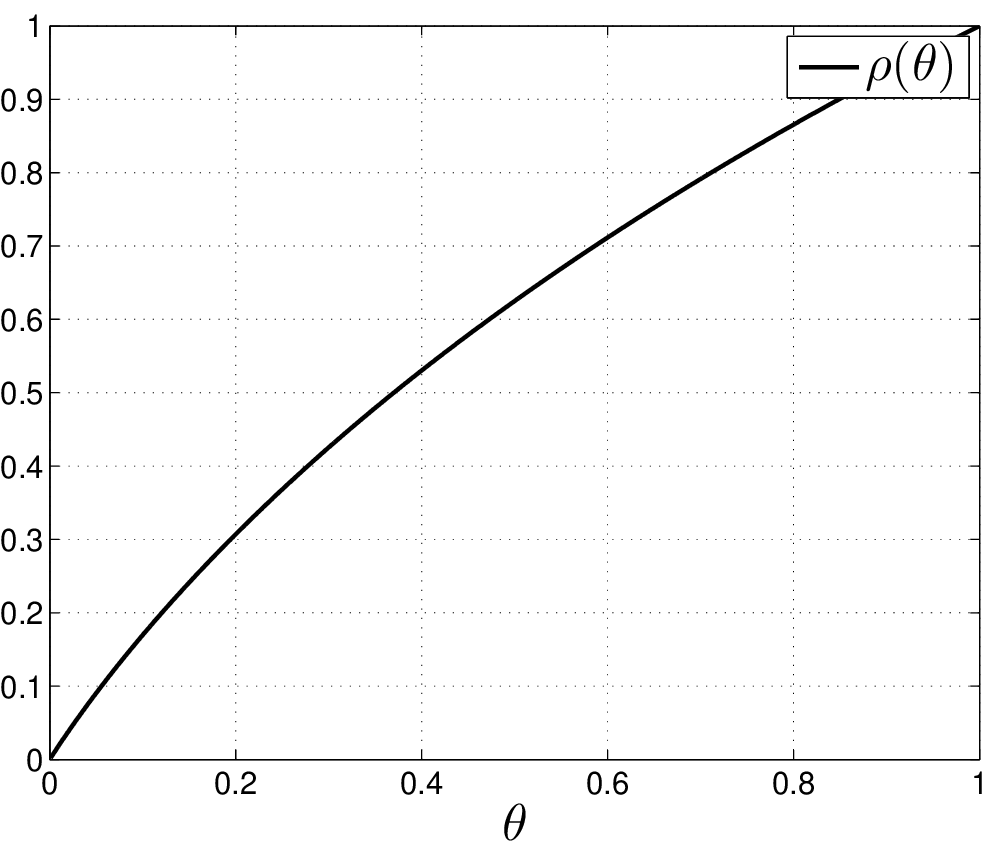}
  \caption{}{\small{Correlation coefficient $\rho$ for the Z channel as a function of 
	crossover probability $\theta$}.}
  \label{fig:Z-rho}
\end{minipage}
\end{figure}

The constants $\overline{A}_\epsilon$ and $\underline{A}_\epsilon$ can be evaluated by plugging
the expressions above into \eqref{eq:Aeps+} and \eqref{eq:Aeps-}. It it seen from Figs.~\ref{fig:Z-eps} and \ref{fig:Z-theta}
that the gap between the lower and upper bound ranges from approximately 1 to 3 nats over
the range of $\epsilon$ and $\theta$ considered.

\subsection{Binary Symmetric Channel}
\label{sec:BSC}

The BSC with crossover error probability $\lambda$ has input and output alphabets $\calX=\calY=\{0,1\}$,
and channel law $W(1|0)=W(0|1)=\lambda$.
For $\lambda \notin \{0, \frac{1}{2}, 1\}$, Polyanskiy {\em et al} \cite[Theorem~52, p.~2332]{Polyanskiy10} gave
\[ \log M_\avg^*(n,\epsilon) = nC(\lambda) - \sqrt{nV(\lambda)} t_\epsilon + \frac{1}{2} \log n + O(1) \]
where 
\[ C(\lambda) = \log 2 - h_2(\lambda), \quad V(\lambda) = \lambda(1-\lambda) \log^2 \frac{1-\lambda}{\lambda} . \]
The capacity-achieving input distribution is unique, and both $P^*$ and $Q^*$ are uniform over $\{0,1\}$.

\clearpage

The loglikelihood ratio $L = \log \frac{W(Y|X)}{Q^*(Y)}$ takes value $\log(2(1-\lambda))$ with probability
$1-\lambda$, and value $\log(2\lambda)$ with probability $\lambda$. Hence $L$ takes its values on a lattice $\Omega$
with span $d = |\log\frac{1-\lambda}{\lambda}|$ and offset $\log(2\lambda)$.
The third central moment of $L$ is given by
\begin{eqnarray}
   T(\lambda) & = & (1-\lambda) [\log (2(1-\lambda)) - C(\lambda)]^3 + \lambda [\log (2\lambda) - C(\lambda)]^3 \nonumber \\
	 & = & (1-\lambda) [\log(1-\lambda) + h_2(\lambda)]^3 + \lambda [\log \lambda + h_2(\lambda)]^3 \nonumber \\
	 & = & (1-\lambda) [-\lambda \log \lambda + \lambda \log(1-\lambda)]^3 
			+ \lambda [(1-\lambda) \log \lambda - (1-\lambda) \log(1-\lambda)]^3 \nonumber \\
	 & = & (1-\lambda) \lambda^3 \left( \log \frac{1-\lambda}{\lambda} \right)^3 
			- \lambda (1-\lambda)^3 \left( \log \frac{1-\lambda}{\lambda} \right)^3 \nonumber \\
	 & = & \lambda (1-\lambda) (2\lambda - 1) \left( \log \frac{1-\lambda}{\lambda} \right)^3 \le 0 .
\label{eq:T-BSC}
\end{eqnarray}
The skweness of $L$ is equal to
\begin{equation}
   S(\lambda) = \frac{T(\lambda)}{[V(\lambda)]^{3/2}} = - \frac{|1-2\lambda|}{\sqrt{\lambda (1-\lambda)}} .
\label{eq:S-BSC}
\end{equation}

Denote by $[x]$ the rounding of real $x$ to the nearest integer. Define the integer $k = \max\{1, [\frac{1}{d}]\}$ and the function
\[ f(d) \triangleq \log \frac{d}{1-e^{-d}} \ge 0 , \quad d > 0,
\]
where $f(d) \sim \frac{d}{2}$ as $d \to 0$ ($\lambda \to \frac{1}{2}$) and $f(d) \to \log k$ as $d \to \infty$ ($\lambda \to 0$). 

\begin{theorem}
For the BSC with crossover probability $\lambda \notin \{0, \frac{1}{2}, 1\}$, the upper and lower bounds of \eqref{eq:main} hold,
with 
\begin{eqnarray}
   \overline{A}_\epsilon & = & A_\epsilon^* - f(d) - \frac{d}{2} \nonumber \\
   \underline{A}_\epsilon & = & A_\epsilon^* - 2f(d) - kd + \log(kd)
\label{eq:Aeps-BSC}
\end{eqnarray}
where
\begin{equation}
   A_\epsilon^* = - \frac{2\lambda-1}{6} \left( \log \frac{1-\lambda}{\lambda} \right) (t_\epsilon^2 - 1) 
	+ \frac{t_\epsilon^2}{2} + \frac{1}{2} \log \left( 2\pi \lambda (1-\lambda) \log^2 \frac{1-\lambda}{\lambda} \right) .
\label{eq:A*-BSC}
\end{equation}
\label{thm:BSC}
\end{theorem}
The proof builds on the results of Secs.~\ref {sec:strongLD-HT}, \ref{sec:achieve}, and \ref{sec:converse},
and is given in Appendix~\ref{app:BSC}.

\begin{figure}[h!]
\centering
\begin{tabular}{cc}
\includegraphics[width=2.2in]{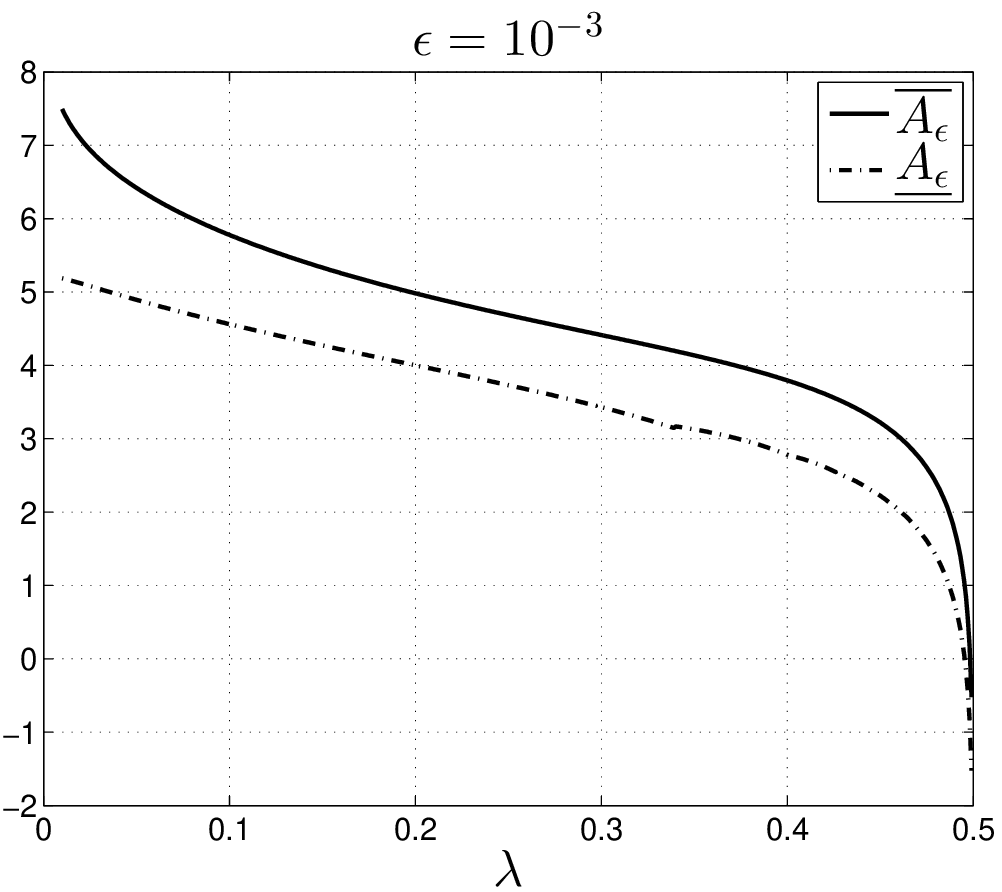} &
\includegraphics[width=2.2in]{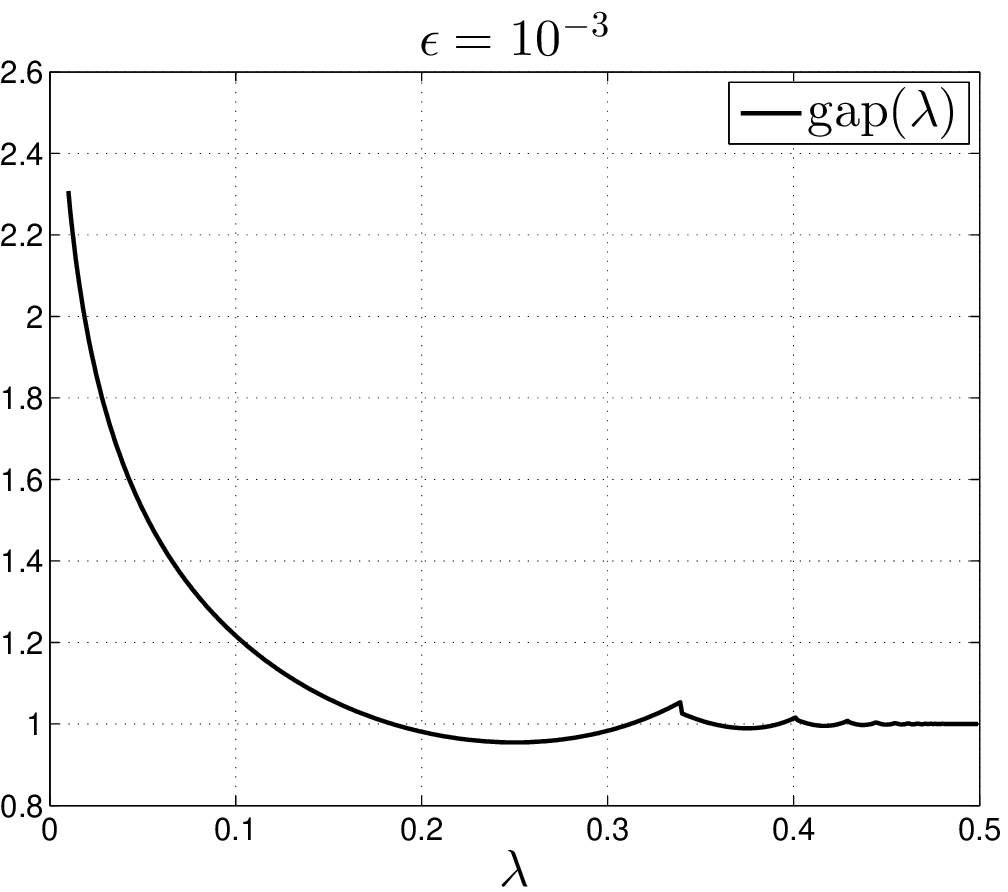} \\
(a) & (b) 
\end{tabular}
\caption{\small{The constants $\overline{A}_\epsilon$ and $\underline{A}_\epsilon$ of \eqref{eq:Aeps-BSC}
	for the BSC($\lambda$) channel with $\lambda \in (0, \frac{1}{2})$ and error probability $\epsilon = 10^{-3}$.
    The corresponding log-volume gap $\overline{A}_\epsilon - \underline{A}_\epsilon$ of \eqref{eq:BSC-gap} is shown in (b).}}
\label{fig:BSC-eps}
\end{figure}

Note that \eqref{eq:A*-BSC} coincides with the expression \eqref{eq:Aeps+} for $\overline{A}_\epsilon$ in the nonlattice, symmetric case. 
The asymptotic gap between the lower and upper bound is
\begin{equation}
   \overline{A}_\epsilon - \underline{A}_\epsilon = f(d) + \left( k-\frac{1}{2} \right) d - \log(kd)
\label{eq:BSC-gap}
\end{equation}
independently of $\epsilon$.
Fig.~\ref{fig:BSC-eps} shows that this gap is approximately one nat
for $\lambda > 0.2$, and less than two nats for $\lambda > 0.02$.

\section{Optimization Lemmas}
\setcounter{equation}{0}
\label{sec:optimization}

Our achievability and converse theorems rely on several optimization lemmas which are stated in this section.
The proofs of Lemmas~\ref{lem:zetan-P}---\ref{lem:game} 
make use of a $|\calX| \times |\calX|$ matrix $\Jcap \succeq 0$ defined as follows. 
If $\calX=\Xcap$, then $\Jcap = J^\calX$.
Else $\Jcap$ coincides with $J^\calX (P;W)$ for $x,x'\in\Xcap$ and is zero elsewhere. 
This is analogous to the definition of the matrix $\sfJ$ above \eqref{eq:hJ2}, with $\Xcap$ in place of $\calX^*$.
Analogously to \eqref{eq:LX*}, we define
\[ \scrL(\Xcap) \triangleq \{ h \in \scrL(\calX) : \mathrm{supp}(h) \subseteq \Xcap \} . \]

The set \eqref{eq:Pi} of capacity-achieving input distributions may be written as
\begin{equation}
   \Pi = (P_0 + \mathrm{ker}(\Jcap)) \cap \scrP(\calX)
\label{eq:Pi-space}
\end{equation}
for any capacity-achieving $P_0$.
\begin{lemma} (Saddlepoint for conditional KL divergence.)\\
The conditional KL divergence function $D(W\|Q|P), \,P\in\scrP(\calX),\,Q\in\scrP(\calY)$ 
is linear in $P$ and convex in $Q$, and admits a saddlepoint $(P_0, Q^*)$ for each $P_0 \in \Pi$:
\begin{equation}
   D(W\|Q^*|P) \le D(W\|Q^*|P_0) \le D(W\|Q|P_0) , \quad \forall P\in\scrP(\calX) ,Q\in\scrP(\calY) .
\label{eq:D-SP}
\end{equation}
The value of the saddlepoint is $I(P_0;W)= C$. The left inequality holds with equality for all $P \in \scrP(\Xcap)$.
\label{lem:SP+cost}
\end{lemma}
{\em Proof}. The statement is well known \cite{Topsoe67,Csiszar95}. The left inequality follows from
the property stated below \eqref{eq:Delta}. The right inequality follows from
the decomposition $D(W\|Q|P_0) = I(P_0;W) + D(Q^*\|Q) \ge I(P_0;W)$. 
\hfill $\Box$

Define the shorthands $I(P) = I(P;W)$, $G(P) = - \sqrt{V(P;W)} \,t_\epsilon$, $F(P) = F_\epsilon(P;W)$,
and $\zeta_n(P) \triangleq nI(P) + \sqrt{n} G(P) + F(P)$.
The following lemma is a refinement of \cite[Lemma~64]{Polyanskiy10} under different assumptions.
Application of the aforementioned Lemma~64 yields $\max_P \zeta_n(P) = nC - \sqrt{nV_\epsilon} t_\epsilon + O(1)$.
For clarity, separate statements are given for the Taylor expansion (i) and for the optimization results
(ii) and (iii). The proof is given in Appendix~\ref{app:zetan-P}. The general idea is that $\zeta_n(P)$ decays 
locally quadratically in $P$ (and linearly in $n$) away from $\Pi$; 
and $\zeta_n(P), P\in\Pi$ decays locally linearly in $P$ (and in $\sqrt{n}$) away from $\Pi^*$.

\begin{lemma}
Assume {\bf (A1)} holds and fix any positive vanishing sequence $\delta_n$. Then

\noindent
{\bf (i)} Consider any $P'\in\Pi$. The following Taylor expansion holds uniformly over the closed $\delta_n$-neighborhood
$\{P\in\scrP(\calX) :\,\|P-P'\|_\infty \le \delta_n\}$ and $P'\in\Pi$:
\begin{equation}
  \zeta_n(P) = \zeta_n(P') + n (P-P')^\top \Delta
	- \frac{n}{2} \|P-P'\|_{\sfJ^\calX}^2 + \sqrt{n} \,(P-P')^\top \nabla G(P') + o(n\delta_n^2) + o(\sqrt{n} \delta_n) .
\label{eq:zetan*-taylor}
\end{equation}

\noindent
{\bf (ii)} Consider any $P'\in\Pi^*$. Let $g \triangleq \nabla G(P')$ and define
\begin{equation}
   P_n' \triangleq P' + \frac{1}{\sqrt{n}} \sfJ^+ g .
\label{eq:Pn*-lem}
\end{equation}
Then
\begin{equation}
   \zeta_n(P_n') = \zeta_n(P') + \frac{1}{2} \|g\|_{\sfJ^+}^2 + o(1) .
\label{eq:zetan*-lem}
\end{equation}
Moreover, if $\scrP(\Xcap) \subseteq \Pi + \scrL(\calX^*)$, then
\begin{equation}
	\zeta_n(P_n') = \sup_{P:\,\|P-P'\| \le \delta_n} \zeta_n(P) + o(1) . 
\label{eq:zetan-local-opt}
\end{equation}

\noindent
{\bf (iii)} There exists a constant $c_1 > 0$ such that for any sequences $n^{-1/2} \ll \delta_n \ll \epsilon_n \ll 1$,
\begin{equation}
  \zeta_n(P) \le nC - \sqrt{nV_\epsilon} \,t_\epsilon - c_1 \sqrt{n} \delta_n [1+o(1)] ,
\label{eq:zetan-UB2}
\end{equation}
uniformly over $P \in \Pi(\epsilon_n) \setminus \Pi^*(\delta_n)$.
\label{lem:zetan-P}
\end{lemma}

\begin{remark}
Our random codes in Sec.~\ref{sec:achieve} are drawn from the distribution $P_n' = P' + \frac{1}{\sqrt{n}} h$ associated with 
an optimized $P' \in \Pi^*$ and $h = \sfJ^+ g$. The dimensionality condition above \eqref{eq:zetan-local-opt} holds trivially when
$\Xcap = \calX^*$. If the condition does not, a potentially better direction vector $h$ could be derived by solving 
the $|\Xcap|$-dimensional convex program \eqref{eq:zetan-local-opt-general}.
\end{remark}

\begin{remark}
Using the same method as in the proof of Lemma~\ref{lem:zetan-P}(iii), one can show there exists a constant
$c_2 > 0$ such that for any sequences $n^{-1/2} \ll \delta_n \ll \epsilon_n \ll 1$ and 
any $P \in \Pi(\epsilon_n) \setminus \Pi(\delta_n)$,
\begin{equation}
  \zeta_n(P) \le nC - \sqrt{nV_\epsilon} \,t_\epsilon - c_2 n\delta_n^2 [1+o(1)] .
\label{eq:zetan-UB3}
\end{equation}
\end{remark}

\begin{lemma}
Assume {\bf (A1)} holds.
Consider any $P'\in\Pi^*$ and two positive vanishing sequences $\delta_n \ge \tilde{\delta}_n \ge n^{-1/2}$,
and let $\tilde{P}_n = P' + \tilde{\delta}_n \tilde{h}^*$ where $\tilde{h}^* \in \scrL(\calX^*)$. 
Then $Q_n \triangleq (\tilde{P}_n W) = Q^* + \delta_n (\tilde{h}^* W)$ is independent of $P'$. Let \footnote{
	Recall the definitions of $\breve{v}$, $\tilde{v}$, $\Delta$ in \eqref{eq:vbreve},
	\eqref{eq:vtilde}, and \eqref{eq:Delta}, respectively.}
\begin{equation}
   \breve{g} = - \frac{t_\epsilon}{2\sqrt{V_\epsilon}} \,\breve{v}^{(P')},
	\quad \tilde{g} = - \frac{t_\epsilon}{2\sqrt{V_\epsilon}} \,\tilde{v} ,
	\quad g = - \frac{t_\epsilon}{2\sqrt{V_\epsilon}} \,v^{(P')} .
\label{eq:g-breve-tilde}
\end{equation}
{\bf (i)} The following Taylor expansion applies to the function $\zeta_n$ of \eqref{eq:zeta-PQ} and 
holds uniformly over the closed $\delta_n$-neighborhood $\{P\in\scrP(\calX) :~\|P-P'\|_\infty \le \delta_n\}$ and $P'\in\Pi^*$:
\begin{eqnarray}
  \zeta_n(P,Q_n) & = & \zeta_n(P';W) 
		+ n(P-P')^\top \Delta + \frac{n}{2} \|\tilde{P}_n - P'\|_\sfJ^2 + \sqrt{n} \,\breve{g}^\top (\tilde{P}_n - P') \nonumber \\
	& & \quad - n (P-P')^\top \sfJ^\calX (\tilde{P}_n - P') + \sqrt{n} \tilde{g}^\top (P-P') 
	+ O(\delta_n^2 \sqrt{n}) .
\label{eq:zeta-asymp-general}
\end{eqnarray}
{\bf (ii)} If $\tilde{\delta}_n = n^{-1/2}$, 
\eqref{eq:zeta-asymp-general} simplifies to
\begin{eqnarray}
  \zeta_n(P,Q_n) & = & \zeta_n(P';W) 
		+ n(P-P')^\top \Delta + \frac{1}{2} \tilde{h}^{*\top} \sfJ \tilde{h}^* + \breve{g}^\top \tilde{h}^*
		- \sqrt{n} \,(P-P')^\top (\sfJ^\calX \tilde{h}^* - \tilde{g}) \nonumber \\
	&& \hspace*{3in} + O(\delta_n^2 \sqrt{n}) .
\label{eq:zeta-asymp}
\end{eqnarray}
\label{lem:zetan-PQ}
\end{lemma}

The proof of the lemma is given in Appendix~\ref{app:zetan-PQ}. 
The lemma will be used to prove Prop.~\ref{prop:maxP1} 
(with $\tilde{\delta}_n = n^{-1/2} \ll \delta_n = \frac{1}{n^{1/4} \log n})$.

The following dimensionality lemma~\ref{lem:rank} and the elements of geometry below it are used
to prove the saddlepoint lemma~\ref{lem:game} in the case $\Pi$ is not a singleton ($r < |\Xcap|$).
In that analysis, it is crucial to characterize feasible perturbation directions $h$ away from
any $P' \in \Pi^*$. When $\calX^* = \Xcap$, the feasible set for $h$ is simply $\scrL(\Xcap)$.
However when $\calX^*$ is a strict subset of $\Xcap$, we must have $h(x) \ge 0$ for all $x\in\Xcap\setminus\calX^*$,
and the feasible set is $\Cin$ defined  in \eqref{eq:Cin}.

\begin{lemma} (Dimensionality lemma).
Let $r = \mathrm{rank}(\Jcap)$. Then
\begin{eqnarray}
   \mathrm{dim}(\Pi) & = & \mathrm{nullity}(\Jcap) = |\Xcap| - r ,  \label{eq:nullity-Jcap} \\
   \Pi^* & = & \scrP(\calX^*) \cap \Pi ,		      				\label{eq:Pi^*-Pi} \\
   \mathrm{dim}(\Pi^*) & = & \mathrm{nullity}(\sfJ) , 				\label{eq:nullity-J} \\
   |\calX^*| & \le & r + \mathrm{dim}(\Pi^*) .		  				\label{eq:dimensions}
\end{eqnarray}
\label{lem:rank}
\end{lemma}
\vspace*{-0.5in}

\noindent
{\em Proof}.

\noindent
{\bf (i)} By \eqref{eq:Pi-space}, $\mathrm{dim}(\Pi) \le \mathrm{nullity}(\Jcap)$.
Moreover, since $C > 0$, $\Pi$ may not contain any vertex of the probability simplex $\scrP(\calX)$,
and therefore no edge or face either. The affine space $P_0 + \mathrm{ker}(\Jcap)$ may therefore not 
be tangent to the simplex. Hence $\mathrm{dim}(\Pi) = \mathrm{nullity}(\Jcap)$, and \eqref{eq:nullity-Jcap} holds.

\noindent
{\bf (ii)} By definition of $\Pi^*$ and $\calX^*$ in \eqref{eq:Pi*} and \eqref{eq:X*}, we have 
$\Pi^* \subseteq \scrP(\calX^*) \cap \Pi$. We now show that $\scrP(\calX^*) \cap \Pi \subseteq \Pi^*$. 
Let $\ell$ = nullity($\sfJ$). If $\ell= 0$, pick $P^* \in \Pi^*$. Then 
\[ P \in \scrP(\calX^*) \cap \Pi \,\;\Leftrightarrow\,\; 
	P-P^* \in \scrL(\calX^*) \cap \mathrm{ker}(\Jcap) = \scrL(\calX^*) \cap \mathrm{ker}(\sfJ) = \{0\} 
	\,\;\Leftrightarrow\,\; P = P^* \in \Pi^* , 
\] 
hence \eqref{eq:Pi^*-Pi} holds.
If $\ell \ge 1$, let $u_1, \cdots, u_\ell$ be a basis
for $\scrL(\calX^*) \cap \mathrm{ker}(\Jcap)$ and pick $P^* \in \mathrm{int}(\Pi^*)$.
Hence $\supp\{P^*\} = \calX^*$, $\supp\{u_i\} \subseteq \calX^*$ $\forall i$, and there exists $\delta > 0$ 
such that $P_i \triangleq P^* + a_i u_i \in \scrP(\calX^*) \cap \Pi$ for all $i$ and $|a_i| < \delta$. 
Moreover $V(P_i;W) = P_i^\top \tilde{v} = V(P^*;W) + a_i u_i^\top \tilde{v}$.
Since $P^* \in \Pi^*$, $P_i \in \Pi$, and the signs of $\{a_i\}$ are arbitrary, 
we must have $u_i^\top \tilde{v} = 0$ for $i=1,\cdots,\ell$.
Now any $P \in \scrP(\calX^*) \cap \Pi$ can be expressed as $P = P^* + \sum_{i=1}^\ell a_i u_i$
for some $\{a_i\}_{i=1}^\ell$, hence $V(P;W) = P^\top \tilde{v} = V(P^*;W)$, hence $P\in\Pi^*$, 
hence \eqref{eq:Pi^*-Pi} holds.

\noindent
{\bf (iii)} Follows from the proof of {\bf (ii)} above.

\noindent
{\bf (iv)} The claim \eqref{eq:dimensions} follows from \eqref{eq:Pi^*-Pi} and the rank-nullity theorem:
\begin{equation}
   \underbrace{\mathrm{dim}(\scrP(\calX^*))}_{=|\calX^*|} + \underbrace{\mathrm{dim}(\Pi)}_{= |\Xcap| - r}
	= \underbrace{\mathrm{dim}(\scrP(\calX^*) \cap \Pi)}_{= \mathrm{dim}(\Pi^*)} + \underbrace{\mathrm{dim}(\scrL(\calX^*) + \Pi)}_{\le |\Xcap|} .
\label{eq:dimensions2} 
\end{equation}
\hfill $\Box$

\begin{figure}[h!]
\begin{center}
\begin{tabular}{cc}
\includegraphics[width=3in]{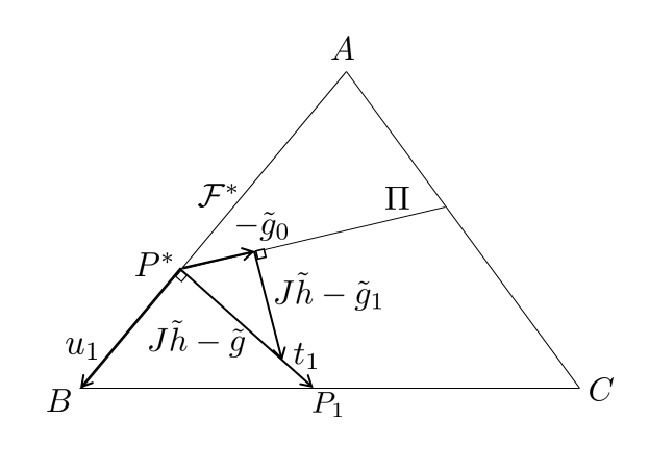} &
\includegraphics[width=3in]{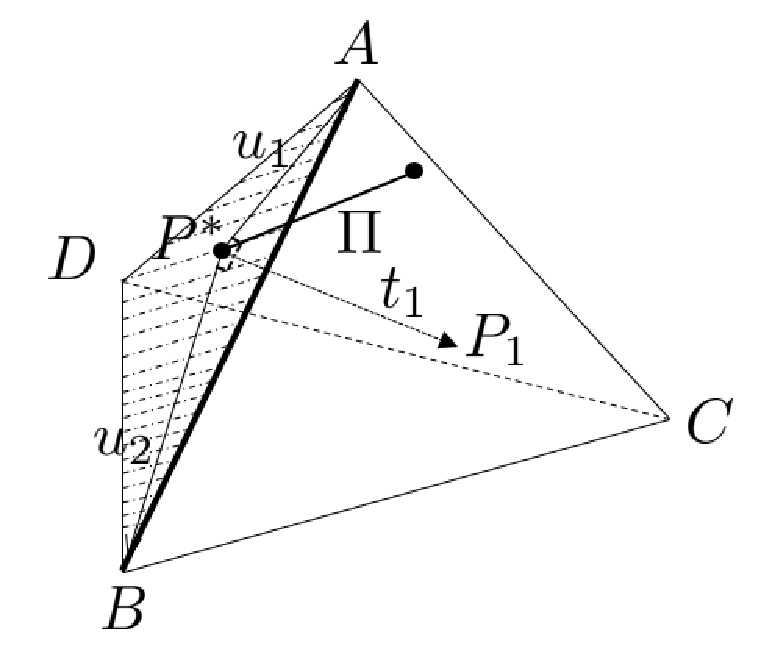} \\
(a) $|\Xcap| = 3$, $r=2$, $\calF^* = [A,B]$, $|\calX^*|=2$ & (b) $|\Xcap| = 4$, $r=3$, $\calF^* = ABD$, $|\calX^*|=3$ \\
\includegraphics[width=3in]{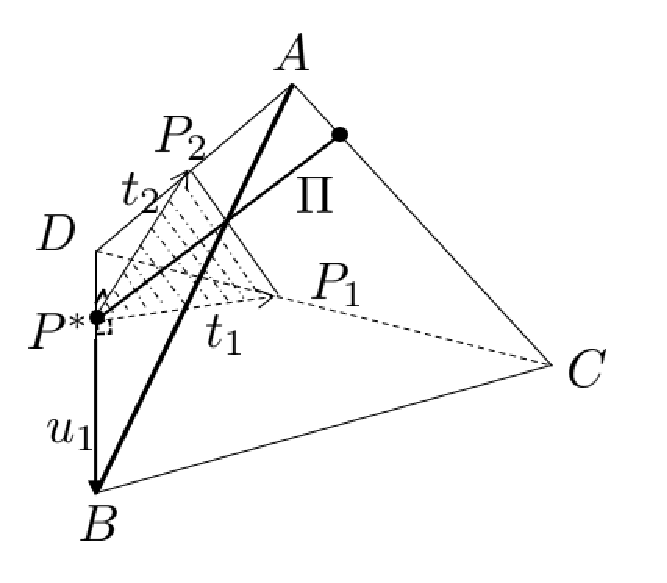} &
\includegraphics[width=3in]{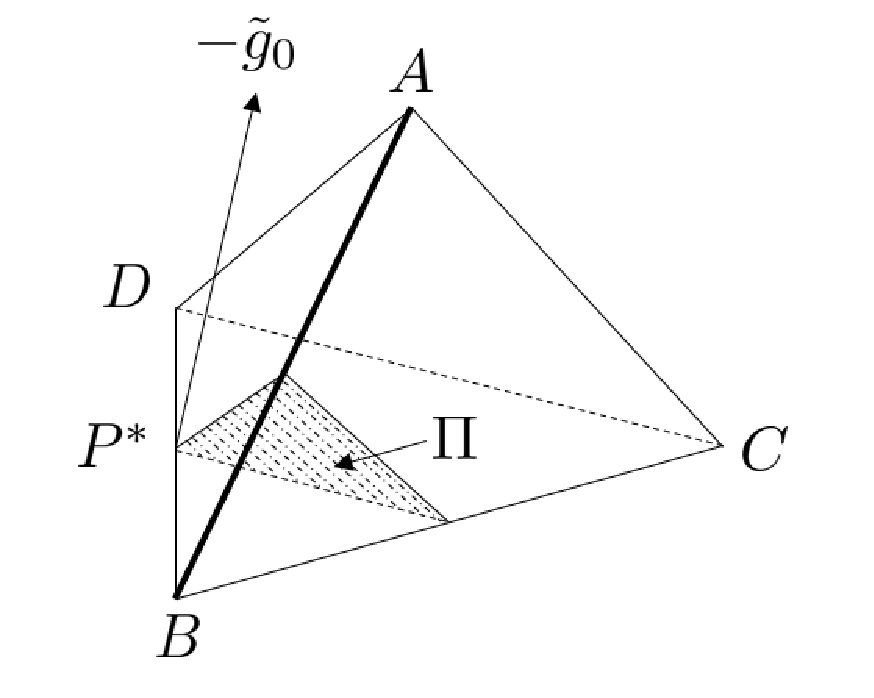} \\
(c) $|\Xcap| = 4$, $r=3$, $\calF^* = [B,D]$, $|\calX^*|=2$ & (d) $|\Xcap| = 4$, $r=2$, $\calF^* = [B,D]$, $|\calX^*|=2$
\end{tabular}
\end{center}
\caption{Geometric illustration of saddlepoint property of Lemma~\ref{lem:game} in case $\mathrm{dim}(\Pi) > \mathrm{dim}(\Pi^*)$.
	In all cases depicted here, $\Pi^* = \{P^*\}$ is a singleton, hence $\calX^* = \mathrm{supp}(P^*)$.}
\label{fig:saddlepoint}
\end{figure}

{\bf Geometry of $\Pi$ and $\Pi^*$}.
(See Fig.~\ref{fig:saddlepoint} for an illustration). Assume $\mathrm{dim}(\Pi) > \mathrm{dim}(\Pi^*)$
(otherwise $\Pi$ is simply equal to $\Pi^*$).
Denote by $\calF^*$ the face of the probability simplex associated with distributions in $\scrP(\calX^*)$.
Fix any $P^* \in \Pi^*$ and let $u_1, \cdots, u_{|\calX^*|-1}$ form a basis for $\scrL(\calX^*)$.
Denote by $\calT_0$ the $|\Xcap|-|\calX^*|$ dimensional subspace of $\scrL(\Xcap)$ that is orthogonal to $\scrL(\calX^*)$,
namely,
\[ \calT_0 = \{ h\in\scrL(\Xcap) :~h(x) = \mathrm{constant~for~} x\in\calX^*\} .\]
The intersection of the affine subspace $P^* + \calT_0$ with $\scrP(\Xcap)$ defines a (scaled) 
standard simplex $\calS \subset \rR^{|\Xcap|-|\calX^*|+1}$ that has $P^*$ as a vertex as well as
$|\Xcap|-|\calX^*|$ other vertices denoted by $P_1,\cdots, P_{|\Xcap|-|\calX^*|}$. 
Define $t_i = P_i - P^* \in \scrL(\Xcap)$ for each $i$. 
Hence $t_1, \cdots, t_{|\Xcap|-|\calX^*|}$ form a basis for $\calT_0$ and
are orthogonal to $\scrL(\calX^*)$.

Augmenting $\calT$ to include vectors with nonzero mean, we define the following $|\Xcap|-|\calX^*|+1$ dimensional subspace of $\rR^\Xcap$:
\begin{eqnarray*}
   \calT & \triangleq & \{ z = h + \alpha 1 ,\;h\in\calT_0 , \,\alpha\in\rR\} \\
	& = & \{ z\in\rR^\Xcap :~z(x) = \mathrm{constant~for~} x\in\calX^*\}
\end{eqnarray*}
and the infinite cones
\begin{eqnarray}
   \CTin & \triangleq & \left\{h\in\calT_0 :\;h = \sum_i \alpha_i t_i ,\;\alpha_i \ge 0 ,\;i=1,2,\cdots,|\Xcap|-|\calX^*|\right\} \nonumber \\
   	& = & \{ h\in\scrL(\Xcap) :~h(x) = \mathrm{constant~for~} x\in\calX^*, \;h(x) \ge 0 \;\forall x\in\Xcap\setminus\calX^*\}
\label{eq:CTin}
\end{eqnarray}
and
\begin{eqnarray}
   \CTout & \triangleq & \{z\in\calT : z^\top t_i \ge 0 ,\;i=1,2,\cdots,|\Xcap|-|\calX^*|\} \nonumber \\
   	& = & \{ z\in\rR^\Xcap :~z(x) = \mathrm{constant~for~} x\in\calX^*, \;z(x) \ge 0 \;\forall x\in\Xcap\setminus\calX^*\} .
\label{eq:CTout}
\end{eqnarray}
Also define the polytopes
\begin{eqnarray}
   \Cin & \triangleq & \left\{h\in\scrL(\Xcap) : \mathrm{Proj}_{\calT_0} (h) \in \CTin \right\} \nonumber \\
	& = & \{ h\in\scrL(\Xcap) :~h(x) \ge 0 \;\forall x\in\Xcap\setminus\calX^*\} \label{eq:Cin} \\
  \Cout & \triangleq & \left\{z\in\rR^\Xcap : \mathrm{Proj}_\calT (z) \in \CTout \right\} \nonumber \\
	& = & \{ z\in\rR^\Xcap :~z(x) \ge 0 \;\forall x\in\Xcap\setminus\calX^*\} . \label{eq:Cout}
\end{eqnarray}
Note that $P^* + h \in \scrP(\Xcap) \;\Rightarrow\; h\in\Cin$.

Denote by $\tilde{g}_0$ and $\tilde{g}_1$ respectively the projections of $\tilde{g}$ of \eqref{eq:g-breve-tilde} onto 
$\mathrm{ker}(\Jcap)$ and $\mathrm{ker}(\Jcap)^\perp$.
By the extremal property of $P^*$ in $\Pi$, we have $- \tilde{g}_0^\top (P-P^*) \ge 0$ for all $P\in\Pi$,
hence,
\begin{equation}
   - \tilde{g}_0 \in \Cout .
\label{eq:g0}
\end{equation}

The following properties follow immediately from the symmetry properties of the standard simplex.
\begin{itemize}
\item Any $P\in\scrP(\Xcap)$ satisfies $\mathrm{Proj}_{\calT_0} (P-P^*) \in \CTin$, i.e., $P-P^* \in \Cin$.
\item For all $z \in \calT$,
\begin{equation}
   z \in \CTout \quad \Longleftrightarrow \quad \mathrm{Proj}_{\mathrm{ker}(\Jcap)} (z) \in \Cout .
\label{eq:Proj-kerJ}
\end{equation}
\item For all $h\in\Cin$ and $z\in\CTout$,
	\begin{equation}
	   h^\top z \ge 0 .
	\label{eq:hz+}
	\end{equation}
\end{itemize}
If $\mathrm{dim}(\Pi) = \mathrm{dim}(\Pi^*)$, we define $\Cin = \scrL(\Xcap)$.

{\bf An optimization game}.
Now consider the game with payoff function
\begin{equation}
   \Gamma(h,\tilde{h}) \triangleq \frac{1}{2} \tilde{h}^\top \Jcap \tilde{h} 
		+ \tilde{h}^\top \breve{g} - h^\top (\Jcap \tilde{h}  - \tilde{g}) ,
	\quad \begin{cases} h, \tilde{h} \in \scrL(\Xcap) & :~r = |\Xcap| \\
		h \in \Cin, \;\tilde{h} \in \calH & :~\mathrm{else,} \end{cases}
\label{eq:gameE}
\end{equation}
to be maximized over $h$ and minimized over $\tilde{h}$. 
The payoff function is linear in $h$ and convex quadratic in $\tilde{h}$.
Recall $\calH = \scrL(\calX^*) \cap \mathrm{ker}(\sfJ)^\perp$ was defined in \eqref{eq:H}.

Lemma~\ref{lem:game} below shows that the game \eqref{eq:gameE} admits a saddlepoint
and will be used to prove Prop.~\ref{prop:maxP1}. Fig.~\ref{fig:saddlepoint} provides
a geometrical illustration of the saddlepoint property in the case of rank-deficient $\Jcap$. 

One might expect that $\inf_h \Gamma(h,\tilde{h}) = -\infty$ because the payoff function is linear in $h$. This is true
for all $\tilde{h}$, except for a special choice $\tilde{h}^*$. The key idea 
in this game is most easily understood in the case $\Pi^*$ is a singleton ($\sfJ$ has rank $|\calX^*|$). 
Then the variable $\tilde{h} \in \calH \perp \mathrm{ker}(\Jcap)$ has $r-1 \ge |\calX^*|-1$ degrees of freedom, 
hence one can construct $\tilde{h}^*$ to ``cancel'' $\tilde{g}$ in the sense that $z = \Jcap \tilde{h}^* - \tilde{g} \in \calT$ 
is orthogonal to $\scrL(\calX^*)$ (i.e., $z$ satisfies $|\calX^*|-1$ orthogonality conditions). 
Moreover $z \in \CTout$, hence by \eqref{eq:hz+},
$h^\top (\Jcap \tilde{h}^* - \tilde{g}) = h^\top z \ge 0$ for all $h\in\Cin$.
The argument can be extended to the case $\Pi^*$ is not a singleton, by exploiting the fact that $\tilde{g}$ is orthogonal to $\mathrm{ker}(\sfJ)$.

\begin{lemma}
The game (\ref{eq:gameE}) admits a saddlepoint $(h^*, \tilde{h}^*)$ where
$\tilde{h}^* = \sfJ^+ \tilde{g}$ and $h^* = \sfJ^+ g \in \calH$.
The value of the game is $\Gamma^* = \frac{1}{2} \|g\|_{\sfJ^+}^2 - \frac{1}{2} \|\breve{g}\|_{\sfJ^+}^2$,
where $g=\tilde{g}+\breve{g}$. 
The saddlepoint satisfies the equalizer property
\begin{equation}
   \Gamma(h,\tilde{h}^*) = \Gamma(h^*,\tilde{h}^*) \le \Gamma(h^*,\tilde{h}) \quad \forall \mathrm{~feasible~} h, \tilde{h} .
\label{eq:gameE-SP}
\end{equation}
\label{lem:game}
\end{lemma}

{\em Proof}. 
First consider the maximization of $\Gamma(h,\tilde{h})$ with respect to $h$. We have
\begin{eqnarray}
   \sup_{\mathrm{feasible~}h} \Gamma(h,\tilde{h}) & = & \frac{1}{2} \tilde{h}^\top \Jcap \tilde{h} + \tilde{h}^\top \breve{g} 
	- \inf_{\mathrm{feasible~}h} [h^\top (\Jcap \tilde{h} - \tilde{g})] 
\label{eq:gameE-sup-h}
\end{eqnarray}
If $\tilde{h} = \sfJ^+ g = \tilde{h}^*$, we claim that the infimum above is zero. 
\begin{itemize}
\item
If the capacity-achieving input distribution is unique, this is straightforward: $\calX^* = \Xcap$, $\Jcap = \sfJ$ has rank $|\Xcap|$,
and $\Jcap \tilde{h}^* - \tilde{g}$ is equal to the mean of the vector $- \tilde{g}$ times the vector $\mathds1\{x\in\Xcap\}$. 
Since this vector is orthogonal to the feasible space $\scrL(\Xcap)$ for $h$, the infimum of \eqref{eq:gameE-sup-h} is zero
and is achieved by any $h\in\scrL(\Xcap)$.
For any other choice of $\tilde{h}$, the infimum would be $-\infty$.
\item
If the capacity-achieving input distribution is not unique, then $\Jcap$ has rank $r < |\Xcap|$.
\begin{itemize} 
\item {\em Case~I: $\Pi^*$ is a singleton}.
   By \eqref{eq:nullity-J}, $\sfJ$ has rank $|\calX^*|$, and by \eqref{eq:dimensions}, $|\calX^*| \le r$.
   We have $(\sfJ \tilde{h}^* - \tilde{g})(x) = (\Jcap \tilde{h}^* - \tilde{g})(x) =$ 
   constant for all $x\in\calX^*$ or equivalently, $z \triangleq \Jcap \tilde{h}^* - \tilde{g} \perp \scrL(\calX^*)$, hence $z \in \calT$.
   Moreover $z + \tilde{g}_0 = \Jcap \tilde{h}^* - \tilde{g}_1 \perp \mathrm{ker}(\Jcap)$, 
   hence $\mathrm{Proj}_{\mathrm{ker}(\Jcap)} (z) = - \tilde{g}_0$,
   hence \eqref{eq:g0} and \eqref{eq:Proj-kerJ} imply $z \in \CTout$.
   Therefore
   \[
      \inf_{h\in\Cin} [h^\top (\Jcap \tilde{h}^* - \tilde{g})] 
	   = \inf_{h\in\Cin} h^\top z = 0 
   \]
   where the last equality follows from \eqref{eq:hz+}.
\item {\em Case~II: $\mathrm{dim}(\Pi^*) =k \ge 1$}. 
   We have $z = \Jcap \tilde{h}^* - \tilde{g} \perp \mathrm{ker}(\Jcap)$. 
   By \eqref{eq:nullity-J}, nullity$(\sfJ) = k$.
   Let $u_1, \cdots, u_k$ be basis vectors for $\scrL(\calX^*) \cap \mathrm{ker}(\sfJ)$ 
   and $u_{k+1}, \cdots, u_{|\calX^*|-1}$ be basis vectors for $\scrL(\calX^*) \cap \mathrm{ker}(\sfJ)^\perp = \calH$.
   By the extremal property of $\Pi^*$, we have $\tilde{g}^\top u_1 = \cdots = \tilde{g}^\top u_k = 0$.
   Since $u_1, \cdots, u_k \in \mathrm{ker}(\sfJ) \subseteq \mathrm{ker}(\Jcap)$, we obtain $z^\top u_1 = \cdots = z^\top u_k = 0$.
   On the other hand, the minimizing variable $\tilde{h} \in \calH$ is in the range of $\sfJ$ and therefore of $\Jcap$, 
   and has $r-1$ degrees of freedom which can be used to satisfy the $|\calX^*| - k -1 \le r-1$ equations 
   $z^\top u_{k+1} = \cdots = z^\top u_{|\calX^*|-1} = 0$. Therefore choosing $\tilde{h} = \tilde{h}^*$ 
   results in $z \perp \scrL(\calX^*)$ again, and the rest of the proof follows as in Case~I above.
\end{itemize}
\end{itemize}

Next, consider the minimization of $\Gamma(h,\tilde{h})$ with respect to $\tilde{h}$. 
Fix $h = h^* $. Then
\begin{eqnarray}
   \Gamma(h^*,\tilde{h}) 
	& = & \frac{1}{2} \tilde{h}^\top \Jcap \tilde{h} - \tilde{h}^\top (\Jcap h^* - \breve{g}) + h^{*\top} \tilde{g} \nonumber \\
	& = & \frac{1}{2} \tilde{h}^\top \sfJ \tilde{h} - \tilde{h}^\top (\sfJ h^* - \breve{g}) + h^{*\top} \tilde{g}
\label{eq:gameE-sup-htilde}
\end{eqnarray}
which is a convex quadratic function of $\tilde{h}$. By \eqref{eq:J+}, its minimizer is obtained as 
$\sfJ^+ (\sfJ h^* - \breve{g}) = \sfJ^+ (g - \breve{g}) = \sfJ^+ \tilde{g} = \tilde{h}^*$, 
and the inequality holds in (\ref{eq:gameE-SP}).

Hence $(h^*,\tilde{h}^*)$ is a saddlepoint, and the value of the game is 
\[
   \Gamma^* = \Gamma(h^*,\tilde{h}^*) = \frac{1}{2} \tilde{h}^{*\top} \sfJ \tilde{h}^* + \tilde{h}^{*\top} \breve{g} .
\]
Moreover
\begin{equation}
	\Gamma^* = \frac{1}{2} \tilde{g}^\top \sfJ^+ (\tilde{g} + 2 \breve{g}) 
	= \frac{1}{2} (g-\breve{g})^\top \sfJ^+ (g+\breve{g}) 
	= \frac{1}{2} g^\top \sfJ^+ g - \frac{1}{2} \breve{g}^\top \sfJ^+ \breve{g} .
\label{eq:Gamma*-app}
\end{equation}
\hfill $\Box$

\section{Background: Refined Asymptotics}
\label{sec:asymptotics}

This section reviews basic techniques for central limit asymptotics and strong large deviations.

\subsection{Central Limit Asymptotics}
\label{sec:CLT}

We first review some key results on the asymptotics of
a normalized sum of independent random variables. Consider first 
iid random variables $U_i, i \ge 1$, with common cdf $F_U$, finite mean $\mu$,
variance $\sigma^2 > 0$, and skewness $S \triangleq \eE[(U-\mu)^3]/\sigma^{3/2}$.
The normalized random variable
\begin{equation}
  T_n = \frac{\sum_{i=1}^n U_i - n\mu}{\sqrt{n\sigma^2}}
\label{eq:Tn}
\end{equation}
has zero mean and unit variance and by the Central Limit Theorem,
converges in distribution to $\calN(0,1)$. Denote by $F_n$ the cdf of $T_n$.

{\bf iid Nonlattice Random Variables.}
If $\{ U_i \}_{i=1}^\infty$ are iid nonlattice random variables,
the Cram\'{e}r-Ess\'{e}en theorem \cite[p.~49]{Esseen45} \cite[p.~538]{Feller71}
for nonlattice random variables states that
\begin{equation}
   F_n(t) = \Phi(t) - \frac{S}{6\sqrt{n}} (1-t^2) \,\phi(t) + o(1/\sqrt{n})
\label{eq:CE}
\end{equation}
uniformly in $t$, i.e.,
\[
   \lim_{n\to\infty} \left\{ \sqrt{n} \sup_{t\in\rR} \left| F_n(t) - \Phi(t) 
	- \frac{S}{6\sqrt{n}} (1-t^2) \,\phi(t) \right| \right\} = 0.
\]
The $o(1/\sqrt{n})$ remainder in (\ref{eq:CE}) can be strengthened to $O(1/n)$
for {\em strongly nonlattice random variables}, i.e., distributions that satisfy Cram\'{e}r's 
condition on the characteristic function $\hat{f}$ of $U$:
\begin{equation}
  \lim_{|\omega| \to \infty} |\hat{f}(\omega)| < 1 ,
\label{eq:cramer-cond}
\end{equation}
in which case (\ref{eq:CE}) is the first-order Edgeworth expansion of $F_n$ \cite{Wallace58,DasGupta08}.

Higher-order expansions in terms of successive powers of $n^{-1/2}$ can also
be derived \cite{Wallace58,Feller71,Barndorff91,DasGupta08}
but a two-term expansion suffices for our purposes. The Berry-Ess\'{e}en formula 
$|F_n(t) - \Phi(t)| \le \frac{A_3}{6 V^{3/2} \sqrt{n}}$
(where $A_3 \triangleq \eE[|U-\mu|^3]$ is the {\em absolute} third central moment of $U$)
that was used in \cite{Strassen62,Polyanskiy10} does not yield
the sharp asymptotics of interest here.

Let $\epsilon = 1-\alpha$ and $t_\epsilon$ and $t_{\epsilon,n}$ be the upper $\alpha$-quantiles of $\Phi$ 
and $F_n$ respectively, i.e., $\Phi(t_\epsilon) = \alpha$ and $F_n(t_{\epsilon,n}^-) \le \alpha \le F_n(t_{\epsilon,n})$. 
The precise asymptotics of $t_{\epsilon,n}$ are of interest, in particular the Cornish-Fisher inversion
formula \cite{Wallace58,Barndorff91,DasGupta08,Cornish37} yields
\begin{equation}
   t_{\epsilon,n} = t_\epsilon + \frac{S}{6\sqrt{n}} (t_\epsilon^2 -1) + o(n^{-1/2}) .
\label{eq:CF}
\end{equation}

{\bf iid Lattice Random Variables.}
If $\{U_i\}_{i=1}^\infty$ are iid lattice random variables with span $d$, 
the normalized sum $T_n$ of \eqref{eq:Tn} is also a lattice random variable
(with span $d/\sqrt{n\sigma^2}$) and its cdf $F_n(t)$ is piecewise constant with jumps of size $O(n^{-1/2})$
at the lattice points. An $o(n^{-1/2})$ asymptotic approximation to $F_n$ can still be constructed,
analogously to (\ref{eq:CE}) in the nonlattice case \cite[pp.~52---67]{Esseen45} \cite[p.~540]{Feller71}. 
In particular, (\ref{eq:CE}) holds at the midpoints of the lattice. 
At the lattice points, (\ref{eq:CE}) holds with $F_n(t)$ replaced with $\frac{1}{2} [F_n(t) + F_n(t^-)]$.

{\bf Non-iid Random Variables.}
In case $U_i, \,i \ge 1$, are independent but have different distributions,
with respective means $\mu_i$ and variances $\sigma_i^2$ such that
\begin{equation}
  \overline{\sigma}_n^2 \triangleq \frac{1}{n} \sum_{i=1}^n \sigma_i^2
\label{eq:Vn-BE-noniid}
\end{equation}
is bounded away from 0 and $\infty$ as $n \to \infty$, let
\begin{equation}
   \overline{S}_n \triangleq \frac{\frac{1}{n}\sum_{i=1}^n \eE[(U_i-\mu_i)^3]}{\overline{\sigma}_n^3} .
\label{eq:xi-BE-noniid}
\end{equation}
Then the expansions (\ref{eq:CE}) and (\ref{eq:CF}) for the iid nonlattice case hold with $S$ 
replaced by $\overline{S}_n$ \cite[pp.~546, 547]{Feller71}: \footnote{
	Note a typo in \cite[Eqn~(6.1)]{Feller71}, where $s_n^2$ should be replaced with $s_n^3$.}
\begin{eqnarray}
   F_n(t) & = & \Phi(t) - \frac{\overline{S}_n}{6\sqrt{n}} (1-t^2) \,\phi(t) + o(1/\sqrt{n}) \label{eq:CE-noniid} \\
   t_{\epsilon,n} & = & t_\epsilon + \frac{\overline{S}_n}{6\sqrt{n}} (t_\epsilon^2 - 1) + o(n^{-1/2}) \label{eq:CF-noniid}
\end{eqnarray}
provided that the following additional assumptions hold: 
each $U_i$ has bounded fourth moment, and there exists $\delta > 0$ such that
\begin{equation}
   \left| \prod_{i=1}^n \hat{f}_i(\omega) \right| = o(n^{-1/2}) \quad \mathrm{uniformly~in~} \omega > \delta
\label{eq:noniid-feller}
\end{equation}
where $\hat{f}_i (\omega) \triangleq \eE[e^{i\omega U_i}]$ is the characteristic function of $U_i$.
Condition \eqref{eq:noniid-feller} holds when each $U_i$ is a nonlattice random variable,
as well as in many problems where each $U_i$ is a lattice random variable (but $\sum_{i=1}^n U_i$ is not). 
In the case each $\sum_{i=1}^n U_i$ is a lattice random variable with span $d_n$, \eqref{eq:noniid-feller}
does not hold but the $o(n^{-1/2})$ asymptotic approximation (\ref{eq:CE}) to $F_n$
holds at midpoints of the lattice, with $\sigma^2$ and $S$ replaced with $\overline{\sigma}_n^2$ and $\overline{S}_n$ of 
\eqref{eq:Vn-BE-noniid} and \eqref{eq:xi-BE-noniid}.

\subsection{Strong Large-Deviations Asymptotics}
\label{sec:strongLD}

Conventional large-deviations methods characterize the exponential decay of tail probabilities,
in particular, Cram\'{e}r's theorem \cite{Cramer38} states that
\[ - \frac{1}{n} \log P^n \left\{ \sum_{i=1}^n U_i \ge na \right\} \sim \Lambda(a) \quad \mbox{as~} n \to \infty ,
	\quad \forall a > \eE[U] \]
where $U_i , \,i \ge 1$ is a sequence of iid random variables with common distribution $P$,
cumulant generating function (cgf) $\kappa(s) = \log \eE_P[e^{sU}]$, 
and large-deviations function $\Lambda(a) = \sup_s [as - \kappa(s)]$.

Strong large-deviations asymptotics provide arbitrarily accurate approximations
to tail probabilities, using Laplace's method of asymptotic expansion of integrals.
If $U$ is a nonlattice random variable and $U_i , \,i \ge 1$, are iid then the Bahadur-Rao theorem
\cite{Bahadur60} yields
\begin{equation}
   P^n \left\{ \sum_{i=1}^n U_i \ge na \right\} \sim \frac{e^{-n\Lambda(a)}}{s \sqrt{2\pi n \kappa''(s)}} 
	\quad \mbox{as~} n \to \infty
\label{eq:bahadur}
\end{equation}
where $s$ satisfies $\kappa'(s) = a$. Define the normalized random variable 
$W_n = (\sum_{i=1}^n U_i - na)/\sqrt{n\kappa''(s)}$ which has zero mean and unit variance,
and the exponentially tilted distribution $d\tilde{F}_U(u) \triangleq e^{su-\kappa(s)} dF_U(u)$.
Denote by $\widetilde{F}_n$ the cdf for $W_n$ when $U$ has cdf $\tilde{F}_U$.
The asymptotic formula (\ref{eq:bahadur}) is obtained by writing the identity
\begin{equation}
   P^n \left\{ \sum_{i=1}^n U_i \ge na \right\} 
	= e^{-n\Lambda(a)} \int_0^\infty e^{-s\sqrt{n\kappa''(s)} v} d\widetilde{F}_n(v)
\label{eq:esscher-iid}
\end{equation}
and evaluating the asymptotics of the integral in the right side.
Since $U$ is a nonlattice random variable, $\widetilde{F}_n$ does not exhibit $O(n^{-1/2})$ jumps,
and \eqref{eq:bahadur} is derived by applying the Cram\'{e}r-Ess\'{e}en expansion \eqref{eq:CE} 
to $\widetilde{F}_n$.

For lattice random variables, $\widetilde{F}_n$ has jumps of size $O(n^{-1/2})$ in the vicinity
of the origin. Blackwell and Hodges \cite{Blackwell59}
showed that the right side of \eqref{eq:bahadur} is multiplied by a bounded, oscillatory
function of $n$. In particular, if $na$ is in the range of the sum $\sum_{i=1}^n U_i$, then
\begin{equation}
   P^n \left\{ \sum_{i=1}^n U_i \ge na \right\} \sim \frac{d\,e^{-n\Lambda(a)}}{(1-e^{-sd}) 
	\sqrt{2\pi n \kappa''(s)}}  \quad \mbox{as~} n \to \infty
\label{eq:blackwell}
\end{equation}
where $d$ is the span of the lattice random variable $U$.
As expected, the asymptotic expansion \eqref{eq:blackwell} coincides with \eqref{eq:bahadur}
in the limit as $d \to 0$.

{\bf Example}: Tail of binomial distribution Bi($n,p$) which is a sum of $n$ iid Be($p$)
random variables where $1/2 \le p < 1$. Then $\kappa(s) = \log (1 + e^s)$ and $\kappa''(s) = e^s / (1+e^s)^2$.
The large-deviations function is $\Lambda(a) = d(a\|p) = a \log \frac{a}{p} + (1-a) \log \frac{1-a}{1-p}$
for $a \in (p,1)$, and $s$ that achieves the supremum in the definition of $\Lambda(a)$ is $s = \log \frac{a}{1-a}$,
corresponding to $\kappa''(s) = a(1-a)$. The lattice span is $d=1$. 
Thus (\ref{eq:blackwell}) yields
\begin{equation}
   \Pr \{\mathrm{Bi}(n,p) \ge na\} \sim \frac{a}{2a-1} \,\frac{e^{-n d(a\|p)}}{\sqrt{2\pi a(1-a) n}} , \quad p < a < 1 ,
\label{eq:binomial-exact}
\end{equation}
for any integer $na \to \infty$.
Compare with the oft-used approximation \cite[p.~115]{Ash65} which states, for $p=1/2$,
\[  \frac{e^{-n d(a\|p)}}{\sqrt{2\pi a(1-a) n}} \le \Pr \{\mathrm{Bi}(n,p) \ge na\} \le  e^{-n d(a\|p)} . \]
Clearly the lower bound is tighter than the upper bound as it captures the $n^{-1/2}$ asymptotic term. 

A generalization of (\ref{eq:bahadur}) to the non-iid case was given by Bucklew and Sadowsky 
\cite{Bucklew93}. 
An even more more general expression for both the nonlattice and the lattice cases was obtained
by Chaganty and Sethuraman \cite[Theorems~3.3 and 3.5]{Chaganty93} in the non-iid case.
The following special case is particularly relevant to this paper: evaluate the asymptotics of $\Pr\{Z_n \ge n a_n\}$
where $a_n$ is a sequence converging to a limit $a$, and $Z_n$ is a sequence of random variables
with respective cdf's $F_n$ and moment generating functions (mgf) $M_{(n)}(s) \triangleq \eE[e^{s Z_n}]$, 
assumed to be nonvanishing and analytic in the region $\Omega = \{ s\in\cC ~:~|s| < \overline{s}\}$ for some $\overline{s} > 0$.
Denote by $n\overline{\kappa}_n(s)$ the cgf for $Z_n$, and by
\begin{equation}
   \overline{\Lambda}_n(a) = \sup_{s\in\rR} [as - \overline{\kappa}_n(s)]
\label{eq:LD}
\end{equation}
the associated large-deviations function. Assume the supremum defining $\overline{\Lambda}_n(a_n)$
is achieved at $s_n \in \Omega$, hence $a_n = \overline{\kappa}_n'(s_n)$. Define the exponentially tilted
distribution $\widetilde{F}_n(t) \triangleq \int_{-\infty}^t \exp\{ s_n u - a_n \kappa_n(s_n)\} \,dF_n(u)$
and the normalized random variable $W_n = (\widetilde{Z}_n - n a_n)/\sqrt{n\kappa_n''(s_n)}$ 
which has zero mean and unit variance when $\widetilde{Z}_n$ has cdf $\widetilde{F}_n$.
Denoting by $\widetilde{G}_n$ the resulting cdf for $W_n$, one obtains \cite[p.~1683]{Chaganty93}
\begin{equation}
  \Pr\{Z_n \ge n a_n\} = e^{-n \overline{\Lambda}_n(a_n)} \int_0^\infty 
		e^{- s_n \sqrt{n\kappa_n''(s_n)} v} d\widetilde{G}_n(v) .
\label{eq:esscher}
\end{equation}
The characteristic function of $W_n$ is obtained from \eqref{eq:esscher} as \cite[Eqn~(3.7)]{Chaganty93}
\begin{equation}
  \hat{f}_n(\omega) = \exp\left\{ - \frac{i\omega n a_n}{\sqrt{n\kappa_n''(s_n)}} \right\} 
	\,\frac{M_{(n)}(s_n + i\omega/\sqrt{n\kappa_n''(s_n)})}{M_{(n)}(s_n)} , \quad \omega \in \rR 
\label{eq:cf-Vn}
\end{equation}
where $i = \sqrt{-1}$. Further assume there exists $n_0 \in \nN$ such that \footnote{
	It was assumed in \cite{Chaganty93} that $n_0=1$ but this restriction is unnecessary.}
\begin{description}
\item[(CS1)] $\exists \tilde{\kappa} < \infty$ such that $|\kappa_n(s)| < \tilde{\kappa}$
	for all $n \ge n_0$ and $s\in\Omega$.
\item[(CS2)] $\exists V_0 > 0$ such that $\kappa_n''(s_n) > V_0$ for all $n \ge n_0$.
\end{description}
If $Z_n$ is a nonlattice random variable, assume that
\begin{description}
\item[(CS3)] $\exists \delta_0 > 0$ such that
	\begin{equation}
	   \sup_{\delta < |\omega| \le \lambda s_n} \left| 
		\frac{M_{(n)}(s_n + i\omega)}{M_{(n)}(s_n)} \right| = o(1/\sqrt{n})
	\label{eq:CS3}
	\end{equation}
	for any given $\delta$ and $\lambda$ such that $0 < \delta < \delta_0 < \lambda$. 
\end{description}
Under Assumptions {\bf (CS1)},{\bf (CS2)},{\bf (CS3)}, $W_n$ converges in distribution to $\calN(0,1)$.
Theorem~3.3 of \cite{Chaganty93} in the nonlattice case 
(with $a_n$, $s_n$, $\overline{\kappa}_n$ and $\overline{\Lambda}_n$ respectively playing the roles of
$m_n$, $\tau_n$, $\psi_n$ and $\gamma_n$ in \cite{Chaganty93}) states that analogously to \eqref{eq:bahadur},
\begin{equation}
   \Pr \left\{Z_n \ge n a_n \right\} 
	\sim \frac{e^{-n \overline{\Lambda}_n(a_n)}}{s_n \sqrt{2\pi n \overline{\kappa}_n''(s_n)}} 
	\quad \mathrm{as~} n \to \infty . 
\label{eq:chaganty:nonlattice}
\end{equation}

In the lattice case, let $Z_n$ have span $d_n$ and $a_n$ be in the range of $Z_n/n$.
Then $W_n$ is lattice-valued with span $d_n/\sqrt{n\kappa_n''(s_n)}$ and zero offset.
Assumption {\bf (CS3)} above is replaced with
\begin{description}
\item[(CS3')] $\exists \delta_1 > 0$ such that for all $0 < \delta < \delta_1$,
	\[ \sup_{\delta < |\omega| \le \pi/d_n} \left| \frac{M_{(n)}(s_n + i\omega)}{M_{(n)}(s_n)} \right| = o(1/\sqrt{n}) . \]
\end{description}
Under Assumptions {\bf (CS1)},{\bf (CS2)},{\bf (CS3')},
Theorem~3.5 of \cite{Chaganty93} in the lattice case states that analogously to \eqref{eq:blackwell},
\begin{equation}
   \Pr \left\{Z_n \ge n a_n \right\} 
	\sim \frac{d_n \,e^{-n \overline{\Lambda}_n(a_n)}}{(1-e^{-s_n d_n}) \,\sqrt{2\pi n \overline{\kappa}_n''(s_n)}} 
	\quad \mathrm{as~} n \to \infty . 
\label{eq:chaganty:lattice}
\end{equation}
Again \eqref{eq:chaganty:lattice} coincides with \eqref{eq:chaganty:nonlattice}
in the limit as $d_n \to 0$.
In all cases considered, to obtain sharp asymptotics of the form \eqref{eq:chaganty:nonlattice}
or \eqref{eq:chaganty:lattice}, the analysis begins with the identity \eqref{eq:esscher},
and the key step is to verify the asymptotics of the normalized random variable $W_n$.

\section{Strong Large Deviations for Likelihood Ratios}
\label{sec:strongLD-HT}
\setcounter{equation}{0}

This section presents new results on strong large deviations for likelihood ratios.
Of particular interest are NP tests, which are instrumental
in deriving the strong converse to the channel coding theorem.
The full asymptotics of deterministic NP tests for distributions with iid components were obtained 
by Strassen \cite[Theorem~1, p.~690]{Strassen62}. He also derived somewhat looser bounds for the case of 
distributions with independent but not identically distributed components \cite[Theorem~3, p.~702]{Strassen62}, 
as did Polyanskiy {\em et al.} \cite[Lemma~58, p.~2340]{Polyanskiy10} allowing randomized tests.

In this section we derive the full asymptotics of such NP tests. The starting point is a strong large-deviations 
analysis of likelihood ratios (Sec.~\ref{sec:LR-vary}), followed by an application to NP testing (Sec.~\ref{sec:NP}).

\subsection{Likelihood Ratio with Varying Component Distributions}
\label{sec:LR-vary} 

The basic probabilistic model is as follows.
Let $Y_i, \,1 \le i \le n$, be independent random variables with respective distributions
$P_i$ and $Q_i$ under hypotheses $H_1$ and $H_0$, respectively. 
Write $\pP_n \triangleq \prod_{i=1}^n P_i$ and $\qQ_n \triangleq \prod_{i=1}^n Q_i$. 
Assume the following holds:
\begin{description}
\item[(LR1)] For each $i \ge 1$: $P_i \ll Q_i$ ($P_i$ is dominated by $Q_i$).
\item[(LR2)] Under $P_i$, the loglikelihood ratio $L_i \triangleq \log dP_i(Y_i)/dQ_i(Y_i)$ has finite mean
	$D_i = D(P_i\|Q_i)$, finite variance $V_i = V(P_i\|Q_i)$, 
	and finite third central moment $T_i = T(P_i\|Q_i)$. Let 
	\begin{equation}
	   \overline{D}_n = \frac{1}{n} \sum_{i=1}^n D_i , \quad  \overline{V}_n = \frac{1}{n} \sum_{i=1}^n V_i ,
		\quad \overline{T}_n = \frac{1}{n} \sum_{i=1}^n T_i , 
		\quad \overline{S}_n = \frac{\overline{T}_n}{[\overline{V}_n]^{3/2}} .
	\label{eq:LR2-averages}
	\end{equation}
	There exist $n_0 \in \nN$ and $V_0 > 0$ such that $\overline{V}_n \ge V_0$ for all $n \ge n_0$.
\item[(LR3)] Under $Q_i$, the loglikelihood ratio $L_i$ has cgf $\kappa_i(s) = \log \eE_{Q_i}[e^{sL_i}] = (s-1) D_s(P_i\|Q_i)$. 
	There exists an open neighborhood of $s=1$ in which the averaged cgf $\overline{\kappa}_n(s) \triangleq \frac{1}{n} \sum_{i=1}^n \kappa_i(s)$
	is analytic for all $n \ge 1$.
	
\end{description}
The {\em lattice case} is defined by the condition that $\sum_{i=1}^n L_i$ is a lattice random variable
for all $n \ge 1$. The {\em strongly nonlattice case} could be similarly defined as the case where 
the Cram\'{e}r condition \eqref{eq:cramer-cond} on the mgf of $\sum_{i=1}^n L_i$ holds for each $n \ge 1$. 
Unfortunately Cram\'{e}r's condition is inapplicable to finite alphabets, so we weaken it as follows.
The following general definition seems general enough to apply to most problems.

\begin{definition} (Semistrong nonlattice case.)
A sequence $\{L_i\}_{i \ge 1}$ of independent random variables with respective marginal distributions $\{\pi_i\}_{i \ge 1}$
satisfies the semistrong nonlattice condition if it satisfies at least one of the following three conditions: \footnote{
	$W_1(\cdot,\cdot)$ denotes the first Wasserstein metric ({\em aka} earth mover's distance) between 
	two probability distributions on $\rR$, and $\cdot \star \cdot$ denotes the convolution of two distributions on $\rR$.}
\begin{description}
\item[(NL1)] $\forall d > 0$, $\exists n_0 \in \nN$, $\delta > 0$ such that $\forall n>n_0$:
	$\displaystyle \min_{1 \le i \le n} \inf_{P\in\scrP_\mathrm{lat}(d)} W_1(\pi_i, P) \ge \delta$.
\item[(NL2)] $\forall d > 0$, $\exists n_0 \in \nN$, $\delta \in (0,1]$ such that $\forall n>n_0$:
	$\displaystyle \min_{i\in\calI_\delta} \inf_{P\in\scrP_\mathrm{lat}(d)} W_1(\pi_i, P)\ge \delta$
	for some subset $\calI_\delta$ of $\{1,2,\cdots,n\}$ of size $|\calI_\delta| \ge \delta n$.
\item[(NL3)] $\forall d > 0$, $\exists n_0 \in \nN$, $\delta \in (0,\frac{1}{2}]$ such that $\forall n>n_0$:
	$\displaystyle \min_{i\in\calI_\delta, j\in\calJ_\delta} \inf_{P\in\scrP_\mathrm{lat}(d)} W_1(\pi_i \star \pi_j, P) \ge \delta$
	for some $\calI_\delta$ and $\calJ_\delta$, disjoint subsets of $\{1,2,\cdots,n\}$ of size $\ge \delta n$ each.
\end{description}
\label{def:semistrong}
\end{definition}
Condition {\bf (NL1)} is stronger than merely requiring that each $L_i$ be a nonlattice random variable:
no subsequence of $\{\pi_i\}$ converges to a lattice distribution under the $W_1$ metric.
Condition {\bf (NL2)} is weaker as it allows a fraction ($1-\delta$) of the $\pi_i$'s to be lattice distributions
or converge to a lattice distribution.
Condition {\bf (NL3)} is even weaker as it applies to problems such as the Z channel of Sec.~\ref{sec:Z} where each
$L_i$ is a lattice random variable but the sum $\log \frac{W_0(Y)}{Q_n(Y)} + \log \frac{W_1(Y)}{Q_n(Y)}$ is not.
Finally there are somewhat pathological problems (not considered in this paper)
which pertain neither to the lattice nor to the semistrong nonlattice case, for instance $\sum_{i=2}^n L_i$ 
is a lattice random variable for all $n \ge 2$ but $L_1$ is not.

\begin{lemma}
({\em Second-order Taylor expansion of large-deviations function for loglikelihood ratio}.)
Consider two probability measures $P$ and $Q$ over a common space and assume that $P$ is dominated
by $Q$ ($P \ll Q$). Assume the cgf $\kappa(s) = \log \eE_Q [(\frac{dP}{dQ})^s] = (s-1) D_s(P\|Q)$ 
for $L \triangleq \log \frac{dP}{dQ}$ (under $Q$) is finite and thrice differentiable in an open neighborhood of $s=1$. 
Assume $D=D(P\|Q)$ and $V=V(P\|Q)$ are positive and finite, and $T=T(P\|Q)$ is finite. 
Then the large-deviations function
\begin{equation}
   \Lambda(a) \triangleq \sup_s [as-\kappa(s)]
\label{eq:Lambda}
\end{equation}
for $L$ satisfies the Taylor expansion
\begin{equation}
   \Lambda(a) = a + \frac{(a-D)^2}{2V} + O((a-D)^3) \quad \mathrm{as~} a \to D .
\label{eq:HT-Taylor}
\end{equation}
\label{lem:HT-Taylor}
\end{lemma}
{\em Proof:} see Appendix~\ref{app:HT-Taylor}.

\begin{proposition}
Let $t$ and $c \in \rR$ be arbitrary constants and define the sequence
\begin{equation}
   a_n = \overline{D}_n - t \sqrt{\frac{\overline{V}_n}{n}} + \frac{c}{n} .
\label{eq:an0}
\end{equation}
Assume {\bf (LR1)}, {\bf (LR2)}, and {\bf (LR3)} hold. \\
{\bf (i)} If  $\{L_i\}_{i \ge 1}$ is a sequence of random variables of the semistrong nonlattice type, then
\begin{eqnarray}
   \qQ_n \left\{ \sum_{i=1}^n \log \frac{dP_i(Y_i)}{dQ_i(Y_i)} \ge n a_n\right\}
	& = & \frac{\exp\{-n a_n - \frac{t^2}{2} + o(1)\}}
		{\sqrt{2\pi n \overline{V}_n}} . 
\label{eq:nonlattice}
\end{eqnarray}
Moreover \eqref{eq:nonlattice} holds if the inequality ``$\,\ge na_n$" is replaced with a strict inequality.\\
{\bf (ii)} If $\sum_{i=1}^n L_i$ is a lattice random variable for each $n \ge 1$, denote by $d_n$ its span and
by $\Omega_n$ its range. Then
\begin{eqnarray}
   \qQ_n \left\{ \sum_{i=1}^n \log \frac{dP_i(Y_i)}{dQ_i(Y_i)} \ge na_n \right\}
	& = & \frac{d_n}{1-e^{-{d_n}}} \,\frac{\exp\{-na_n - \frac{t^2}{2} + o(1)\}}
		{\sqrt{2\pi n \overline{V}_n}} , \quad n a_n \in \Omega_n . 
\label{eq:lattice}
\end{eqnarray}
In particular, if $d_n \to 0$ as $n \to \infty$, the expressions \eqref{eq:lattice} 
and \eqref{eq:nonlattice} coincide.
\label{prop:strongLD}
\end{proposition}
{\em Proof}: See Appendix \ref{app:strongLD}.

\subsection{Neyman-Pearson Test}
\label{sec:NP}

Let $P$ and $Q$ be two probability distributions over a set $\calZ$ and consider
the Neyman-Pearson test for $P$ versus $Q$ at significance level $\alpha$.
Let $Z = \log \frac{dP}{dQ}$ and
denote by $\delta~:~\calZ \to [0,1]$ a randomized decision rule returning
$\delta(z) = \Pr\{\mathrm{Say}~P |Z=z\}, \;z\in\calZ$.
The type-II error probability of the NP test at significance level $\alpha$ is denoted by
\begin{equation}
   \beta_\alpha(P,Q) = \min_{\delta\,:\,\eE_P[\delta(Z)] \ge \alpha} \eE_Q[\delta(Z)]
\label{eq:beta-NP}
\end{equation}
and is a convex function of $\alpha$ \cite{Lehman86}.

The following theorem gives the exact asymptotics of a Neyman-Pearson test between $\pP_n = \prod_{i=1}^n P_i$
and $\qQ_n = \prod_{i=1}^n Q_i$ at significance level $\alpha=1-\epsilon$.

\begin{theorem} (NP Asymptotics.)
Assume {\bf (LR1)}, {\bf (LR2)}, and {\bf (LR3)} hold. Define
\begin{equation}
   a_n \triangleq \overline{D}_n - \sqrt{\frac{\overline{V}_n}{n}} \,t_\epsilon 
		- \frac{1}{6n} \overline{S}_n \sqrt{\overline{V}_n} (t_\epsilon^2 - 1) .
\label{eq:an}
\end{equation}
\\
{\bf (i)} In the semistrong nonlattice case,
\begin{equation}
   \beta_\alpha(\pP_n,\qQ_n) = \frac{\exp\{-n a_n - \frac{1}{2} t_\epsilon^2 + o(1) \} }{\sqrt{2\pi\overline{V}_n n}} .
\label{eq:beta-NP-nonlattice}
\end{equation}
Moreover \footnote{
   The $o(n^{-1/2})$ residual in (\ref{eq:NP}) can be strengthened to $O(n^{-1})$ in the strong lattice
	case where Cram\'{e}r's condition is satisfied.}
\begin{equation}
   \pP_n \left\{ \sum_{i=1}^n \log \frac{dP_i(Y_i)}{dQ_i(Y_i)} \ge n a_n \right\} = \alpha + o(n^{-1/2})
	\quad \mathrm{as~} n \to \infty . 
\label{eq:NP}
\end{equation}
{\bf (ii)} In the lattice case, 
\begin{equation}
   \beta_\alpha(\pP_n,\qQ_n) = \frac{d_n \,e^{d_n/2}}{1-e^{-d_n}}
	\,\frac{\exp\{-n a_n - \frac{1}{2} t_\epsilon^2 + o(1) \} }{\sqrt{2\pi\overline{V}_n n}}
\label{eq:beta-NP-lattice}
\end{equation}
for all $n,\alpha$ such that $na_n + \frac{d_n}{2} \in \Omega_n$. 
The function $\beta_\alpha(\pP_n,\qQ_n)$ is piecewise linear in $\alpha$, with breaks at the points
satisfying $na_n + \frac{d_n}{2} \in \Omega_n$.
If $d_n \to 0$ as $n \to \infty$, the asymptotic expressions \eqref{eq:beta-NP-nonlattice}
and \eqref{eq:beta-NP-lattice} coincide.
\label{thm:NP}
\end{theorem}

{\em Proof}. The type-II error probability $\beta_\alpha(\pP_n,\qQ_n)$ of the NP test is achieved by 
a randomized likelihood ratio test (LRT) with test statistic 
$Z_n \triangleq \sum_{i=1}^n \log \frac{dP_i(Y_i)}{dQ_i(Y_i)}$ \cite{Lehman86}:
\[ \deltaNP(\bY) = \left\{ \begin{array}{ll} 1 & :~ Z_n > n b_n \\ 
	\rho_n & :~ Z_n = n b_n \\ 0 & :~ Z_n < n b_n \end{array} \right. \]
where the threshold $n b_n$ and randomization parameter $\rho_n$ are chosen
to satisfy the constraint
\begin{equation}
   \alpha = \eE_{\pP_n} [\deltaNP(\bY)] = \pP_n\{Z_n > n b_n\} + \rho_n \pP_n\{Z_n = n b_n\}
\label{eq:alpha-NP}
\end{equation}
hence
\begin{equation}
   \pP_n\{Z_n > n b_n\} \le \alpha \le \pP_n\{Z_n \ge n b_n\} .
\label{eq:alpha-sandwich}
\end{equation}
The type-II error probability of the NP test is
\begin{equation}
   \beta_\alpha(\pP_n,\qQ_n) = \eE_{\qQ_n} [\deltaNP(\bY)] = \qQ_n\{Z_n \ge n b_n\} - (1-\rho_n) \qQ_n\{Z_n = n b_n\}
\label{eq:beta-NP2}
\end{equation}
hence satisfies the following lower and upper bounds:
\begin{equation}
   \qQ_n\{Z_n > n b_n\} \le \beta_\alpha(\pP_n,\qQ_n) \le \qQ_n\{Z_n \ge n b_n\} .
\label{eq:beta-sandwich}
\end{equation}

By \eqref{eq:LR2-averages}, we have $\eE_{\pP_n} [Z_n] = n\overline{D}_n$ and $\Var_{\pP_n} [Z_n] = n\overline{V}_n$.
Denote by $F_n$ the cdf of the normalized random variable
\[ T_n \triangleq \frac{Z_n - n\overline{D}_n}{\sqrt{n\overline{V}_n}} . \]
By \eqref{eq:alpha-sandwich} and the definition of the quantile $t_{\epsilon,n}$ above \eqref{eq:CF}, we have
\begin{equation}
   t_{\epsilon,n} = \frac{n b_n - n\overline{D}_n}{\sqrt{n\overline{V}_n}} .
\label{eq:talpha-n}
\end{equation}

{\bf (i)} {\em Semistrong nonlattice case.} If the cdf $F_n$ is continuous at $b_n$, then $\pP_n\{Z_n = n b_n\} = 0$,
and so from \eqref{eq:alpha-NP} we have $F_n(t_{\epsilon,n}) = \alpha$. Else 
$F_n$ may have a jump of size $o(n^{-1/2})$. Then $\pP_n\{Z_n = n b_n\} = o(n^{-1/2})$ and
$F_n(t_{\epsilon,n}) = \alpha + o(n^{-1/2})$.
Applying successively \eqref{eq:CF-noniid} and \eqref{eq:an}, we have
\begin{eqnarray}
  t_{\epsilon,n} & = & t_\epsilon + \frac{\overline{S}_n}{6\sqrt{n}} (t_\epsilon^2 - 1) + o(n^{-1/2}) \nonumber \\
	& = & \frac{n a_n - n\overline{D}_n}{\sqrt{n\overline{V}_n}} + o(n^{-1/2}) .
\label{eq:talpha-n2}
\end{eqnarray}
Comparing with \eqref{eq:talpha-n}, we obtain $n b_n = n a_n + o(1)$.

Applying Prop.~\ref{prop:strongLD}(i) with $t=t_\epsilon$ and 
$c = - \frac{1}{6} \overline{S}_n \sqrt{\overline{V}_n} (t_\epsilon^2 - 1)$, we obtain
\[ \qQ_n\{Z_n \ge n b_n\} = \qQ_n\{Z_n \ge n a_n + o(1)\}
	= \frac{\exp\{-n a_n - \frac{1}{2} t_\epsilon^2 + o(1) \} }{\sqrt{2\pi\overline{V}_n n}} 
\]
and the same asymptotics hold for $\qQ_n\{Z_n > n b_n\}$. The claim then follows from \eqref{eq:beta-sandwich}.

{\bf (ii)} In the lattice case, $\beta_\alpha(\pP_n,\qQ_n)$, viewed as a function of $\alpha$, 
is convex and piecewise linear, with breaks at the points 
where randomization is not needed ($\rho_n = 1$). At any such point $\alpha$ we have
$\beta_\alpha(\pP_n,\qQ_n) = \qQ_n\{Z_n \ge n b_n\}$ and $\alpha = \pP_n\{Z_n \ge n b_n\}$.
Recalling the central limit asymptotics for non-iid lattice random variables at the end of Sec.~\ref{sec:CLT},
\eqref{eq:talpha-n2} holds if $na_n$ is a midpoint of the lattice, i.e., $na_n + \frac{1}{2} d_n \in \Omega_n$. 
Hence $na_n = nb_n - \frac{1}{2} d_n$, $\pP_n\{Z_n \ge n a_n\} = \alpha + o(n^{-1/2})$, 
and by Prop.~\ref{prop:strongLD}(ii),
\begin{eqnarray*}
   \qQ_n\{Z_n \ge n b_n\} & = & \frac{d_n}{1-e^{-d_n}} 
		\,\frac{\exp\{-n b_n - \frac{1}{2} t_\epsilon^2 + o(1) \} }{\sqrt{2\pi\overline{V}_n n}} \\
	& = & \frac{d_n \,e^{d_n/2}}{1-e^{-d_n}} 
		\,\frac{\exp\{-n a_n - \frac{1}{2} t_\epsilon^2 + o(1) \} }{\sqrt{2\pi\overline{V}_n n}} .
\end{eqnarray*}
\hfill $\Box$

\begin{remark}
Theorem~\ref{thm:NP}(i) reduces to Theorem~1 of Strassen \cite[p.~690]{Strassen62} 
in the case of iid components and deterministic rules. For the non-iid case, 
Strassen \cite[Theorem~3, p.~702]{Strassen62} used the Berry-Ess\'{e}en theorem to show
\begin{equation}
   | \log \beta_\alpha(\pP_n,\qQ_n) + n \overline{D}_n - \sqrt{n\overline{V}_n} \,t_\epsilon + \frac{1}{2} \log n |
	< \frac{140}{\delta^8}
\label{eq:beta-strassen}
\end{equation}
where $\delta$ is any number satisfying the four inequalities
\[ 0 < \delta < 1-\alpha, \quad \sqrt{n} \ge \frac{140}{\delta^8} , 
	\quad \delta \le \sqrt{\overline{V}_n} , \quad \delta \le \overline{T}_n^{-1/3} . \]
The right side of \eqref{eq:beta-strassen} is at least equal to $140/(1-\alpha)^8$,
and the inequality holds for $n$ at least equal to $140^2/(1-\alpha)^{16}$.
\end{remark}

\begin{remark}
Polyanskiy {\em et al.} \cite[Lemma~14, p.~33]{Polyanskiy10b} derived
\begin{eqnarray}
   \lefteqn{-n \overline{D}_n - \sqrt{n\overline{V}_n} \,\Phi^{-1} \left( 
		\alpha - \frac{6\overline{S}_n + \delta}{\sqrt{n}} \right) 
		+ \log \delta - \frac{1}{2} \log n \le \log \beta_\alpha(\pP_n,\qQ_n)} \nonumber \\
	& \le & -n \overline{D}_n - \sqrt{n\overline{V}_n} \,\Phi^{-1} 
	\left( \alpha - \frac{6\overline{S}_n}{\sqrt{n}} \right) - \frac{1}{2} \log n
		+ \log \left( \frac{2 \log 2}{\sqrt{2\pi\overline{V}_n}} + 24 \overline{S}_n \right)
\label{eq:beta-polyanskiy}
\end{eqnarray}
valid for any $\delta > 0$. (A looser upper bound, without the $\frac{1}{2} \log n$,
is given in \cite[Lemma~58, p.~2340]{Polyanskiy10}). The lower bound is asymptotic to
\[ -n \overline{D}_n - \sqrt{n\overline{V}_n} \,\Phi^{-1} (\alpha) - \frac{1}{2} \log n
		- c_\alpha \sqrt{\overline{V}_n} (6\overline{S}_n + \delta) + \log \delta + o(1)
\]
where $c_\alpha > 0$ is the derivative of the function $- \Phi^{-1}$ at $\alpha$.
Maximizing this expression over $\delta$, one obtains $\delta = 1/(c_\alpha \sqrt{\overline{V}_n})$ 
hence the asymptotic lower bound
\begin{equation}
  \log \beta_\alpha(\pP_n,\qQ_n) \ge -n \overline{D}_n - \sqrt{n\overline{V}_n} \,\Phi^{-1} (\alpha) 
	- \frac{1}{2} \log n - \log \left( c_\alpha \sqrt{\overline{V}_n} \right) 
	- 6 c_\alpha \sqrt{\overline{V}_n} \overline{S}_n  - 1 + o(1) .
\label{eq:beta-PPV-asym}
\end{equation}
\end{remark}

\begin{remark}
Using Chebyshev's inequality, Polyanskiy {\em et al.} \cite[Lemma~59, p.~2341]{Polyanskiy10} 
also derived the following nonasymptotic lower bound:
\begin{eqnarray}
   \log \beta_\alpha(\pP_n,\qQ_n) \ge -n \overline{D}_n 
	- \sqrt{\frac{2n\overline{V}_n}{1-\epsilon}} + \log \frac{1-\epsilon}{2} .
\label{eq:beta-PPV-lemma59}
\end{eqnarray}
A closely related inequality was derived by Kemperman \cite[Theorem~7.1]{Kemperman62}.
\end{remark}

\section{Achievability}
\label{sec:achieve}
\setcounter{equation}{0}

The direct coding theorem~\ref{thm:direct} is proved using iid random codes and ML decoding.

Fix $P^*$ achieving the maximum in \eqref{eq:Aeps-}.
From (\ref{eq:U}), (\ref{eq:u-skew}), and \eqref{eq:rho}, we define $\Vu \triangleq \Vu(P^*,W)$,
$\Su \triangleq \Su(P^*,W)$ and $\rho \triangleq \rho(P^*,W)$.
The function
\begin{equation}
   \omega(h,t) \triangleq - \frac{1}{2} \|h\|_{\sfJ^+}^2 - \frac{t}{2\sqrt{\Vu}} h^\top  \nabla \Vu(P^*) , 
	\quad h \in \scrL(\calX), \,t\in\rR
\label{eq:omega(h)}
\end{equation}
is quadratic in $h$ and linear in $t$.
It follows from \eqref{eq:delU=delV} and $\sum_{x\in\calX} h(x) = 0$ that
\begin{equation}
   h^\top  \nabla \Vu(P^*) =  h^\top  \nabla V(P^*) .
\label{eq:Vbar=Ubar}
\end{equation}
Recall that $\Vu = V_\epsilon > 0$ by {\bf (A1)} and that $\Su=S$. In the remainder of this section,
the subscript ``u'' could be dropped everywhere.

Fixing $t=t_\epsilon$. Recalling \eqref{eq:J+}, \eqref{eq:normJ+}, and (\ref{eq:Ans}), we have
\begin{equation}
   \sup_{h\in\calH} \omega(h,t_\epsilon) 
	= \frac{t_\epsilon^2}{8\Vu} \|\nabla \Vu(P^*)\|_{\sfJ^+}^2
	= \frac{t_\epsilon^2}{8} \Ans^{(P^*)}
\label{eq:omega-max}
\end{equation}
where the supremum is achieved at
\begin{equation}
   h^* \triangleq - \frac{t_\epsilon}{2\sqrt{V_\epsilon}}  \,\sfJ^+ v^{(P^*)} \quad \in \calH .
\label{eq:h*}
\end{equation}
Next, define
\begin{equation}
   \hat{t}_{\epsilon,n} \triangleq t_\epsilon + \frac{\Su}{6\sqrt{n}} (t_\epsilon^2 - 1) .
\label{eq:tepsn}
\end{equation}

Fix any $h\in\calH$ and draw random codes iid from
\begin{equation}
   P_n = P^* + n^{-1/2} h \quad \in \scrP(\calX) .
\label{eq:Pn}
\end{equation}
Shannon's codes are obtained as a special case with $h=0$, but our achievability bound
will be obtained when $h=h^*$.
The number of codewords is $M_n = \lfloor e^{nR_n} \rfloor$ where
\begin{eqnarray}
   nR_n & = & nC - \sqrt{n\Vu} \,t_\epsilon + \frac{1}{2} \log n + \underline{A}_\epsilon 
			+ \omega(h,t_\epsilon) - \frac{t_\epsilon^2}{8} \Ans^{(P^*)} \nonumber \\
		& = & nC - \sqrt{n\Vu} \,t_\epsilon + \frac{1}{2} \log n \nonumber \\
		& & \quad - \frac{1}{6} \Su \sqrt{\Vu} (t_\epsilon^2 - 1) + \frac{t_\epsilon^2}{2} \frac{1-\rho}{1+\rho}
			+ \frac{1}{2} \log (2\pi (1-\rho^2)\Vu) + \omega(h,t_\epsilon) - 1 
\label{eq:nRn-1}
\end{eqnarray}
where the last equality follows from the definition of $\underline{A}_\epsilon$ in (\ref{eq:Aeps-}).

{\bf Random coding scheme.}
The codeword symbols $\{ X_i(m), \,1 \le i \le n, \,1 \le m \le M_n\}$
are drawn iid from the distribution $P_n$ of \eqref{eq:Pn}.
Define the random variables
\begin{equation}
   Z_{m,n} \triangleq \log \frac{W^n(\bY|\bX(m))}{(P_n W)^n(\bY)}
	= \sum_{i=1}^n \log \frac{W(Y_i|X_i(m))}{(P_n W)(Y_i)} , \quad 1 \le m \le M_n .
\label{eq:Zm}
\end{equation}
Hence $Z_{m,n}$ is the same as the information density $i(\bX(m);\bY)$. 
The ML decoding rule is deterministic and can be written as
\begin{equation}
   \hat{m} = \arg\max_{1 \le m \le M_n} Z_{m,n} .
\label{eq:ML}
\end{equation}
In case of a tie, an error is declared. \footnote{
	Decoding performance could potentially be improved using randomization for tie breaking \cite{Haim13}.} 

{\bf Overview of error probability analysis.}
By symmetry of the codebook construction and the decoding rule, the conditional error
probability (given message $m$) is independent of $m$. We show that the average error probability over the random
ensemble is $\epsilon + o(n^{-1/2})$, hence there exists a deterministic code
with at most the same average error probability. This proves the claim, 
since by Taylor's formula, $\sqrt{nV_\epsilon} Q^{-1}(\epsilon + o(n^{-1/2})) = \sqrt{nV_\epsilon} Q^{-1}(\epsilon) + o(1)$.

For the calculation below, we assume without loss of generality that $m=1$ was sent. 
We start from a standard random coding union bound \footnote{
	This first step is ubiquitous in analyses of random codes and has also been used
	in particular to state the RCU bound of \cite[Theorem~16]{Polyanskiy10}.}
\begin{eqnarray}
   \Pr\{\mathrm{Error}\} & = & \Pr \left\{\max_{m \ge 2} Z_{m,n} \ge Z_{1,n}\right\} \nonumber \\
   & \le & \Pr \left\{ Z_{1,n} \le z_n^*\right\} + \Pr \left\{\max_{m \ge 2} Z_{m,n} \ge Z_{1,n} \ge z_n^* \right\} \nonumber \\
   & \le & \Pr \left\{ Z_{1,n} \le z_n^*\right\} + (M_n - 1) \Pr \left\{ Z_{2,n} \ge Z_{1,n} \ge z_n^* \right\} .
\label{eq:UnionBound}
\end{eqnarray}
In the last line, $z_n^*$ is arbitrary but will be chosen
so that the second probability is $\frac{\phi(t_\epsilon)}{\sqrt{n\Vu}} [1+o(1)]$.
We then use precise asymptotics for the two probabilities of (\ref{eq:UnionBound})
to derive the desired result. 

Thus \eqref{eq:UnionBound} upperbounds the error probability as the sum of two terms. 
The first is the probability that the correct message scores below some threshold $z_n^*$,
and the second upperbounds the probability that some incorrect message scores at least as well as the correct
message, conditioned on the score of the latter being at least equal to $z_n^*$. There is a somewhat weak but
possibly significant dependency (via $\bY$) between the score of an incorrect message and that of the correct message.
As shown in the derivations below, this dependency is quantified by the correlation coefficient $\rho$
and contributes to the $O(1)$ term in the achievable rate.

{\bf Central Limit Asymptotics of $Z_{1,n}$.}
For each $m=1,2,\cdots,M_n$, the pairs $(X_i(m), Y_i)$, $1 \le i \le n$ are iid,
hence $Z_{m,n}$ is the sum of conditionally (given $m$) iid random variables.
The mean of $Z_{1,n}$ is given by
\[
   \eE[Z_{1,n}]
		= n \eE_{P_n \times W} \left[ \log \frac{W(Y|X)}{(P_n W)(Y)} \right] 
		= n I(P_n;W) .
\]
By \eqref{eq:I-taylor} and \eqref{eq:Pn} we have
\begin{eqnarray}
   \overline{D}_n = I(P_n;W) & = & I(P^*;W) - \frac{1}{2} \|P^* - P_n\|_\sfJ^2 + o(1/n) \nonumber \\
		& = & C - \frac{1}{2n} \|h\|_\sfJ^2 + o(1/n) .
\label{eq:Dbar-n}
\end{eqnarray}
The variance of $Z_{1,n}$ is equal to $n$ times the unconditional information variance
\[
  \overline{V}_n = \Vu(P_n;W) = \Vu + \frac{1}{\sqrt{n}} h^\top  \nabla \Vu(P^*) + o(n^{-1/2}) \]
where the last equality follows from \eqref{eq:G-taylor} and \eqref{eq:Pn}.
Since $\Vu > 0$, we obtain
\begin{equation}
  \sqrt{\overline{V}_n} = \sqrt{\Vu} + \frac{1}{2\sqrt{n\Vu}} h^\top  \nabla \Vu(P^*) + o(n^{-1/2}) .
\label{eq:Vbar-n}
\end{equation}
Combining (\ref{eq:omega(h)}), (\ref{eq:Dbar-n}), and (\ref{eq:Vbar-n}) we obtain
\begin{equation}
   n\overline{D}_n - \sqrt{n\overline{V}_n} \,t = nC - \sqrt{n\Vu} \,t + \omega(h,t) + o(1) , \quad \forall t\in\rR .
\label{eq:+theta}
\end{equation}
The skewness of $Z_{1,n}$ is equal to
\begin{equation}
  \overline{S}_n = \Su(P_n;W) = \Su + O(n^{-1/2}) .
\label{eq:Sbar-n}
\end{equation}

Define the normalized random variable
\begin{equation}
   T_n \triangleq - \frac{Z_{1,n} - n\overline{D}_n}{\sqrt{n\overline{V}_n}}
\label{eq:Tn-achieve}
\end{equation}
which has zero mean and unit variance and by the Central Limit Theorem, converges in distribution to $\calN(0,1)$. 
By the Cornish-Fisher inversion formula \eqref{eq:CF}, $\hat{t}_{\epsilon,n}$ of \eqref{eq:tepsn} satisfies
\begin{equation}
   F_{T_n}(\hat{t}_{\epsilon,n}) = 1-\epsilon + o(n^{-1/2}) .
\label{eq:cornish}
\end{equation}

Now fix $z_n^* = n\overline{D}_n - \sqrt{n\overline{V}_n} \,t_n^*$ where \footnote{
	To obtain the best $O(1)$ term, we could choose $t_n^* = \hat{t}_{\epsilon,n} + a/\sqrt{n\Vu}$
	and optimize $a$. The optimal choice of $a$ turns out to be 1, hence the choice of \eqref{eq:t*-teps}.}
\begin{equation}
   t_n^* \triangleq \hat{t}_{\epsilon,n} + \frac{1}{\sqrt{n\Vu}} 
	= t_\epsilon + \frac{1}{\sqrt{n}} \left(\frac{\Su}{6} (t_\epsilon^2 - 1) + \frac{1}{\sqrt{\Vu}} \right) .
\label{eq:t*-teps}
\end{equation}
Hence
\begin{equation}
   z_n^* = nC - \sqrt{n\Vu} \,t_\epsilon - \frac{\Su\sqrt{\Vu}}{6} (t_\epsilon^2 - 1) + \omega(h,t_\epsilon) - 1.
\label{eq:zn*}
\end{equation}
We have
\begin{eqnarray}
   P_{Z_{1,n}} \{Z_{1,n} < z_n^*\}
    & \stackrel{(a)}{=} & \int_{t > t_n^*} dP_{T_n}(t) \nonumber \\
	& = & 1 - F_{T_n}(t_n^*) \nonumber \\
	& \stackrel{(b)}{=} & 1 - F_{T_n}(\hat{t}_{\epsilon,n}) - (t_n^* - \hat{t}_{\epsilon,n}) \phi(t_\epsilon) + o(n^{-1/2}) \nonumber \\
	& \stackrel{(c)}{=} & \epsilon - \frac{\phi(t_\epsilon)}{\sqrt{n\Vu}} + o(n^{-1/2}) 
		\quad \mathrm{as~} n \to \infty 
\label{eq:int1}
\end{eqnarray}
where (a), (b), and (c) follow from \eqref{eq:Tn-achieve}, (\ref{eq:t*-teps}), and (\ref{eq:cornish}), respectively.

{\bf Large Deviations for $Z_{2,n}$.}
The random variables $Z_{m,n}, \,m \ge 2, \,n \ge 1$ are identically distributed and are conditionally independent
given $\bY$. By \eqref{eq:Zm}, for $m \ge 2$, the joint distribution of $(Z_{m,n}, Z_{1,n})$ is
the same as that of $(\sum_{i=1}^n L_i', \sum_{i=1}^n L_i)$ where 
\[ L_i' \triangleq \log \frac{W(Y_i|X_i')}{(P_n W)(Y_i)}, \quad
	L_i \triangleq \log \frac{W(Y_i|X_i)}{(P_n W)(Y_i)}, \quad 1 \le i \le n \]
are iid  with joint distribution $P_{L'L}$ induced by 
$P_{X'XY}(x',x,y) \triangleq P_n(x') P_n(x) W(y|x)$.

Denote by $\tilde{P}_{L'L}$ the tilted distribution
\begin{equation}
  d\tilde{P}_{L'L}(l',l) \triangleq e^{l'} \,dP_{L'L}(l',l)
\label{eq:Ptilde-L'L}
\end{equation}
induced by
\begin{equation}
   \tilde{P}_{X'XY}(x',x,y) = \frac{W(y|x')}{(P_n W)(y)} P_{X'XY}(x',x,y) 
	= \frac{P_n(x') W(y|x') W(y|x) P_n(x)}{(P_n W)(y)} . 
\label{eq:Pt-X'XY}
\end{equation}
The marginals $\tilde{P}_{L'}$ and $\tilde{P}_L$ are identical with mean 
$\eE_{\tilde{P}}[L] = \eE_{\tilde{P}}[L'] = \overline{D}_n$
and variance $\Var_{\tilde{P}}[L] = \Var_{\tilde{P}}[L'] = \overline{V}_n$.
Since $X \to Y \to X'$ forms a Markov chain under $\tilde{P}$, we have
\[ \eE_{\tilde{P}} [L'L] = \eE_{(P_n W)} \left\{ \eE[L'|Y] \,\eE[L|Y] \right\} = \eE_{(P_n W)} \eE^2[i(X;Y)|Y] . \]
Hence $\Cov_{\tilde{P}}(L',L) =  \eE_{(P_n W)} \Var[i(X;Y)|Y] = V^r(P_n;W)$, and $\rho_n(P;W) = 1 - V^r(P_n;W)/\overline{V}_n$
is the normalized correlation coefficient between $L'$ and $L$ under $\tilde{P}$.
Hence $(L',L)$ has covariance matrix $\Cov_{\tilde{P}}(L,L')^\top = \overline{V}_n \begin{bmatrix} 1 & \rho_n \\ \rho_n & 1 \end{bmatrix}$
where $\overline{V}_n = \Vu + O(n^{-1/2})$ and $\rho_n = \rho + O(n^{-1/2})$.

Define the normalized random variables
\begin{equation}
   W_n' \triangleq \frac{\sum_{i=1}^n L_i' - n\overline{D}_n}{\sqrt{n\overline{V}_n}} + t_n^*
	\quad \mathrm{and} \quad W_n \triangleq \frac{\sum_{i=1}^n L_i - n\overline{D}_n}{\sqrt{n\overline{V}_n}} + t_n^* . 
\label{eq:Wn'Wn}
\end{equation}
Hence the second probability in the right side of \eqref{eq:UnionBound} is
\begin{eqnarray}
   \Pr \left\{ Z_{2,n} \ge Z_{1,n} \ge z_n^* \right\} 
	& = & P_{L'L}^n \left\{ \sum_{i=1}^n L_i' \ge \sum_{i=1}^n L_i \ge n\overline{D}_n - \sqrt{n\overline{V}_n} t_n^* \right\} \nonumber \\
	& = & P_{L'L}^n \{0 \le W_n \le W_n' \}
\end{eqnarray}

Under the tilted distribution $\tilde{P}_{L'L}^n$, the random variables $W_n'$ and $W_n$ 
have means equal to $t$, unit variance, and correlation coefficient $\rho_n$. 
Denote their joint cdf by $F_n$ and the two-dimensional Gaussian cdf with the same mean
and covariance by $\Phi_{t_n^*,\rho_n}$. Recall that $t_n^* = t_\epsilon + O(n^{-1/2})$ and $\rho_n = \rho + O(n^{-1/2})$,
and denote the pdf associated with $\Phi_{t_\epsilon,\rho}$ by $\phi_{t_\epsilon,\rho}$.
Since the pairs $\{(L_i',L_i), \,1 \le i \le n\}$ are iid, by the Central Limit Theorem, 
$F_n$ converges pointwise to $\Phi_{t_\epsilon,\rho}$.
More precisely, analogously to \eqref{eq:CE}, the following expansion holds \cite[Sec.~6.3]{Barndorff91}:
\begin{equation}
   F_n(w',w) = \Phi_{t_n^*,\rho_n}(w',w) + \frac{1}{\sqrt{n}} f(w',w) \,\phi_{t_n^*,\rho_n}(w',w) + o(n^{-1/2})
\label{eq:Wn'Wn-CLT}
\end{equation}
where convergence is uniform over $\rR^2$, and $f(w',w)$ is a polynomial function of degree 3.

We have
\begin{eqnarray}
   P_{L'L}^n \{0 \le W_n \le W_n'\}
	& = & \int_{\rR^n \times \rR^n} dP_{L'L}^n(\bl',\bl) \,\mathds1\left\{ 0 \le W_n \le W_n' \right\} \nonumber \\
	& \stackrel{(a)}{=} & e^{-z_n^*} \,\int_{\rR^n \times \rR^n} d\tilde{P}_{L'L}^n(\bl',\bl) 
		e^{- \left[\sum_{i=1}^n l_i' - z_n^*\right]} \,\mathds1\left\{ 0 \le W_n \le W_n' \right\} \nonumber \\
	& \stackrel{(b)}{=} & e^{-z_n^*} \iint_{0 \le w \le w'} dF_n(w',w) \,e^{- \sqrt{n\overline{V}_n} \,w'} \label{eq:PL'L-0} \\
	& \stackrel{(c)}{=} & e^{-z_n^*} \sqrt{n\overline{V}_n} \int_{w=0}^\infty dF_n(w) 
		\int_{w'=w}^\infty (F_n(w'|w) - F_n(w|w)) \,e^{- \sqrt{n\overline{V}_n} \,w'} \,dw' \nonumber \\
\label{eq:PL'L}
\end{eqnarray}
where (a) follows from (\ref{eq:Ptilde-L'L}), (b) from (\ref{eq:Wn'Wn}), and (c) is obtained using integration by parts
for $w'$. Applying \eqref{eq:Wn'Wn-CLT} yields, after the same algebra as in \cite{Bahadur60} \cite[pp.~110---112]{Dembo98}
\begin{eqnarray}
   P_{L'L}^n \{0 \le W_n \le W_n'\} 
	& \sim & e^{-z_n^*}  \sqrt{n\Vu} \iint_{0 \le w \le w'} (\Phi_{t_\epsilon,\rho}(w',w) - \Phi_{t_\epsilon,\rho}(w,w)) \,e^{- \sqrt{n\overline{V}_n} \,w'} \,dwdw' \nonumber \\
	& = & \frac{e^{-z_n^*}}{\sqrt{n\Vu}} \int_{u \ge 0} \int_{u' \ge u} \left[ \Phi_{t_\epsilon,\rho} \left( \frac{u'}{\sqrt{n\Vu}} , \frac{u}{\sqrt{n\Vu}} \right) 
		- \Phi_{t_\epsilon,\rho} \left( \frac{u}{\sqrt{n\Vu}} , \frac{u}{\sqrt{n\Vu}} \right) \right] \,e^{-u'} \,dudu' \nonumber \\
	& \sim & \frac{e^{-z_n^*}}{\sqrt{n\Vu}} \int_{u \ge 0} \int_{u' \ge u} \frac{u'-u}{\sqrt{n\Vu}}
		\,\phi_{t_\epsilon,\rho} \left( \frac{u}{\sqrt{n\Vu}} , \frac{u}{\sqrt{n\Vu}} \right) \,e^{-u'} \,dudu' \nonumber \\
	& = & \frac{e^{-z_n^*}}{n\Vu} \int_{u \ge 0} \,\phi_{t_\epsilon,\rho} \left( \frac{u}{\sqrt{n\Vu}} , \frac{u}{\sqrt{n\Vu}} \right) \,e^{-u} \,du \nonumber \\
	& \sim & \frac{e^{-z_n^*}}{n\Vu} \phi_{t_\epsilon,\rho}(0,0) \label{eq:PL'L-1} \\
	& = & \frac{e^{-z_n^*}}{n\Vu} \frac{\phi(t_\epsilon) \exp\{ - \frac{1-\rho}{2(1+\rho)} t_\epsilon^2\} }{\sqrt{2\pi(1-\rho^2)}} . \nonumber
\end{eqnarray}
Hence the second term in the right side of \eqref{eq:UnionBound} is
\begin{eqnarray}
   \lefteqn{(M_n-1) P_{L'L}^n \{0 \le W_n \le W_n'\} } \nonumber \\
	& = & \frac{\phi(t_\epsilon)}{\sqrt{n\Vu}} \,\exp\left\{ 
		nR_n - z_n^* - \frac{1-\rho}{2(1+\rho)} t_\epsilon^2 - \frac{1}{2} \log (2\pi(1-\rho^2)n\Vu) + o(1) \right\} \nonumber \\
	& = & \frac{\phi(t_\epsilon)}{\sqrt{n\Vu}} [1+o(1)]
\label{eq:int2}
\end{eqnarray}
where the last equality follows from \eqref{eq:nRn-1} and \eqref{eq:zn*}

Substituting (\ref{eq:int2}) and (\ref{eq:int1}) into \eqref{eq:UnionBound}, we obtain
\begin{eqnarray*}
   \Pr\{\mathrm{Error}\} & \le & \epsilon - \frac{\phi(t_\epsilon)}{\sqrt{n\Vu}} 
							+ \frac{\phi(t_\epsilon)}{\sqrt{n\Vu}} + o(n^{-1/2}) \\
	& = & \epsilon + o(n^{-1/2}) 
\end{eqnarray*}
which proves the claim.
\hfill $\Box$

\section{Converse}
\label{sec:converse}
\setcounter{equation}{0}

To prove Theorem~\ref{thm:converse}, we begin in Sec.~\ref{sec:converse-background} 
with some background from Polyanskiy {\em et al.} \cite{Polyanskiy10}.
Sec~\ref{sec:converse-CC} presents a converse for constant-composition codes under both
the maximum and average error probability criteria. Sec~\ref{sec:converse-max} presents a converse
for general codes under the maximum error probability criterion, and Sec~\ref{sec:converse-avg}
presents the converse under the average error probability criterion. 
Each section builds on the results from the previous section.

\subsection{Background}
\label{sec:converse-background}

Here we review upper bounds on $M_\avg^*(n,\epsilon)$ and  $M_{\max}^*(n,\epsilon)$ that are expressed in terms
of the type-II error probability of NP tests at significance level $\alpha=1-\epsilon$.

\begin{theorem}
\cite[Theorem~31 p.~2319]{Polyanskiy10}
The volume $M_{\sf F}$ of any code with codewords in ${\sf F} \subseteq \calX^n$
and {\em maximum} error probability $\epsilon$ satisfies
\begin{equation}
   M_{\sf F} \le \inf_{Q_\bY} \sup_{\bx\in \sf F} \frac{1}{\beta_{1-\epsilon}(W^n(\cdot|\bx), Q_{\bY})}
\label{eq:poly-converse-max}
\end{equation}
where the supremum is over all feasible codewords,
and the infimum is over all probability distributions on $\calY^n$.
\label{thm:poly-converse-max}
\end{theorem}

\begin{theorem}
\cite[Theorem~27 p.~2318]{Polyanskiy10}
The volume $M_{\sf F}$ of any code with codewords in ${\sf F} \subseteq \calX^n$
and {\em average} error probability $\epsilon$ satisfies
\begin{equation}
   M_{\sf F} \le \sup_{P_\bX} \inf_{Q_\bY}  \frac{1}{\beta_{1-\epsilon}(P_{\bX} \times W^n, P_{\bX} \times Q_{\bY})}
\label{eq:poly-converse-avg}
\end{equation}
where the supremum is over all probability distributions on $\sf F$,
and the infimum is over all probability distributions on $\calY^n$.
\label{thm:poly-converse-avg}
\end{theorem}
It was later shown that the order of sup and inf can be exchanged 
in (\ref{eq:poly-converse-avg}) \cite{Polyanskiy13}; this property
is unrelated to the saddlepoint property of Prop.~\ref{prop:maxP1} 
presented later in this section.

In some cases $\beta_{1-\epsilon}(P_{\bY|\bX=\bx}, Q_{\bY})$ is constant for all $\bx \in {\sf F}$,
e.g., when $Q_\bY = Q_Y^n$ is the $n$-fold product of a distribution $Q_Y$ over $\calY$,
and when all $\bx \in {\sf F}$ have the same composition.
With some abuse of notation, we write $\beta_{1-\epsilon}(Q_Y^n) = \beta_{1-\epsilon}(P_{\bY|\bX=\bx}, Q_Y^n)$.
Then the following result holds.

\begin{theorem} \cite[Theorem~28 p.~2318]{Polyanskiy10}.
Fix a distribution $Q_Y$ over $\calY$ and assume $\beta_{1-\epsilon}(P_{\bY|\bX=\bx}, Q_Y^n)$ $= \beta_{1-\epsilon}(Q_Y^n)$
for each $\bx \in {\sf F} \subseteq \calX^n$. Then the volume $M_{\sf F}$ of any code with codewords 
in ${\sf F} \subseteq \calX^n$ and average error probability $\epsilon$ satisfies
$M_{\sf F} \le 1/\beta_{1-\epsilon}(Q_Y^n)$.
\label{thm:poly-beta-CC}
\end{theorem}

This result was used in Polyanskiy {\em et al.} \cite{Polyanskiy10} to derive a converse theorem
for constant-composition codes. Denote by $M_{\mathrm{cc}}^*(n,\epsilon)$ the maximal number of codewords
for any constant-composition code over the DMC $W$, with {\em maximal error probability} $\epsilon$.

\begin{theorem} \cite[Theorem~48 p.~2331]{Polyanskiy10}.
If $0 < \epsilon \le 1/2$, there exists $n_0 \in \nN$ and a constant $F > 0$ such that
\[ \forall n > n_0 : \quad \log M_{\mathrm{cc}}^*(n,\epsilon) \le nC - \sqrt{nV_\epsilon} t_\epsilon + \frac{1}{2} \log n + F . \]
\label{thm:poly-converse-CC}
\end{theorem}

The key idea in the proof of this theorem (Eqn (494) p.~2350) is to use a result on
the asymptotics of NP tests (Lemma~58 p.~2340) that provides a lower bound on $\beta_{1-\epsilon}(Q_Y^n)$.

Finally, applying \cite[Eqn~(284) p.~2332]{Polyanskiy10}
\[ M_{\max}^*(n,\epsilon) \le M_{\avg}^*(n,\epsilon) \le \frac{\tau}{\tau - 1} M_{\max}^*(n,\tau\epsilon) \]
with $\tau = 1+ 1/\sqrt{n}$, we obtain $\calQ^{-1}(\tau\epsilon) = \calQ^{-1}(\epsilon) + O(n^{-1/2})$.
Hence $M_{\max}^*(n,\tau\epsilon)$ and $M_{\max}^*(n,\epsilon)$ asymptotically differ only by a constant and
\begin{equation}
   M_{\max}^*(n,\epsilon) \le M_{\avg}^*(n,\epsilon) \le \sqrt{n} \,[M_{\max}^*(n,\epsilon) + O(1)] .
\label{eq:Mmax-sandwich}
\end{equation}
Hence the gap between the maximum log-volumes under the average and maximum error probability criteria
is at most $\frac{1+o(1)}{2} \log n$.

\subsection{Converse for Constant-Composition Codes}
\label{sec:converse-CC}

Fix some positive vanishing sequence $\delta_n$. 
Consider $Q=(\tilde{P}W)$ where $\tilde{P} \in \scrP(\calX)$ and define
\begin{equation}
  \hat{\zeta}_n(P,Q;\delta_n) \triangleq \left\{ \begin{array}{l}
	n D(W\|Q|P) - \sqrt{n V(W\|Q|P)} \,t_\epsilon + F_\epsilon(W\|Q|P) \\
		\quad \mathrm{if~} \exists P_0\in \Pi :~\|P-P_0\|_\infty \le \delta_n , \,\|\tilde{P}-P_0\|_\infty \le \delta_n , \\
	n D(W\|Q|P) + \sqrt{\frac{2n V(W\|Q|P)}{1-\epsilon}} - \log \frac{1-\epsilon}{2} \\
		\quad \mathrm{else.} 
	\end{array} \right.
\label{eq:zeta-hat-PQ} 
\end{equation}
For $Q=(PW)$ this function is equal to
\begin{equation}
  \hat{\zeta}_n(P;W;\delta_n) \triangleq \left\{ \begin{array}{ll}
		nI(P;W) - \sqrt{nV(P;W)} t_\epsilon + F_\epsilon(P;W) & \mathrm{if~} \exists P_0\in \Pi :~\|P-P_0\|_\infty \le \delta_n  \\
		nI(P;W) + \sqrt{\frac{2n V(P;W)}{1-\epsilon}} - \log \frac{1-\epsilon}{2} & \mathrm{else.}
		\end{array} \right.
\label{eq:zeta-hat-PW}
\end{equation}
In the case $\epsilon > \frac{1}{2}$, note that 
$\sqrt{\frac{2n V(P;W)}{1-\epsilon}} \le - \frac{1}{2} \sqrt{nV_{\max}} t_\epsilon$ for all $P\in\Pi$.

\begin{proposition}
Assume {\bf (A1)} and {\bf (A3)} hold.
Fix any positive vanishing sequence $\delta_n$. 
Consider any sequence $\bx \in \calX^n$ and any distribution $\tilde{P} \in \scrP(\calX)$ such that 
$\tilde{P} \in \Pi(\delta_n)$ if $\hat{P}_\bx \in \Pi(\delta_n)$ (and $\tilde{P}$ is unrestricted otherwise).
Let $Q=(\tilde{P}W)$.
The following asymptotic inequality holds uniformly over $\bx$ and $Q$ satisfying the conditions above:
\begin{equation}
   - \log \beta_{1-\epsilon}(W^n(\cdot|\bx), Q^n) \le \hat{\zeta}_n(\hat{P}_\bx,Q;\delta_n) + \frac{1}{2} \log n + o(1) .
\label{eq:beta-converse}
\end{equation}
\label{prop:beta-xQ}
\end{proposition}

{\em Proof}. 
Define
\begin{eqnarray}
   \overline{D}_n & \triangleq & \frac{1}{n} \sum_{i=1}^n D(W_{x_i}\|Q) = \sum_{x\in\calX} P(x) D(W_x\|Q) = D(W\|Q|P) ,
		\label{eq:Dbar-converse-CC} \\
   \overline{V}_n & \triangleq & \frac{1}{n} \sum_{i=1}^n V(W_{x_i}\|Q) 
				= \sum_{x\in\calX} P(x) V(W_x\|Q) = V(W\|Q|P) ,
		\label{eq:Vbar-converse-CC} \\
   \overline{T}_n & \triangleq & \frac{1}{n} \sum_{i=1}^n T(W_{x_i}\|Q) 
				= \sum_{x\in\calX} P(x) T(W_x\|Q)= T(W\|Q|P) ,
		\label{eq:Tbar-converse-CC} \\
   \overline{S}_n & \triangleq & \frac{\overline{T}_n}{[\overline{V}_n]^{3/2}} 
		\label{eq:Sbar-converse-CC}
\end{eqnarray}
and
\[ na_n \triangleq n\overline{D}_n - \sqrt{n\overline{V}_n} t_\epsilon 
	- \frac{1}{6} \overline{S}_n \sqrt{\overline{V}_n} (t_\epsilon^2 - 1) . 
\]
Similarly $\frac{1}{n} \sum_{i=1}^n D_s(W_{x_i}\|Q) = D_s(W\|Q|P)$.

{\bf Case~I:}
$P=\hat{P}_\bx \in \Pi(\delta_n)$ and $\tilde{P} \in \Pi(\delta_n)$.
Then \eqref{eq:zeta-PQ} and \eqref{eq:zeta-hat-PQ} yield $\hat{\zeta}_n(P,Q;\delta_n) = \zeta_n(P,Q)$ where $Q = (\tilde{P}W)$.
Assumption {\bf (LR1)} of Sec.~\ref{sec:LR-vary} holds because $\supp\{Q^*\} = \calY$, hence
$\supp\{(\tilde{P}W)\} = \calY$ for all $\tilde{P} \in \Pi(\delta_n)$,
and $n$ large enough. Assumption {\bf (LR2)} holds because $\calX$ is finite, $V_{\min} = \min_{P\in\Pi} V(P;W) > 0$ by {\bf (A1)},
and $V(W_x\|\cdot)$ is continuous at $Q^*$, hence 
$\overline{V}_n = V(W\|Q|P) > \frac{1}{2} V_{\min}$ 
for all $P,\tilde{P} \in \Pi(\delta_n)$, $Q=(\tilde{P}W)$, and $n$ large enough. 
Assumption {\bf (LR3)} holds because $\calX$ and $\calY$ are finite alphabets.

First assume $Q$ is such that $L = \log \frac{W(Y|X)}{Q(Y)}$ is a nonlattice random variable.
Then the sequence of loglikelihoods $L_i \triangleq \log [W(Y_i|x_i)/Q(Y_i)],\,i \ge 1$ satisfies
the semistrong nonlattice condition (Def.~\ref{def:semistrong}), where the marginal distribution $\pi_i$ is a function of $x_i$. 
Further note that $\{L_i\}$ satisfies a stronger property in which the statements after the semicolons in {\bf (NL1)}---{\bf (NL3)}
hold not only $\forall n > n_0$ but also $\forall \bx$ such that $\hat{P}_\bx \in \Pi(\delta_n)$.
The Taylor expansions of Lemma~\ref{lem:zetan-PQ} (with $P=\hat{P}_\bx$) and \eqref{eq:nan} (with $\pP_n = W^n(\cdot|\bx)$)
hold uniformly over $\bx$ such that $\hat{P}_\bx \in \Pi(\delta_n)$.
The asymptotics of $\beta_{1-\epsilon}(W^n(\cdot|\bx), Q^n)$ are then obtained from Theorem~\ref{thm:NP}(i):
\begin{eqnarray}
   \lefteqn{- \log \beta_{1-\epsilon}(W^n(\cdot|\bx), Q^n)} \nonumber \\
	& = & n a_n + \frac{1}{2} \log n + \frac{1}{2} t_\epsilon^2 
		+ \frac{1}{2} \log (2\pi\overline{V}_n) + o(1) \nonumber \\
	& = & n D(W\|Q|P) - \sqrt{n V(W\|Q|P)} \,t_\epsilon 
		+ \frac{1}{2} \log n + F_\epsilon(W\|Q|P) + o(1) \nonumber \\
	& = & \zeta_n(P,Q) + \frac{1}{2} \log n + o(1)
\label{eq:beta-converse-GoodTypes}
\end{eqnarray}
uniformly over $\hat{P}_\bx, \tilde{P} \in \Pi(\delta_n)$.

The case where $Q$ is such that $L$ is a lattice random variable with span $d_n$ yields the same solution
\eqref{eq:beta-converse-GoodTypes} because under {\bf (A3)}, $d_n$ must tend to 0 as $n \to \infty$. 
Indeed our condition on $Q$ may be written as $Q = Q^* + (\tilde{h}W)$ where $\|\tilde{h}\|_\infty \le 1$.
Hence
\[ L = L_0 - \log \left( 1 + \delta_n \frac{(\tilde{h}W)(Y)}{Q^*(Y)} \right) \]
where $L_0 = \log \frac{W(Y|X)}{Q^*(Y)}$ is a nonlattice random variable,
and $\frac{(\tilde{h}W)(Y)}{Q^*(Y)}$ is bounded. Hence $L = L_0 + O(\delta_n)$
with probability 1, and $d_n \to 0$. The asymptotics of $\beta_{1-\epsilon}(W^n(\cdot|\bx), Q^n)$ 
are then obtained from Theorem~\ref{thm:NP}(ii), and \eqref{eq:beta-converse-GoodTypes} holds again.
Thus \eqref{eq:beta-converse} holds as well.

{\bf Case~II:}
$\tilde{P} \notin \Pi(\delta_n)$ or $\hat{P}_\bx \notin \Pi(\delta_n)$.
We use the nonasymptotic bound of \eqref{eq:beta-PPV-lemma59} 
with $\pP_n = W^n(\cdot|\bx)$ and $\qQ_n = Q^n$:
\begin{eqnarray}
   - \log \beta_{1-\epsilon}(W^n(\cdot|\bx), Q^n)
	& \le & n \overline{D}_n + \sqrt{\frac{2n\overline{V}_n}{1-\epsilon}} - \log \frac{1-\epsilon}{2} \nonumber \\
	& = & n D(W\|Q|P) + \sqrt{\frac{2nV(W\|Q|P)}{1-\epsilon}} - \log \frac{1-\epsilon}{2} \nonumber \\
	& = & \hat{\zeta}_n(P,Q;\delta_n)
\label{eq:beta-converse-BadTypes}
\end{eqnarray}
and therefore \eqref{eq:beta-converse} holds again.
\hfill $\Box$

\begin{theorem}
Assume {\bf (A1)} and {\bf (A3)} hold.

\noindent
{\bf (i)} The volume $M[P]$ of any code of constant composition $P \in \scrP_n(\calX)$ and
(maximum or average) error probability $\epsilon$ satisfies
\begin{equation}
  \log M[P] \le \hat{\zeta}_n(P;W;\delta_n) + \frac{1}{2} \log n + o(1)
\label{eq:logM-UB-cc}
\end{equation}
uniformly over $P$, for any positive vanishing sequence $\delta_n$. 
There exist constants $c_1, c_2 > 0$ such that the following statements hold.

\noindent
{\bf (ii)} {\em (Bad Types}). 
If $\delta_n \gg n^{-1/4}$, there exists $n_0 \ge 1$ such that
\begin{equation}
  \log M[P] \le nC - c_2 n\delta_n^2 \quad \forall P \notin \Pi(\delta_n), \;n \ge n_0 .
\label{eq:logM-UB-bad-cc}
\end{equation}

\noindent
{\bf (iii)} {\em (Mediocre Types}). If $\delta_n \gg n^{-1/2} \log n$ and $P \notin \Pi^*(\delta_n)$ then 
\begin{equation}
  \log M[P] \le nC - \sqrt{nV_\epsilon} \,t_\epsilon - c_1 \sqrt{n} \delta_n [1+o(1)] .
\label{eq:logM-UB-bad2-cc}
\end{equation}

\noindent
{\bf (iv)} {\em (All Types)}.\footnote{
	The result \eqref{eq:logM-UB2-cc} is not used in this paper but is stated for completeness.}
\begin{equation}
 \log M[P] \le nC - \sqrt{nV_\epsilon} \,t_\epsilon + \frac{1}{2} \log n + \overline{A}_\epsilon + o(1) .
\label{eq:logM-UB2-cc}
\end{equation}
\label{thm:converse-CC}
\end{theorem}

{\em Proof}. 
{\bf (i)} Fix the distributions $Q=(PW)$ and $Q_\bY = Q^n$, and any positive vanishing sequence $\delta_n$.
Then $\hat{\zeta}_n(P,Q;\delta_n) = \hat{\zeta}_n(P;W;\delta_n)$.
Under the maximum-error probability criterion, \eqref{eq:logM-UB-cc}
follows from Theorem~\ref{thm:poly-converse-max} and Prop.~\ref{prop:beta-xQ}.
Since $\beta_{1-\epsilon}(W^n(\cdot|\bx), Q^n)$ is the same for all $\bx$ of composition $P$, 
\eqref{eq:logM-UB-cc} also holds under the average-error probability criterion
\cite[Lemma~29 p.~2318]{Polyanskiy10}. 

{\bf (ii)} If $\delta_n \gg n^{-1/4}$ and $P \notin \Pi(\delta_n)$, the Taylor expansion \eqref{eq:I-taylor}
implies that $I(P;W) \le C - 2c_2 \delta_n^2$ $\forall n \ge n_1$ for some $c_2 > 0$ and $n_1 \ge 1$. 
Let $\overline{V} \triangleq \max_{P\in\scrP(\calX)} V(P;W)$ which is finite.
Then \eqref{eq:logM-UB-cc} and \eqref{eq:zeta-hat-PW} yield
\begin{eqnarray*} 
   \log M[P] & \le & nI(P;W) + \sqrt{\frac{2nV(P;W)}{1-\epsilon}} - \log \frac{1-\epsilon}{2} \\
	& \le & nC - 2c_2 \delta_n^2 + \sqrt{\frac{2n\overline{V}}{1-\epsilon}} - \log \frac{1-\epsilon}{2} \\
	& \le & nC - c_2 n\delta_n^2 \quad \forall n > n_0, \,P \notin \Pi(\delta_n)
\end{eqnarray*}
for some $n_0 \ge n_1$. This establishes \eqref{eq:logM-UB-bad-cc}.

{\bf (iii)} If $P \notin \Pi(n^{-1/5})$, the claim follows directly from Part {\bf (ii)}.
If $n^{-1/2} \log n \ll \delta_n < n^{-1/5}$ and $P \in \Pi(n^{-1/5}) \setminus \Pi^*(\delta_n)$, 
then applying successively \eqref{eq:logM-UB-cc}, \eqref{eq:zeta-hat-PW}, and Lemma~\ref{lem:zetan-P}(iii)
with $\epsilon_n = n^{-1/5}$, we obtain
\begin{eqnarray*}
   \log M[P] \le \hat{\zeta}_n(P;W;\epsilon_n) + \frac{1}{2} \log n + o(1) 
	& = & \zeta_n(P;W) + \frac{1}{2} \log n + o(1) \\
	& \le & nC - \sqrt{nV_\epsilon} \,t_\epsilon - c_1 \sqrt{n} \delta_n [1+o(1)] ,
\end{eqnarray*}
hence \eqref{eq:logM-UB-bad2-cc} holds.

{\bf (iv)} If $P \notin \Pi^*(n^{-1/2} \log n)$, the claim follows directly from Part {\bf (iii)}. 
For $P \in \Pi^*(n^{-1/2} \log n)$, the claim follows from \eqref{eq:MP1} which is proved in the next section.
\hfill $\Box$

\subsection{Converse for General Codes, Maximum Error Probability Criterion}
\label{sec:converse-max}

In this section we derive an upper bound
on the volume of arbitrary codes with maximum error probability $\epsilon$.

\begin{theorem} 
Assume {\bf (A1)} and {\bf (A3)} hold.
The volume $M_{\calX^n}$ of any code with codewords in $\calX^n$ 
and maximum error probability $\epsilon  \in (0,1)$ satisfies
\begin{equation}
   \log M_{\calX^n} \le nC - \sqrt{nV_\epsilon} t_\epsilon + \frac{1}{2} \log n + \overline{A}_\epsilon + o(1) .
\label{eq:converse-max}
\end{equation}
\label{thm:converse-max}
\end{theorem}

The theorem is proved at the end of this section.
The analysis involves an optimization over $\scrP(\calX)$.
A different treatment is needed for a ``good set'' $\scrP_1$ of distributions
that are close to $\Pi^*$ and for the complement set $\scrP_1^c$. 

First we provide some motivation for the method of proof and give some definitions and two propositions 
(\ref{prop:M-P0-converse} and \ref{prop:maxP1}).
Fix any ${\sf F} \subseteq \calX^n$ and let $\scrP_0$ be any subset of $\scrP(\calX)$
such that $\bx \in {\sf F} \,\Rightarrow\, \hat{P}_\bx \in \scrP_0$.
Let $\calQ \triangleq \{Q=(PW), \,P\in\scrP(\calX)\}$.
The following proposition follows immediately from Theorem~\ref{thm:poly-converse-max} 
and Prop.~\ref{prop:beta-xQ}.

\begin{proposition}
Assume {\bf (A1)} and {\bf (A3)} hold. Fix any positive vanishing sequence $\delta_n$,
and consider any subset $\scrP_0$ (possibly dependent of $n$) of $\scrP(\calX)$.
The volume $M[\scrP_0]$ of any code with codewords of composition $\in \scrP_0$
and maximum error probability $\epsilon \in (0,1)$ satisfies
\begin{equation}
   \log M[\scrP_0] \le \inf_{Q\in\calQ} \sup_{P\in\scrP_0} \hat{\zeta}_n(P,Q;\delta_n) + \frac{1}{2} \log n + o(1) .
\label{eq:M-P0-converse}
\end{equation}
\label{prop:M-P0-converse}
\end{proposition}
At first sight this suggests seeking a solution to the minmax game with payoff function $\hat{\zeta}_n(P,Q;\delta_n)$
over $\scrP_0 \times \scrP(\calY)$. 

If $\Pi \subset \scrP_0$, a version of the above game with payoff
\[ \lim_{n \to \infty} \frac{1}{n} \zeta_n(P,Q) = D(W\|Q|P) = \sum_{x\in\calX} P(x) D(W_x\|Q) \]
admits the well-known equalizer saddlepoint solution $(P,Q) = (P^*,Q^*)$ for each $P^* \in \Pi$ \cite{Topsoe67},
as given in Lemma~\ref{lem:SP+cost}(i).
Owing to the equalizer property in \eqref{eq:D-SP}, we have
\begin{eqnarray*}
   \inf_{Q\in\calQ} \sup_{P\in\scrP_0} D(W\|Q|P)
	& = & \sup_{P\in\scrP_0} \inf_{Q\in\calQ}  D(W\|Q|P) = C
\end{eqnarray*}
even if $\scrP_0$ is not a convex set.

For finite $n$ there is generally no saddlepoint because  unlike $D(W\|Q|P)$, 
the payoff function $\zeta_n(P,Q)$ of (\ref{eq:zeta-PQ}) is not linear nor even concave in $P$ 
for $\epsilon < \frac{1}{2}$ due to the $O(\sqrt{n})$ variance term.
However
\begin{itemize}
\item In the CT weakly symmetric case where $\Pi^* = \{P^*\}$ (singleton), $V(W_x\|Q^*) = V(P^*;W)$ 
and $F_\epsilon(W_x\|Q^*) = F_\epsilon(P^*;W)$ for all $x\in\calX$, 
the capacity-achieving input and output distributions $P^*$ and $Q^*$ are uniform.
The game clearly admits an {\em asymptotic saddlepoint} solution $(P,Q) = (P^*,Q^*)$ in the sense that
	\begin{equation}
	   \zeta_n(P,Q^*) + o(1) \le \zeta_n(P^*,Q^*) \le \zeta_n(P^*,Q) + o(1) 
		\quad \forall P \in \scrP(\calX), \,\forall Q\in\calQ ,
    \label{eq:zeta-sym-SP}
    \end{equation}
	and the asymptotic value of the game is $\zeta_n(P^*,Q^*) = \zeta_n(P^*;W)$.
\item In the nonsymmetric case with $\Pi^* = \{P'\}$ (singleton), using Taylor expansion of $\zeta_n(\cdot,\cdot)$
	around $(P',Q^*)$ (Lemmas~\ref{lem:zetan-PQ} and \ref{lem:game}), we shall see there still exists 
	an {\em asymptotic saddlepoint} $(P_n',\tilde{Q}_n^*)$ if the maximization over $P$ is restricted to 
	a vanishing neighborhood of $P'$, of size $\delta_n = \frac{1}{n^{1/4} \log n}$:
	\begin{equation}
	   \zeta_n(P,\tilde{Q}_n^*) + o(1) \le \zeta_n(P_n',\tilde{Q}_n^*) \le \zeta_n(P_n',Q) ,
		\quad \forall P:\,\|P-P'\| \le \delta_n ,\,\forall Q .
    \label{eq:zeta-ns-SP}
    \end{equation}
	The asymptotic value of the game is
	$\zeta_n(P_n',\tilde{Q}_n^*) = nC - \sqrt{nV_\epsilon} \,t_\epsilon + \overline{A}_\epsilon + o(1)$,
	and the asymptotic saddlepoint is given in (\ref{eq:Pn*}), \eqref{eq:tQn*}.
\item If $\Pi^*$ is not a singleton, the saddlepoint property \eqref{eq:zeta-ns-SP} holds for each $P' \in \Pi^*$,
	$\tilde{Q}_n^*$ is independent of $P'$, and 
	$\zeta_n(P_n',\tilde{Q}_n^*) \le nC - \sqrt{nV_\epsilon} \,t_\epsilon + \overline{A}_\epsilon + o(1)$.
	The upper bound is achieved when $P'$ is the maximizer of \eqref{eq:Aeps+}.
\end{itemize}

Instead of applying Theorem~\ref{thm:poly-converse-max} with ${\sf F} = \calX^n$ directly, we
define a set of subcodes with maximum error probability $\epsilon$ each, and derive the asymptotics 
for these subcodes. The upper bound on $M$ is the sum of the upper bounds on the volume of the subcodes.

Define the ``good'' class of codewords
\begin{equation}
  {\sf F}_1 \triangleq \left\{ \bx \in \calX^n ~:~ \hat{P}_\bx \in \Pi^*\left(\frac{1}{n^{1/4} \log n}\right) \right\}
\label{eq:F1}
\end{equation}
and the corresponding ``good'' class of distributions over $\calX$
\begin{equation}
   \scrP_1 \triangleq \Pi^*\left(\frac{1}{n^{1/4} \log n}\right) .
\label{eq:P1}
\end{equation}
Hence $\bx \in {\sf F}_1 \,\Leftrightarrow\, \hat{P}_\bx \in \scrP_1$.

The following proposition shows that for each $P' \in \Pi^*$, the payoff function $\zeta_n(P,\tilde{Q}_n^*)$ 
is constant (up to a vanishing term) over the neighborhood $\{P\in\scrP(\calX) : \,\|P-P'\|_\infty \le \frac{1}{n^{1/4} \log n}\}$. 
Define the two {\em correction vectors}
\begin{eqnarray}
  h' & \triangleq & - \frac{t_\epsilon}{2\sqrt{V_\epsilon}}  \,\sfJ^+ v^{(P')} \quad \in \calH \label{eq:h'} \\
  \tilde{h}^* & \triangleq & - \frac{t_\epsilon}{2\sqrt{V_\epsilon}}  \,\sfJ^+ \tilde{v} \qquad \in \calH \label{eq:htilde*}
\end{eqnarray}
to which we associate modifications $P_n'$ and $\tilde{Q}_n^*$ of the capacity-achieving
input and output distributions as follows:
\begin{eqnarray}
  P_n' & \triangleq & P' + n^{-1/2} h' \qquad \quad \;\;\;\;\in \scrP(\calX) \label{eq:Pn*} \\
  \tilde{Q}_n^* & \triangleq & Q^* + n^{-1/2} (\tilde{h}^* W) \quad \quad \in \scrP(\calY) . \label{eq:tQn*}
\end{eqnarray}
Note that $\tilde{Q}_n^*$ does not depend on $P'$.

\begin{proposition} 
Assume {\bf (A1)} holds.
The pair of distributions $(P_n',\tilde{Q}_n^*)$ of \eqref{eq:Pn*}, \eqref{eq:tQn*}
satisfies the following asymptotic inequality, uniformly over $P' \in \Pi^*$ and 
$P:\,\|P-P'\|_\infty \le \frac{1}{n^{1/4} \log n}$:
\begin{eqnarray} 
	\zeta_n(P,\tilde{Q}_n^*)  
	& \le & \zeta_n(P_n',\tilde{Q}_n^*) + O \left( \frac{1}{\log^2 n} \right) \label{eq:maxP1-0} \\
	& = & \zeta_n(P';W) + \frac{t_\epsilon^2}{8} (\Ans^{(P')} - \Abns^{(P')})
		+ O \left( \frac{1}{\log^2 n} \right) \label{eq:maxP1-1} \\
	& \le & nC - \sqrt{nV_\epsilon} t_\epsilon + \overline{A}_\epsilon + O \left( \frac{1}{\log^2 n} \right) .
\label{eq:maxP1}
\end{eqnarray}
Hence the upper bound \eqref{eq:maxP1} holds uniformly over $P\in\scrP_1$.
Equality holds in \eqref{eq:maxP1-0} if $\supp\{P\} \subseteq \Xcap$, and in \eqref{eq:maxP1} if $P'$
achieves the maximum that defines $\overline{A}_\epsilon$ in \eqref{eq:Aeps+}.
\label{prop:maxP1}
\end{proposition}

{\em Proof:} Applying Lemma~\ref{lem:zetan-PQ}(ii) with $\delta_n = \frac{1}{n^{1/4} \log n}$, we obtain
\begin{eqnarray*}
  \zeta_n(P,\tilde{Q}_n^*) 
	& = & \zeta_n(P';W) + n(P-P')^\top \Delta - \sqrt{n} (P-P')^\top (\sfJ^\calX \tilde{h}^*-\tilde{g}) \\
	&& + \frac{1}{2} \tilde{h}^{*\top} \sfJ \tilde{h}^*
		+ \tilde{h}^{*\top} \breve{g}^{(P')} + O \left( \frac{1}{\log^2 n} \right) .
\end{eqnarray*}
Consider the second and third terms in the right side above.
If $\supp\{P\} \subseteq \Xcap$, then $(P-P')^\top \Delta = 0$, and Lemma~\ref{lem:game} implies
\[ (P-P')^\top (\sfJ^\calX \tilde{h}^*-\tilde{g}) = (P-P')^\top (\Jcap \tilde{h}^*-\tilde{g}) = 0 . \]
Else $\Xcap$ is a strict subset of $\calX$.
In that case, split $P-P'$ into a term $\delta_n  u_1$ that has support within $\Xcap$ and a term $\delta_n u_0$
that has support within $\calX \setminus \Xcap$, where $u_0 + u_1 \in \scrL(\calX)$, $u_0(x) \ge 0 \,\forall x$,
and $\|u_0\|_\infty, \|u_1\|_\infty \le 1$.
Therefore $\exists n_0$ such that
\begin{eqnarray*}
   \lefteqn{n(P-P')^\top \Delta - \sqrt{n} (P-P')^\top (\sfJ^\calX \tilde{h}^*-\tilde{g})} \\
	& \le & \sup_{u_0,u_1} [n\delta_n u_0^\top \Delta - \sqrt{n}\delta_n (u_0 + u_1)^\top (\sfJ^\calX \tilde{h}^*-\tilde{g})] \\
	& \stackrel{(a)}{=} & \sqrt{n} \delta_n \sup_{u_1 \in \scrL(\Xcap)} - u_1^\top (\Jcap \tilde{h}^*-\tilde{g}) \quad \forall n > n_0 \\
	& \stackrel{(b)}{=} & 0 ,
\end{eqnarray*}
where (a) holds because $\displaystyle \sup_{x\notin\Xcap} \Delta(x) < 0$ and $\|u_0\|_\infty, \|u_1\|_\infty \le 1$,
and (b) holds owing to Lemma~\ref{lem:game}.
In both cases,
\begin{eqnarray}
  \zeta_n(P,\tilde{Q}_n^*) 
	& \stackrel{(a)}{\le} & \zeta_n(P';W) + \frac{1}{2} \tilde{h}^{*\top} \sfJ \tilde{h}^*
		+ \tilde{h}^{*\top} \breve{g}^{(P')} + O \left( \frac{1}{\log^2 n} \right) \nonumber \\
	& \stackrel{(b)}{=} & \zeta_n(P';W) + \frac{1}{2} \|g^{(P')}\|_{\sfJ^+}^2 - \frac{1}{2} \|\breve{g}^{(P')}\|_{\sfJ^+}^2 
		 + O \left( \frac{1}{\log^2 n} \right) \nonumber \\
	& \stackrel{(c)}{=} & \zeta_n(P';W) + \frac{t_\epsilon^2}{8} (\Ans^{(P')} - \Abns^{(P')})
		 + O \left( \frac{1}{\log^2 n} \right) \nonumber \\
	& \stackrel{(d)}{=} & nC - \sqrt{nV_\epsilon} \,t_\epsilon + F_\epsilon(P';W) + \frac{t_\epsilon^2}{8} (\Ans^{(P')} - \Abns^{(P')})
		+ O \left( \frac{1}{\log^2 n} \right) \nonumber \\
	& \le & nC - \sqrt{nV_\epsilon} \,t_\epsilon + \max_{P'\in\Pi^*} \left\{ F_\epsilon(P';W) + \frac{t_\epsilon^2}{8} (\Ans^{(P')} - \Abns^{(P')})
		\right\} + O \left( \frac{1}{\log^2 n} \right) \nonumber \\
	& \stackrel{(e)}{=} & nC - \sqrt{nV_\epsilon} \,t_\epsilon + \overline{A}_\epsilon + O \left( \frac{1}{\log^2 n} \right) \nonumber
\end{eqnarray}
where (a) holds with equality if $\supp\{P\} \subseteq \Xcap$, 
(b) holds by Lemma~\ref{lem:game},
(c) by \eqref{eq:g-breve-tilde} and \eqref{eq:Ans}, 
(d) by \eqref{eq:zeta-PW}, 
and (e) by \eqref{eq:F} and \eqref{eq:Aeps+}. 
Hence \eqref{eq:maxP1-1} and \eqref{eq:maxP1} hold.
Moreover, since $\delta_n \gg n^{-1/2}$, we have $P_n' \in \scrP_1$ for sufficiently large $n$.
The right side of equality (c) equals the right side of \eqref{eq:maxP1-0}, and
inequality (a) holds for $P=P_n'$. Hence \eqref{eq:maxP1-0} holds.
\hfill $\Box$ \\

{\em Proof of Theorem~\ref{thm:converse-max}.}
Denote by $M[\scrP_1]$ the volume of a subcode with codewords in ${\sf F}_1$ and maximum error
probability $\epsilon$. Fix $\delta_n = \frac{1}{n^{1/4} \log n}$. Then \eqref{eq:zeta-hat-PQ}
yields $\hat{\zeta}_n(P,\tilde{Q}_n^*;\delta_n) = \zeta_n(P,\tilde{Q}_n^*)$ for all $P \in \scrP_1$ and thus
\begin{eqnarray}
   M[\scrP_1] & \stackrel{(a)}{\le} & \sup_{\bx\in {\sf F}_1} \frac{1}{\beta_{1-\epsilon}(W^n(\cdot|\bx), (\tilde{Q}_n^*)^n)} \nonumber \\
	& \stackrel{(b)}{=} & \exp\left\{ \max_{P \in \scrP_1} \zeta_n(P,\tilde{Q}_n^*) + \frac{1}{2} \log n + o(1) \right\} \nonumber \\
	& \stackrel{(c)}{\le} & \exp\left\{ nC - \sqrt{nV_\epsilon} \,t_\epsilon + \frac{1}{2} \log n + \overline{A}_\epsilon + o(1) \right\}
\label{eq:MP1}
\end{eqnarray}
where inequality (a) and equalities (b) and (c) follow from Theorem~\ref{thm:poly-converse-max},
Prop.~\ref{prop:beta-xQ}, and Prop.~\ref{prop:maxP1}, respectively.

We now upperbound $M[\scrP_1^c]$.
For the codewords not in ${\sf F}_1$ we have up to $(n+1)^{|\calX|-1}$ types. 
By Theorem~\ref{thm:converse-CC}(iii), the cardinality of each constant-composition subcode 
with type $P \notin \scrP_1$ is upper-bounded by \eqref{eq:logM-UB-bad2-cc}.
Hence
\begin{eqnarray*}
	M[P] & \le & \exp \left\{ nC - \sqrt{nV_\epsilon} \,t_\epsilon - \frac{c_1 \,n^{1/4}}{\log n} \,[1+o(1)] \right\} ,
	\quad \forall P \notin \scrP_1 .
\end{eqnarray*}
The cardinality of the union of such subcodes is therefore upper bounded by
\begin{eqnarray}
   M[\scrP_1^c] = \sum_{P \notin \scrP_1} M[P] 
	& \le & (n+1)^{|\calX|-1} \,\max_{P \notin \scrP_1} M[P] \nonumber \\
	& \le & \exp\left\{ nC - \sqrt{nV_\epsilon} \,t_\epsilon - \frac{c_1 \,n^{1/4}}{\log n} \,[1+o(1)] \right\} .
\label{eq:MP1c}
\end{eqnarray}

Finally, combining \eqref{eq:MP1} and \eqref{eq:MP1c}, we obtain
\begin{eqnarray}
   M_{\calX^n} & \le & M[\scrP_1] + M[\scrP_1^c] \nonumber \\
	& = & \exp \left\{ nC - \sqrt{nV_\epsilon} \,t_\epsilon + \frac{1}{2} \log n + \overline{A}_\epsilon + o(1) \right\} .
\label{eq:M-union-finiteX}
\end{eqnarray}
This completes the proof of Theorem~\ref{thm:converse-max}.
\hfill $\Box$

\subsection{Converse for General Codes, Average Error Probability}
\label{sec:converse-avg}

The key tools in this section are Theorem~\ref{thm:poly-converse-avg} \cite{Polyanskiy10} which involves
an optimization over $P_{\bX}$, and the same subcode decomposition that was presented in Sec.~\ref{sec:converse-max}.
We now leverage the analysis of Sec.~\ref{sec:converse-max} to derive the converse under
the average error probability criterion.

Consider a subset $\scrP_0$ of $\scrP(\calX)$ and the set ${\sf F}_0$ of sequences 
$\bx \in \calX^n$ that have composition $\hat{P}_{\bx} \in \scrP_0 \cap \scrP_n(\calX)$. 
For the constant-composition codes of Sec.~\ref{sec:converse-CC}, $\hat{P}_{\bx}$ is the same for all codewords.
For a more general code, $\hat{P}_{\bx}$ is not fixed but has a (nondegenerate) empirical
distribution $\pi$ over $\scrP_0 \cap \scrP_n(\calX)$. That is,
\begin{eqnarray}
   \pi (P) & = & \frac{1}{M} \sum_{1 \le m \le M} \sum_{\bx \in {\sf F}_0 : \,\hat{P}_{\bx} = P} P_{\bX(m)}(\bx) \nonumber \\
	& = & \sum_{\bx \in {\sf F}_0 : \,\hat{P}_{\bx} = P} P_\bX(\bx) ,
		\quad \forall P \in \scrP_0 \cap \scrP_n(\calX) 
\label{eq:pi-n}
\end{eqnarray}
where each probability distribution $P_{\bX(m)},\,1 \le m \le M$ is degenerate for deterministic codes, and
$P_\bX = \frac{1}{M} \sum_{1 \le m \le M} P_{\bX(m)}$.
Define $\Pi_n(\scrP_0)$ as the image of $\scrP({\sf F}_0)$ under the linear mapping that maps
each $P_{\bX} \in \scrP({\sf F}_0)$ to a distribution $\pi$ according to \eqref{eq:pi-n}.

Consider the set of sequences $\bx\in\calX^n$ that have the same type
$P \in \scrP_0 \cap \scrP_n(\calX)$ (this set is a type class).
Define $U_{\bX|P}$ as the uniform distribution over this type class.
Clearly $U_{\bX|P}$ is permutation-invariant for each $P \in \scrP_0 \cap \scrP_n(\calX)$,
and so is $P_{\bX} = \sum_{P \in \scrP_0 \cap \scrP_n(\calX)} \pi(P) \,U_{\bX|P}$ for any $\pi \in \Pi_n(\scrP_0)$.
Define
\[ \scrP^\mathrm{p.i.}({\sf F}_0) \triangleq \left\{ P_\bX = \sum_{P \in \scrP_0 \cap \scrP_n(\calX)} \pi(P) \,U_{\bX|P} ,
	\,\pi \in \Pi_n(\scrP_0) \right\} ,
\]
the set of all permutation-invariant distributions over ${\sf F}_0$.
With some abuse of notation, define the error probability
\begin{eqnarray}
   \beta_{1-\epsilon}(\pi,Q_\bY) & \triangleq & \beta_{1-\epsilon}(P_\bX \times W^n , P_\bX \times Q_\bY ) \nonumber \\
	\mathrm{with~} \pi \in \Pi_n(\scrP_0) \mathrm{~~and~~} P_{\bX} & = & \sum_{P \in \scrP_0 \cap \scrP_n(\calX)} \pi(P) \,U_{\bX|P}  
	\quad \in \scrP^\mathrm{p.i.}({\sf F}_0) .
\label{eq:beta-exch}
\end{eqnarray}

One difficulty with Theorem~\ref{thm:poly-converse-avg} \cite{Polyanskiy10} is that the minimization over
all probability distributions $P_\bX \in \scrP({\sf F}_0)$ is apparently intractable. 
However the minimization problem can be considerably simplified if one considers only permutation-invariant distributions
$Q_\bY$, as stated in the proposition below
\cite[Prop.~4.4]{Moulin12} \cite{Polyanskiy13}.
	
\begin{proposition}
Let ${\sf F}_0$ be a permutation-invariant subset of $\calX^n$. Then
\begin{eqnarray*}
   \lefteqn{\inf_{P_\bX \in \scrP({\sf F}_0)} \sup_{Q_\bY \in \scrP^\mathrm{p.i.}(\calY^n)} 
		\beta_{1-\epsilon}(P_{\bX} \times W^n, P_{\bX} \times Q_\bY)} \\
	& = & \inf_{P_\bX \in \scrP^\mathrm{p.i.}({\sf F}_0)}  \sup_{Q_\bY \in \scrP^\mathrm{p.i.}(\calY^n)} 
		\beta_{1-\epsilon}(P_{\bX} \times W^n, P_{\bX} \times Q_\bY) \\
	& = & \inf_{\pi\in\Pi_n(\scrP_0)} \sup_{Q_\bY \in \scrP^\mathrm{p.i.}(\calY^n)} \beta_{1-\epsilon}(\pi,Q_\bY) .
\end{eqnarray*}
\label{prop:pi}
\end{proposition}
{\em Proof}.
Consider a random variable $\Omega$ that is uniformly distributed over the set of all $n!$
permutations of the set $\{1,2,\cdots,n\}$. Denote by $\omega\bx$ the sequence obtained by
applying permutation $\omega$ to a sequence $\bx\in{\sf F}_0$. Hence
$\scrP^\mathrm{p.i.}({\sf F}_0) = \{ P_{\Omega\bX} ,\,P_{\bX} \in \scrP({\sf F}_0) \}$.

For any permutation-invariant $Q_\bY$, the error probability
$\beta_{1-\epsilon}(P_{\omega\bX} \times W^n, P_{\omega\bX} \times Q_\bY)$
is independent of $\omega$. Hence, by the same arguments as in \cite[Lemma~29]{Polyanskiy10},
we have
\begin{equation}
   \forall \omega : \quad \beta_{1-\epsilon}(P_{\omega\bX} \times W^n, P_{\omega\bX} \times Q_\bY) 
	= \beta_{1-\epsilon}(P_{\Omega\bX} \times W^n, P_{\Omega\bX} \times Q_\bY) 
\label{eq:beta-quasi}
\end{equation}
The claim follows by first applying \eqref{eq:beta-quasi} to the identity permutation
($\omega \bx \equiv \bx$) and then applying $\inf_{P_\bX} \sup_{Q_\bY}$ to both sides of the equation.
\hfill $\Box$

\begin{proposition}
The volume $M_{{\sf F}_0}$ of any code with codewords in a permutation-invariant subset
${\sf F}_0 \subseteq \calX^n$ and average error probability $\epsilon$ satisfies
\begin{equation}
   M_{{\sf F}_0} \le \sup_{\pi\in\Pi_n(\scrP_0)} \inf_{Q_\bY \in \scrP^\mathrm{p.i.}(\calY^n)} \frac{1}{\beta_{1-\epsilon}(\pi,Q_\bY)} .
\label{eq:M-pi}
\end{equation}
\label{prop:pi2}
\end{proposition}
{\em Proof}. Follows directly from Theorem~\ref{thm:poly-converse-avg} and Prop.~\ref{prop:pi}.
\hfill $\Box$

\begin{theorem}
Assume {\bf (A1)} and {\bf (A3)} hold.
The volume $M[\scrP_1]$ of any code with codewords in the set ${\sf F}_1$ of (\ref{eq:F1})
and average error probability $\epsilon$ satisfies
\[ \log M[\scrP_1] \le nC - \sqrt{nV_\epsilon} \,t_\epsilon + \frac{1}{2} \log n + \overline{A}_\epsilon + o(1) . \]
\label{thm:converse-CC2}
\end{theorem}
{\em Proof}.
Fix $Q_{\bY} = (\tilde{Q}_n^*)^n$, the $n$-fold product of $\tilde{Q}_n^*$ defined in (\ref{eq:tQn*}).
By Prop.~\ref{prop:pi2} we have
\begin{equation}
  M[\scrP_1] \le \sup_{\pi\in\Pi_n(\scrP_1)} \frac{1}{\beta_{1-\epsilon}(P_{\bX} \times W^n, P_{\bX} \times (\tilde{Q}_n^*)^n)}
\label{eq:pi2-P1}
\end{equation}
with $P_\bX = \sum_{P \in \scrP_1 \cap \scrP_n(\calX)} \pi(P) \,U_{\bX|P}$. 
Define the shorthand
\[ \beta_{1-\epsilon,n,P} \triangleq \beta_{1-\epsilon}(W^n(\cdot|\bx), (\tilde{Q}_n^*)^n) \]
and let
\begin{equation}
   \hat{\beta}_{1-\epsilon,n} \triangleq \exp\left\{ - nC + \sqrt{nV_\epsilon} \,t_\epsilon - \frac{1}{2} \log n - \overline{A}_\epsilon \right\} .
\label{eq:beta-epsilon-n}
\end{equation}
For any $P \in \scrP_1 \cap \scrP_n(\calX)$ and $\epsilon \in (0,1)$, we have
\begin{eqnarray}
  \beta_{1-\epsilon}(U_{\bX|P} \times W^n, U_{\bX|P} \times (\tilde{Q}_n^*)^n) 
	& \stackrel{(a)}{=} & \beta_{1-\epsilon}(W^n(\cdot|\bx), (\tilde{Q}_n^*)^n) \quad \forall \bx~:~ \hat{P}_\bx = P \nonumber \\
	& = & \beta_{1-\epsilon,n,P} \label{eq:beta-quasi2} \\
	& \stackrel{(b)}{=} & \exp\{ - \zeta_n(P,\tilde{Q}_n^*) - \frac{1}{2} \log n + o(1) \} \nonumber \\
	& \stackrel{(c)}{\ge} & \exp\left\{ - nC + \sqrt{nV_\epsilon} t_\epsilon - \frac{1}{2} \log n - \overline{A}_\epsilon + o(1) \right\} \nonumber \\
	& \stackrel{(d)}{=} & \hat{\beta}_{1-\epsilon,n} \,[1+o(1)]
\label{eq:beta-quasi3}
\end{eqnarray}
where (a) follows from \cite[Lemma~29]{Polyanskiy10},
(b) from \eqref{eq:zeta-hat-PW} and Prop.~\ref{prop:beta-xQ} with $\delta_n = \frac{1}{n^{1/4} \log n}$, 
(c) from Prop.~\ref{prop:maxP1}, and (d) from \eqref{eq:beta-epsilon-n}.
Inequality (c) holds with equality if $\Pi^*$ is a singleton and $\supp\{P\} \subseteq \calX^*$.

Now we show that for any $\pi\in\Pi_n(\scrP_1)$, we have
\begin{equation}
  \beta_{1-\epsilon}(P_{\bX} \times W^n, P_{\bX} \times (\tilde{Q}_n^*)^n) 
	\ge \hat{\beta}_{1-\epsilon,n} \,[1 + o(1)]
\label{eq:beta-P1-LB}
\end{equation}
with $P_\bX = \sum_{P \in \scrP_1 \cap \scrP_n(\calX)} \pi(P) \,U_{\bX|P}$. 
To show this lower bound, consider any test $Z$ that satisfies the type-I error probability constraint
\[ (P_{\bX} \times W^n)\{Z=1\} = \sum_{P \in \scrP_1 \cap \scrP_n(\calX)} \pi(P) \,U_{\bX|P} \times W^n \{Z=1\} \ge 1-\epsilon . \]
Let $\epsilon(P) \triangleq 1 - U_{\bX|P} \times W^n \{Z=1\}$, then 
we have
\begin{eqnarray}
  U_{\bX|P} \times (\tilde{Q}_n^*)^n \{Z=1\} 
	& \ge & \beta_{1-\epsilon(P)}(U_{\bX|P} \times W^n, U_{\bX|P} \times (\tilde{Q}_n^*)^n) , \nonumber \\
  \epsilon & \ge & \sum_{P \in \scrP_1 \cap \scrP_n(\calX)} \pi(P) \,\epsilon(P) ,
\label{eq:beta-epsilonP}
\end{eqnarray}
and
\begin{eqnarray}
  P_{\bX} \times (\tilde{Q}_n^*)^n \{Z=1\} 
	& = & \sum_{P \in \scrP_1 \cap \scrP_n(\calX)} \pi(P) \,U_{\bX|P} \times (\tilde{Q}_n^*)^n \{Z=1\} \nonumber \\
	& \stackrel{(a)}{\ge} & \sum_{P \in \scrP_1 \cap \scrP_n(\calX)} \pi(P) 
		\,\beta_{1-\epsilon(P)}(U_{\bX|P} \times W^n, U_{\bX|P} \times (\tilde{Q}_n^*)^n) \nonumber \\
	& \stackrel{(b)}{=} & \sum_{P \in \scrP_1 \cap \scrP_n(\calX)} \pi(P) \,\beta_{1-\epsilon(P),n,P}
\label{eq:NP-P1}
\end{eqnarray}
where (a) follows from \eqref{eq:beta-epsilonP} and (b) from \eqref{eq:beta-quasi2}.

We use the following lemma which is proved in Appendix~\ref{app:2GoodCodes}.
\begin{lemma} \label{lem:2beta}
The following asymptotic inequality holds:
\begin{equation}
  \sum_{P \in \scrP_1 \cap \scrP_n(\calX)} \pi(P) \beta_{1-\epsilon(P),n,P} \ge \hat{\beta}_{1-\epsilon,n} \,[1 + o(1)] .
\label{eq:beta-mix2}
\end{equation}
\end{lemma}

Combining \eqref{eq:NP-P1} with \eqref{eq:beta-mix2}, we obtain \eqref{eq:beta-P1-LB}.
The claim of Theorem~\ref{thm:converse-CC2} follows 
from \eqref{eq:pi2-P1}, \eqref{eq:beta-epsilon-n}, and \eqref{eq:beta-P1-LB}. 
\hfill $\Box$

\begin{remark}
The asymptotic inequality \eqref{eq:beta-P1-LB} can be strengthened to an asymptotic equality
in the case $\calX^* = \calX$ and $\Pi^*$ is a singleton. Per the discussion below \eqref{eq:beta-quasi3},
a sufficient condition for the asymptotic equality
$\beta_{1-\epsilon,n,P} = \hat{\beta}_{1-\epsilon,n} \,[1 + o(1)]$ to hold is that
$\calX^* = \calX$ and $\Pi^*$ is a singleton.
Now consider a collection of optimal tests $Z_P$ (at significance level $1-\epsilon$)
for each pair $U_{\bX|P} \times W^n$ {\em vs.} $U_{\bX|P} \times (\tilde{Q}_n^*)^n$ 
when $P$ ranges over $\scrP_1  \cap \scrP_n(\calX)$, i.e.,
\begin{eqnarray}
   \forall P \in \scrP_1  \cap \scrP_n(\calX): \quad U_{\bX|P} \times W^n \{Z_P =1\} & \ge & 1-\epsilon , \nonumber \\
	U_{\bX|P} \times (\tilde{Q}_n^*)^n \{Z_P =1\}
		& = & \beta_{1-\epsilon}(U_{\bX|P} \times W^n, U_{\bX|P} \times (\tilde{Q}_n^*)^n) \nonumber \\
		& = & \beta_{1-\epsilon,n,P} \nonumber \\
		& = & \hat{\beta}_{1-\epsilon,n} \,[1 + o(1)] .
\label{eq:P1-NP}
\end{eqnarray}
Take $Z_P$ as a test for $P_{\bX} \times W^n$ {\em vs.} $P_{\bX} \times (\tilde{Q}_n^*)^n$.
It then follows from \eqref{eq:P1-NP} that
\begin{equation}
  \beta_{1-\epsilon}(P_{\bX} \times W^n, P_{\bX} \times (\tilde{Q}_n^*)^n) \le \hat{\beta}_{1-\epsilon,n} \,[1 + o(1)] .
\label{eq:beta-P1-UB}
\end{equation}
Combining the asymptotic upper bound \eqref{eq:beta-P1-UB} with the asymptotic lower bound \eqref{eq:beta-P1-LB},
we conclude that asymptotic equality holds in \eqref{eq:beta-P1-LB}.
\end{remark}

Next we need to take care of the ``bad'' subcode whose codewords have composition outside $\scrP_1$.
This is done by splitting the subcode into subexponentially many subcodes $\calC_j, \,1 \le j \le J$
with respective volume $M_j$ and average error probability $\epsilon_j$. An upper bound on the volume
$M[\scrP_1^c]$ of the bad subcode is then derived and holds for all feasible choices
of $\{\epsilon_j\}$ and $\{M_j\}$.

\begin{theorem}
Assume {\bf (A1)} holds.
The volume $M[\scrP_1^c]$ of any code with codewords in $\calX^n \setminus {\sf F}_1 = {\sf F}(\scrP_1^c)$ 
and average error probability $\epsilon \in (0,1)$ satisfies
\[ \log M[\scrP_1^c] \le nC - \sqrt{nV_\epsilon} \,t_\epsilon - \frac{c_1 \,n^{1/4}}{\log n} \,[1+o(1)] . \]
\label{thm:converse-CC3}
\end{theorem}
{\em Proof}: see Appendix~\ref{app:converse-CC3}.

Finally we combine Theorems~\ref{thm:converse-CC2} and \ref{thm:converse-CC3} 
to derive the lower bounds on the volume of arbitrary codes.
The proof of Theorem~\ref{thm:converse} is given in Appendix~\ref{app:converse-avg}.

\section{Conclusion}
\label{sec:discuss}
\setcounter{equation}{0}

This paper has investigated the use of strong large deviations for analyzing the fundamental limits
of coding over memoryless channels, under regularity assumptions. Tight asymptotic lower and upper bounds
on the maximum code log-volume have been obtained by analyzing two related strong large-deviations problems.
The lower bound is obtained using the classical random coding union bound
which is evaluated by solving a strong large-deviations problem for a pair of random variables. 
The upper bound is related to the precise asymptotics of NP tests and is derived by carefully
optimizing the auxiliary distribution in these tests. In particular, Theorems~27 and 31 of 
Polyanskiy \cite{Polyanskiy10} providing upper bounds on $M_\avg^*(n,\epsilon)$ and
$M_{\max}^*(n,\epsilon)$  respectively in terms of the type-II error probability 
of a family of NP binary hypothesis tests were fundamental in this analysis. 
Since the gap between the asymptotic lower and upper bounds is easily computable and small (``a few nats'')
for the channels we numerically evaluated, we conclude that the random coding union bound 
and the aforementioned Theorems~27 and 31, coupled with the ``right'' choice of the auxiliary
distribution, are extremely powerful tools indeed.

Our analysis relies on classical asymptotic expansions of probability distributions
\cite{Wallace58} and on tools that were not used by Strassen and his successors,
namely the Cram\`{e}r-Ess\'{e}en theorem (as opposed to the weaker Berry-Ess\'{e}en theorem)
and Chaganty and Sethuraman's general approach to strong large deviations, which goes
beyond the classical iid setting. 

We conclude that this analytic framework is systematic and powerful.
The same framework could in principle be used to derive the full asymptotics of the $o(1)$
terms in our lower and upper bounds. Indeed we have only used one term of the Edgeworth
expansion beyond the asymptotic normal distribution, but more could be used.
Other extensions of this work include the case of cost constraints, which is treated similarly,
and for which the third-order term remains $\frac{1}{2} \log n$ \cite{Moulin12b,Moulin13b}.

As a final note, even though the lower and upper bounds almost match under
the average error probability criterion, by \eqref{eq:Mmax-sandwich}, 
there remains a $\frac{1+o(1)}{2} \log n$ gap under the maximum error criterion.\\

{\bf Acknowledgements}. The author thanks Patrick Johnstone for implementing and verifying the formulas and
producing the figures in this paper, Profs.~Chandra Nair and Vincent Tan for inspiring discussions, 
and the reviewers for insightful comments and helpful suggestions.

\clearpage

\appendix
\renewcommand{\theequation}{\Alph{section}.\arabic{equation}}

\section{Proof of Lemma~\ref{lem:UV}}
\label{app:UV}
\setcounter{equation}{0}

{\bf (i)} From \eqref{eq:fisher} we have
\begin{equation}
  \forall x,x' \in \calX : \quad \frac{\partial D(W_{x'}\|(PW))}{\partial P(x)} 
	= - \eE_{W_{x'}} \left[ \frac{W_x(Y)}{(PW)(Y)} \right] = - J_{xx'}^\calX (P;W) .
\label{eq:dDx'dpx}
\end{equation}
Moreover
\begin{eqnarray}
   \lefteqn{\frac{\partial V(W_{x'}\|(PW))}{\partial P(x)}} \nonumber \\
	& = & \frac{\partial}{\partial P(x)} \eE_{W_{x'}} \left( \log^2 \frac{W(Y|x')}{(PW)(Y)} \right) 
		- \frac{\partial}{\partial P(x)} D^2(W_{x'}\|(PW)) \nonumber \\
	& = & \eE_{W_{x'}} \frac{\partial}{\partial P(x)} \left( \log^2 \frac{W(Y|x')}{(PW)(Y)} \right) 
		- 2 D(W_{x'}\|(PW)) \,\frac{\partial}{\partial P(x)} D(W_{x'}\|(PW)) \nonumber \\
	& \stackrel{(a)}{=} & - 2 \left[ \eE_{W_{x'}} \left( \frac{W_x(Y)}{(PW)(Y)} \log \frac{W_{x'}(Y)}{(PW)(Y)} \right)
		- \left( \eE_{W_{x'}} \frac{W_x(Y)}{(PW)(Y)} \right) \left( \eE_{W_{x'}} \log \frac{W_{x'}(Y)}{(PW)(Y)} 
		\right) \right] \nonumber \\
	& = & - 2 \,\Cov_{W_{x'}} \left( \frac{W_x(Y)}{(PW)(Y)} , \log \frac{W_{x'}(Y)}{(PW)(Y)} \right) 
\label{eq:dVx'dpx}
\end{eqnarray}
where (a) follows from (\ref{eq:dDx'dpx}).

Taking the expectation of \eqref{eq:dVx'dpx} with respect to $X' \sim P \in \Pi$, we obtain
\begin{equation}
   \sum_{x' \in \calX} P(x') \frac{\partial V(W_{x'}\|(PW))}{\partial P(x)} = \breve{v}^{(P)}(x) . 
\label{eq:E-KXx}
\end{equation}
Averaging with respect to $P$, we obtain $\eE_P[\breve{v}^{(P)}(X)] = 0$.

Next, from \eqref{eq:v}---\eqref{eq:vbreve} and \eqref{eq:V} we obtain, for $P\in\Pi$,
\begin{eqnarray}
   v^{(P)}(x) & = & \frac{\partial V(P;W)}{\partial P(x)} \nonumber \\
	& = & V(W_x\|Q^*) + \sum_{x' \in \calX} P(x') \,\frac{\partial V(W_{x'}\|(PW))}{\partial P(x)} \nonumber \\
	& = & \tilde{v}(x) + \breve{v}^{(P)}(x) , \quad \forall x\in\calX
\label{eq:dVdpx}
\end{eqnarray}
which yields \eqref{eq:v-sum}. Since $\eE_P [\tilde{v}(X)] = V(P;W)$, we obtain
$\eE_P [v^{(P)}(X)] = \eE_P [\tilde{v}(X)] + \eE_P [\breve{v}^{(P)}(X)] = V(P;W)$.

{\bf (ii)} Since $V(P;W) = \sum_x P(x) \tilde{v}(x)$ for all $P\in\Pi$, the vectors $v^{(P)}$ and $\tilde{v}$
have the same projection onto $\calH$ of \eqref{eq:H}.

{\bf (iii)} Equations \eqref{eq:Vu=V}---\eqref{eq:Su=S} follow from 
Part (ii) and the definitions of the functions $V,\Vu,T,\Tu,S,\Su$.

From the definitions (\ref{eq:U}) and (\ref{eq:V}) we have
\begin{equation}
  \Vu(P;W) - V(P;W) = \sum_{x\in\calX} P(x) [I^2(P;W) - D^2(W_x\|PW)] \le 0, \quad \forall P\in\scrP(\calX)
\label{eq:V-U}
\end{equation}
where the inequality holds by Jensen's inequality. 
Differentiating (\ref{eq:V-U}), we have
\begin{eqnarray*}
  \frac{\partial[\Vu(P;W)-V(P;W)]}{\partial P(x)} 
	& = & I^2(P;W) - D^2(W_x\|(PW)) \\
	& & + 2 \sum_{x\in\calX} P(x') \left[ I(P;W) \,\frac{\partial}{\partial P(x)} I(P;W) \right. \\
	& & \quad \left. - D(W_{x'}\|(PW)) \,\frac{\partial}{\partial P(x)} D(W_{x'}\|(PW)) \right] .
\end{eqnarray*}
For $P \in \Pi$, $x \in \mathrm{supp}(P)$, we obtain
\begin{eqnarray*}
  \frac{\partial [\Vu(P;W)-V(P;W)]}{\partial P(x)} 
	& = & 0 + 2 I(P;W) \left[ \frac{\partial I(P;W)}{\partial P(x)} 
			- \sum_{x' \in \calX} P(x') \frac{\partial D(W_{x'}\|PW)}{\partial P(x)} \right] \\
	& = & 2 I(P;W) \,[D(W_x\|Q^*) - 1 + 1] \\
	& = & 2 I^2(P;W)
\end{eqnarray*}
where the second equality follows from \eqref{eq:del-I} and \eqref{eq:dDx'dpx}.
Hence \eqref{eq:delU=delV} holds. We also have
\begin{eqnarray}
  \breve{v}^{(P)}(x) & = & \sum_{x'\in\calX} P(x') \frac{\partial V(W_{x'}\|PW)}{\partial P(x)} \nonumber \\
	& = & -2 \left[ \sum_{x'\in\calX} \sum_{y\in\calY} \frac{W(y|x) \,W(y|x') P(x')}{Q^*(y)} 
		\log \frac{W(y|x')}{Q^*(y)} - I(P;W) \right] \nonumber \\
	& = & - 2 \,\left\{ \eE_{P \times W} \left[ \frac{W_x(Y)}{Q^*(Y)} \log \frac{W(Y|X)}{Q^*(Y)} \right] 
		- \eE_{Q^*} \left[ \frac{W_x(Y)}{Q^*(Y)} \right]  
		\,\eE_{P \times W} \left[ \log \frac{W(Y|X)}{Q^*(Y)} \right] \right\} \nonumber \\
	& = & - 2 \,\Cov_{P \times W} \left( \frac{W_x(Y)}{Q^*(Y)} , \log \frac{W(Y|X)}{Q^*(Y)} \right) 
\label{eq:vbreve0}
\end{eqnarray}
where the first equality is the definition \eqref{eq:vbreve}, and the second uses (b) in (\ref{eq:dVx'dpx}) 
together with $D(W_{x'}\|Q^*) \equiv I(P;W)$ for $x' \in \mathrm{supp}\{P\}$.
Alternatively $\breve{v}^{(P)}(x)$ may be expressed in terms of the reverse channel $\widecheck{W}$ of (\ref{eq:Wreverse}) as
\begin{eqnarray}
   \breve{v}^{(P)}(x) & = & -2 \left[ \sum_{x\in\calX} \sum_{y\in\calY} W(y|x) \,\widecheck{W}(x'|y) 
		\log \frac{\widecheck{W}(x'|y)}{P(x')} - I(P;W) \right] \nonumber \\
	& = & -2 \left[ \sum_{y\in\calY} W(y|x) D(\widecheck{W}_y \| P) - I(P;W) \right] \nonumber \\
	& = & -2 [\eE_{W_x} D(\widecheck{W}_Y \| P) - I(P;W)]
\label{eq:vbreve2}
\end{eqnarray}
which completes the derivation of \eqref{eq:vbreve2-lemma}.
\hfill $\Box$

\clearpage

\section{Proof of Lemma~\ref{lem:zetan-P}}
\label{app:zetan-P}
\setcounter{equation}{0}

{\bf (i)} Application of Taylor's theorem yields the following asymptotics 
for any $P' \in \Pi$ and $\|P-P'\|_\infty \le \delta_n \to 0$:
\begin{eqnarray}
   I(P) & = & I(P') + (P-P')^\top \nabla I(P') - \frac{1}{2} \|P - P'\|_{\sfJ^\calX}^2 + o(\delta_n^2) \label{eq:I-taylor} \\
   G(P) & = & G(P') + (P - P')^\top \nabla G(P') + o(\delta_n) \label{eq:G-taylor} \\
   F(P) & = & F(P') + O(\delta_n)
\label{eq:F-taylor}
\end{eqnarray}
where $P-P' \in \scrL(\calX)$.
By \eqref{eq:I-taylor}---\eqref{eq:F-taylor} and the definitions of $\zeta_n(P)$ and $\Delta$, we have
\begin{eqnarray}
   \zeta_n(P) & = & \zeta_n(P') + n (P-P')^\top \Delta - \frac{n}{2} \|P - P'\|_{\sfJ^\calX}^2 
	+ \sqrt{n} (P - P')^\top \nabla G(P')  \nonumber \\
	& & \quad + o(n\delta_n^2) + o(\sqrt{n} \delta_n) + O(\delta_n) .
\label{eq:zetan-taylor}
\end{eqnarray}
Hence \eqref{eq:zetan*-taylor} holds, and the convergence is uniform.\\

{\bf (ii)} 
For $P' \in \Pi^*$ and $g=\nabla G(P')$, evaluating \eqref{eq:zetan-taylor} at $P_n' = P' + n^{-1/2} \sfJ^+ g$ 
of (\ref{eq:Pn*-lem}), we have $P_n' \in \scrP(\calX^*)$ and thus
\[ \zeta_n(P_n') = \zeta_n(P') - \frac{1}{2} \|\sfJ^+ g\|_\sfJ^2 + \|g\|_{\sfJ^+}^2 + o(1) \]
from which \eqref{eq:zetan*-lem} follows. 
Since $\Delta(x) < 0$ for all $x\notin\Xcap$, there exists $n_0$ such that
\begin{eqnarray}
   \lefteqn{n (P-P')^\top \Delta - \frac{n}{2} \|P - P'\|_{\sfJ^\calX}^2 + \sqrt{n} g^\top (P - P')} \nonumber \\
	& \le & \sup_{P\in\scrP(\Xcap)} \left[ - \frac{n}{2} \|P-P'\|_\Jcap^2 + \sqrt{n} g^\top (P-P') \right] , \quad \forall n > n_0 .
\label{eq:zetan-local-opt-general}
\end{eqnarray}
The maximization problem is convex and admits a simple solution in the case
$\scrP(\Xcap) \subseteq \Pi + \scrL(\calX^*)$. Then any $P\in\scrP(\Xcap)$ is the sum of some $P_0 \in \Pi$ and $h \in \scrL(\calX^*)$.
Hence $\|P-P'\|_\Jcap^2 = \|h\|_\Jcap^2 = \|h\|_\sfJ^2$, and $g^\top (P-P') \le g^\top h$ (since $P' \in \Pi^*$), 
with equality if $P_0 \in \Pi^*$.
Therefore the supremand in \eqref{eq:zetan-local-opt-general} is equal to 
$\sup_{h\in\scrL(\calX^*)} \left[ - \frac{n}{2} \|h\|_\sfJ^2 + \sqrt{n} g^\top h \right] \ge 0$.
Moreover $g^\top h = 0$ for any $h \in \scrL(\calX^*) \cap \mathrm{ker}(\sfJ)$, hence the supremum above
is achieved on $\calH = \scrL(\calX^*) \cap \mathrm{ker}(\sfJ)^\perp$ defined in \eqref{eq:H}.
By Lemma~\ref{lem:J+}, this supremand is equal to $\calE(g) = \frac{1}{2} \|g\|_{\sfJ^+}^2$ 
and is achieved by $h = \sfJ^+ g = P_n'-P'$.
Therefore \eqref{eq:zetan-taylor} yields the local optimality property \eqref{eq:zetan-local-opt}.

{\bf (iii)} Since $V_{\min} = \min_{P\in\Pi} V(P;W) > 0$ owing to Assumption {\bf (A1)}, by continuity of $V(\cdot ;W)$
there exists $\delta > 0$ such that $V(P;W)$ is bounded away from zero for $P\in\Pi(\delta)$, 
hence $F(P) = F_\epsilon(P;W)$ of \eqref{eq:F} is upper-bounded by some finite $F_{\max}$.
For $P\in\Pi(\epsilon_n)\setminus\Pi^*(\delta_n)$, we upperbound $\zeta_n(P)$
by considering the case of full-rank and rank-deficient $\Jcap$ separately.
Let $\lambda > 0$ be the smallest nonzero eigenvalue of $\Jcap$.

(a) If $\Jcap$ has rank $|\Xcap|$, then $\Pi = \Pi^* = \{P^*\}$ is a singleton and $\Jcap = \sfJ$. Let $h = P-P^*$, hence
$\frac{n}{2} \|h\|_\sfJ^2 \gg \sqrt{n} g^\top  h$. Then \eqref{eq:zetan-taylor} yields
\begin{equation}
   \zeta_n(P) \le \zeta_n(P^*) - \frac{n}{2} \|h\|_\sfJ^2 \,[1+o(1)] \le \zeta_n(P^*) - \frac{1}{2} n\delta_n^2 \,[1+o(1)]
\label{eq:opt-a}
\end{equation}
where the last inequality holds because $\|h\|_\sfJ \ge \lambda \|h\|_2 \ge \lambda \|h\|_\infty \ge \lambda \delta_n$.
Hence \eqref{eq:zetan-UB2} holds.

(b) If $\Jcap$ has rank $r < |\Xcap|$, let $P_0$ achieve $\min_{\tilde{P}\in\Pi} \|P-\tilde{P}\|_2$.
We now show that $\zeta_n(P)$ decays as $n\delta_n^2$
away from $\Pi$ and as $\sqrt{n} \delta_n$ inside $\Pi$, away from $\Pi^*$. Three cases are to be considered:

(b1) $P_0 \in \Pi^* \cap \mathrm{int}(\Pi)$. Then $P-P_0$ is orthogonal to ker($\Jcap$). Since
\[ \|P-P_0\|_\Jcap \ge \lambda \|P-P_0\|_2 \ge \lambda \|P-P_0\|_\infty 
	\ge \lambda \min_{\tilde{P}\in\Pi^*} \|P-\tilde{P}\|_\infty \ge \lambda \delta_n . 
\]
we obtain
\begin{equation}
   I(P) \le I(P_0) - \frac{1}{2} \lambda^2 \|P-P_0\|_\infty^2 [1+o(1)], \quad G(P) = G(P_0) + O(\|P-P_0\|_\infty) .
\label{eq:opt-b1}
\end{equation}
Since $I(P_0)=C$, $G(P_0) = -\sqrt{V_\epsilon} \,t_\epsilon$, and $\|P-P_0\|_\infty \ge \delta_n \gg n^{-1/2}$, 
we obtain $\zeta_n(P) \le nC - \sqrt{nV_\epsilon} \,t_\epsilon - \frac{1}{2} \lambda^2 n\delta_n^2$.

(b2) $P_0 \in \Pi^* \cap \partial \Pi$. Denote by $h$ the component of $P-P_0$ orthogonal to ker($\Jcap$),
then $c \|P-P_0\|_2 \le \|h\|_2 \le \|P-P_0\|_2$ for some constant $c > 0$ that depends on the geometry of the polytope $\Pi$.
Proceeding as in (b1) with $h$ in place of $P-P_0$, we obtain $\|h\|_\Jcap \ge c \lambda \delta_n$,  hence
\begin{equation}
   I(P) \le I(P_0) - \frac{1}{2} c^2 \lambda^2 \|P-P_0\|_\infty^2 [1+o(1)], \quad G(P) = G(P_0) + O(\|P-P_0\|_\infty)
\label{eq:opt-b2}
\end{equation}
hence $\zeta_n(P) \le nC - \sqrt{nV_\epsilon} \,t_\epsilon - \frac{1}{2} c^2 \lambda^2 n\delta_n^2$.
For $\Pi = \Pi^*$, either (b1) or (b2) holds, hence \eqref{eq:zetan-UB2} holds.

(b3) $P_0 \notin \Pi^*$. (This case is possible only if $\Pi \ne \Pi^*$). 
Since $V(P;W) = P^\top \tilde{v}$ for all $P\in\Pi$,
the projection of $\nabla G(P')$ onto $\mathrm{ker}(\Jcap)$ is a vector $\tilde{g}_0$
independent of $P'\in\Pi^*$; moreover $\tilde{g}_0 = 0$ if and only if $\Pi = \Pi^*$,
which is excluded by our hypothesis. Hence $\|\tilde{g}_0\|_2 > 0$.
By the extremal property of the polytope $\Pi^*$, there exists a constant $c' > 0$ such that the inequality
\begin{equation}
   (P_0-P')^\top \tilde{g}_0 \le -c' \|\tilde{g}_0\|_2 \,\|P_0-P'\|_2
\label{eq:Pi*-ineq}
\end{equation}
holds for $P'$ achieving $\min_{P'\in\Pi^*} \|P_0 - P'\|_2$. By assumption,
\[ \delta_n \le \|P-P'\|_\infty \le \|P-P_0\|_\infty + \|P_0-P'\|_\infty . \]
Fix the constant
\begin{equation}
   \omega = \frac{1}{2} \frac{c' \|\tilde{g}_0\|_2}{c' \|\tilde{g}_0\|_2 
	+ \sqrt{|\calX|} \max_{P'\in\Pi^*} \|\nabla G(P')\|_2} \in \left( 0,\frac{1}{2} \right) .
\label{eq:omega}
\end{equation}
One of the following statements is true:
\begin{equation}
    \|P-P_0\|_\infty \ge \omega \delta_n \quad \mathrm{or} \quad
	\|P_0-P'\|_\infty \ge (1-\omega) \delta_n \ge \omega \delta_n > \|P-P_0\|_\infty .
\label{eq:zetan-opt-statements}
\end{equation}
In the first case, $P$ is ``sufficiently far'' from $\Pi$; in the second case,
$P$ may be arbitrarily close to $\Pi$ but is ``sufficiently far'' from $\Pi^*$.
Since $\|P-P_0\|_\Jcap \ge \lambda \|P-P_0\|_\infty$, the first statement implies that
\begin{eqnarray}
   I(P) & \le & C - \frac{1}{2} c^2 \lambda^2 \omega^2 \|P-P_0\|_\infty^2 [1+o(1)] , \label{eq:opt-b3a} \\
   G(P) & \le & G(P') + O(\|P-P'\|_\infty)  \label{eq:opt-b3b}
\end{eqnarray}
hence $\zeta_n(P) \le \zeta_n(P') - \frac{1}{2} c^2 \lambda^2 \omega^2 n\delta_n^2 + o(n\delta_n^2) $,
and \eqref{eq:zetan-UB2} holds. The second statement implies that $I(P) \le I(P') = C$ and
\begin{eqnarray}
    G(P) & = & G(P') + (P - P_0 + P_0 - P')^\top \nabla G(P') + o(\delta_n) \nonumber \\
	& \le & G(P') + (P_0 - P')^\top \tilde{g}_0 + \|\nabla G(P')\|_2 \,\|P-P_0\|_2 +  o(\delta_n) \nonumber \\
	& \stackrel{(a)}{\le} & G(P') - \left( c \|P_0 - P'\|_2 \|\tilde{g}_0\|_2 
		- \|P-P_0\|_2 \max_{P'\in\Pi^*} \|\nabla G(P')\|_2 \right) +  o(\delta_n) \nonumber \\
	& \stackrel{(b)}{\le} & G(P') - \left( c(1-\omega) \|\tilde{g}_0\|_2 
		- \omega \sqrt{|\calX|} \max_{P'\in\Pi^*} \|\nabla G(P')\|_2 \right) \,\delta_n + o(\delta_n) \nonumber \\
	& \stackrel{(c)}{=} & G(P') - \frac{c}{2} \|\tilde{g}_0\|_2 \,\delta_n  + o(\delta_n)
\label{eq:opt-b3c}
\end{eqnarray}
where (a) follows from the extremality property \eqref{eq:Pi*-ineq},
(b) from the inequalities $\|P_0-P'\|_2 \ge \|P_0-P'\|_\infty \ge (1-\omega) \delta_n$ and
$\|P-P_0\|_2 \le \sqrt{|\calX|} \|P-P_0\|_\infty \le \sqrt{|\calX|} \omega \delta_n$,
and (c) from our choice of $\omega$.
Therefore
\begin{eqnarray}
    \zeta_n(P) & \le & \zeta_n(P') - \frac{c}{2} \|\tilde{g}_0\|_2 \,\sqrt{n} \delta_n + o(\sqrt{n} \delta_n) \nonumber \\
		& = & nC - \sqrt{nV_\epsilon} \,t_\epsilon - c_1 \sqrt{n} \delta_n [1+o(1)]
\label{eq:zetan-opt-UB}
\end{eqnarray}
for some $c_1 > 0$. We conclude that \eqref{eq:zetan-UB2} holds in both cases.
\hfill $\Box$

\section{Proof of Lemma~\ref{lem:zetan-PQ}}
\label{app:zetan-PQ}
\setcounter{equation}{0}

{\bf (i)} Fix any $P' \in \Pi^*$. We perform a second-order Taylor expansion of $\zeta_n$
around the point $(P', Q^*)$. 
We examine the three components of $\zeta_n$ and compute the corresponding derivatives.

Consider any $h\in\scrL(\calX)$ such that $\|h\|_\infty \le 1$ and let $P_n = P' + \delta_n h$.
Recall that $Q_n = (\tilde{P}_n W)$ where $\tilde{P}_n = P' + \tilde{\delta}_n \tilde{h}^*$
and $\tilde{h}^* \in \scrL(\calX^*)$.
The leading term of
\begin{equation}
   \frac{1}{n} \zeta_n(P_n,Q_n) = D(W\|Q_n|P_n) - \frac{1}{\sqrt{n}} \sqrt{V(W\|Q_n|P_n)} \,t_\epsilon
	+ \frac{1}{n} F_\epsilon(W\|Q_n|P_n)
\label{eq:zetan/n}
\end{equation}
is linear in $P_n$:
\begin{equation}
   D(W\|Q_n|P_n) = \sum_{x\in\calX} P'(x) D(W_x\|Q_n) + \delta_n \sum_{x\in\calX} h(x) D(W_x\|Q_n) .
\label{eq:DWQnPn}
\end{equation}
The following Taylor expansion of the first term in the right side as a function of 
$Q_n = (\tilde{P}_n W)$ holds at $Q^*$:
\begin{eqnarray}
   \sum_{x\in\calX} P'(x) D(W_x\|Q_n) 
	& = & \eE_{P' \times W} \left[ \log \frac{W(Y|X)}{(\tilde{P}_n W)(Y)} \right] \nonumber \\
	& = & \eE_{P' \times W} \left[ \log \frac{W(Y|X)}{Q^*(Y)} \right] 
		- \eE_{Q^*} \left[ \log \frac{(\tilde{P}_n W)(Y)}{Q^*(Y)} \right] \nonumber \\
	& = & I(P';W) - \eE_{Q^*} \log \left( 1 + \tilde{\delta}_n \frac{(\tilde{h}^* W)(Y)}{Q^*(Y)} \right) \nonumber \\
	& = & I(P';W) - \tilde{\delta}_n \eE_{Q^*} \left[ \frac{(\tilde{h}^* W)(Y)}{Q^*(Y)} \right] 
		+ \frac{\tilde{\delta}_n^2}{2} \eE_{Q^*} \left[ \frac{(\tilde{h}^* W)(Y)}{Q^*(Y)} \right]^2 + O(\tilde{\delta}_n^3) \nonumber \\
	& = & I(P';W) + \frac{\tilde{\delta}_n^2}{2} \tilde{h}^{*\top} \sfJ^\calX \tilde{h}^* + O(\tilde{\delta}_n^3) .
\label{eq:DWxQ-int1}
\end{eqnarray}
In the last line, the $O(\tilde{\delta}_n)$ term vanishes because $\sum_{x\in\calX} \tilde{h}^*(x) = 0$,
hence $\sum_{y\in\calY} (\tilde{h}^* W)(y) = 0$. The $O(\tilde{\delta}_n^2)$ term is obtained using (\ref{eq:fisher}).
Since $\supp\{\tilde{h}^*\} \subseteq \calX^*$, we may also write $\tilde{h}^{*\top} \sfJ^\calX \tilde{h}^* = \tilde{h}^{*\top} \sfJ \tilde{h}^*$.
The $O(\tilde{\delta}_n^3)$ remainder term does not depend on $h$.

Similarly, the following Taylor expansion of the second term in the right side of \eqref{eq:DWQnPn} holds:
\begin{eqnarray}
  \delta_n \sum_{x\in\calX} h(x) D(W_x\|Q_n)
	& = & \delta_n \sum_{x\in\calX} h(x) \,\left[ D(W_x\|Q^*) 
		- \eE_{W_x} \log \left( 1 + \tilde{\delta}_n \frac{(\tilde{h}^* W)(Y)}{Q^*(Y)} \right) \right] \nonumber \\
	& = & \delta_n \sum_{x\in\calX} h(x) \,(I(P';W) + \Delta(x)) - \delta_n \sum_{y\in\calY} (hW)(y) 
		\log \left( 1 + \tilde{\delta}_n \frac{(\tilde{h}^* W)(y)}{Q^*(y)} \right) \nonumber \\
	& = & \delta_n h^\top \Delta - \delta_n \tilde{\delta}_n h^\top \sfJ^\calX \tilde{h}^* + O(\delta_n \tilde{\delta}_n^2) .
\label{eq:DWxQ-int2}
\end{eqnarray}
Adding (\ref{eq:DWxQ-int1}) and (\ref{eq:DWxQ-int2}) and multiplying the sum by $n$ yields
\begin{eqnarray}
   n D(W\|Q_n|P_n) & = & nI(P';W) + n\delta_n h^\top \Delta + \frac{n\tilde{\delta}_n^2}{2} \tilde{h}^{*\top} \sfJ^\calX \tilde{h}^* 
	- n \delta_n \tilde{\delta}_n \,h^\top \sfJ^\calX \tilde{h}^* + O(\tilde{\delta}_n^3) + O(n \delta_n \tilde{\delta}_n^2) \nonumber \\
	& = & nI(P';W) + \delta_n h^\top \Delta + \frac{n\tilde{\delta}_n^2}{2} \tilde{h}^{*\top} \sfJ^\calX \tilde{h}^* 
		- n \delta_n \tilde{\delta}_n \,h^\top \sfJ^\calX \tilde{h}^* + O(n \delta_n \tilde{\delta}_n^2) .
\label{eq:zeta-D-asym}
\end{eqnarray}

Next consider the conditional variance term in \eqref{eq:zetan/n}:
\begin{equation}
  V(W\|Q_n|P_n) = \sum_{x\in\calX} P'(x) V(W_x\|Q_n) + \delta_n \sum_{x\in\calX} h(x) V(W_x\|Q_n) .
\label{eq:VWQnPn}
\end{equation}
The first term in the right side admits the Taylor expansion
\begin{eqnarray}
   \sum_{x\in\calX} P'(x) V(W_x\|Q_n) & = & \sum_{x\in\calX} P'(x) V(W_x\|Q^*) \nonumber \\
	& & + \tilde{\delta}_n \sum_{x'\in\calX} \tilde{h}^*(x') 
		\sum_{x\in\calX} P'(x)  \left. \frac{\partial V(W_x\|(PW))}{\partial P(x')} \right|_{P=P'} + O(\tilde{\delta}_n^2) \nonumber \\
	& = & V(P';W) + \tilde{\delta}_n \tilde{h}^{*^\top} \breve{v} + O(\tilde{\delta}_n^2) .
\label{eq:VWxQ-int1}
\end{eqnarray}
The second term in the right side of \eqref{eq:VWQnPn} can be expanded as
\begin{eqnarray}
   \delta_n \sum_{x\in\calX} h(x) V(W_x\|Q_n)
	& = & \delta_n \sum_{x\in\calX} h(x) V(W_x\|Q^*) + O(\delta_n \tilde{\delta}_n) \nonumber \\
	& = & \delta_n h^\top \tilde{v} + O(\delta_n \tilde{\delta}_n)
\label{eq:VWxQ-int2}
\end{eqnarray}
where $\tilde{v}$ was defined in \eqref{eq:vtilde}.
Substituting (\ref{eq:VWxQ-int1}) and (\ref{eq:VWxQ-int2}) into \eqref{eq:VWQnPn} and multiplying by $n$ yields
\begin{eqnarray}
   n V(W\|Q_n|P_n) 
	& = & n \left[ V(P';W) + \tilde{\delta}_n \tilde{h}^{*\top} \breve{v} + \delta_n h^\top \tilde{v} 
		+ O(\delta_n \tilde{\delta}_n) + O(\tilde{\delta}_n^2) \right] \nonumber \\
	& = & n \left[ V(P';W) + \tilde{\delta}_n \tilde{h}^{*\top} \breve{v} + \delta_n h^\top \tilde{v} 
		+ O(\delta_n \tilde{\delta}_n) \right] .
\label{eq:nVWQnPn}
\end{eqnarray}
By Assumption {\bf (A1)},  we have $V(P';W) > 0$. Using the expansion 
$\sqrt{1+\delta} = 1 + \frac{1}{2} \delta + O(\delta^2)$ as $\delta \to 0$, we obtain
\begin{eqnarray}
   \!\!\!\!\!\sqrt{n V(W\|Q_n|P_n)}
	& = &\sqrt{nV(P';W)} \sqrt{ 1 + \frac{1}{V(P';W)} \left( \tilde{\delta}_n \tilde{h}^{*\top} \breve{v} + \delta_n h^\top \tilde{v} 
		+ O(\delta_n \tilde{\delta}_n) \right)} \nonumber \\
	& = &\sqrt{nV(P';W)} + \sqrt{n} \,\frac{\tilde{\delta}_n \tilde{h}^{*\top} \breve{v} + \delta_n \,h^\top \tilde{v} + O(\delta_n \tilde{\delta}_n)}
			{2\sqrt{V(P';W)}} + O(\delta_n^2 \sqrt{n}) .
\label{eq:zeta-sqV-asym}
\end{eqnarray}
Finally,
\begin{eqnarray}
  T(W\|Q_n|P_n) & = & \sum_{x\in\calX} P'(x) T(W_x\|Q_n) + \delta_n \sum_{x\in\calX} h(x) T(W_x\|Q_n) \nonumber \\
	& = & T(P';W) + O(\tilde{\delta}_n) + O(\delta_n ) \nonumber \\
	& = & T(P';W) + O(\delta_n) .
\label{eq:zeta-T-asym}
\end{eqnarray}
It follows from \eqref{eq:nVWQnPn}, \eqref{eq:zeta-T-asym}, \eqref{eq:FQ}, and \eqref{eq:F} that
\begin{eqnarray}
   F_\epsilon(W\|Q_n|P_n) & = & \frac{1}{2} t_\epsilon^2 - \frac{T(W\|Q_n|P_n)}{6V(W\|Q_n|P_n)} (t_\epsilon^2 - 1)
	+ \frac{1}{2} \log (2\pi V(W\|Q_n|P_n)) \nonumber \\
	& = & F_\epsilon(P';W) + O(\delta_n) .
\label{eq:zeta-F-asym}
\end{eqnarray}
Combining \eqref{eq:zetan/n}, (\ref{eq:zeta-D-asym}), (\ref{eq:zeta-sqV-asym}), 
and (\ref{eq:zeta-F-asym}), we obtain
\[ \zeta_n(P_n,Q_n) = \zeta_n(P';W) + n\delta_n h^\top \Delta + \frac{n\tilde{\delta}_n^2}{2} \tilde{h}^{*\top} \sfJ^\calX \tilde{h}^* 
		- n \delta_n \tilde{\delta}_n \,h^\top \sfJ^\calX \tilde{h}^*
		- \frac{t_\epsilon \sqrt{n}[\tilde{\delta}_n \tilde{h}^{*\top} \breve{v} + \delta_n \,h^\top \tilde{v}]}{2\sqrt{V(P';W)}}  
		+ O(\delta_n^2 \sqrt{n}) \] 
which reduces to (\ref{eq:zeta-asymp-general}) using the definitions of $\tilde{g}$ and $\breve{g}$
in \eqref{eq:g-breve-tilde}.\\

{\bf (ii)} Straightforward.
\hfill $\Box$

\section{Proof of Lemma~\ref{lem:HT-Taylor}}
\label{app:HT-Taylor}
\setcounter{equation}{0}

Successive differentiation of the cgf $\kappa$ at $s=1$ yields  $\kappa(1)=0$, $\kappa'(1)=D$,
$\kappa''(1)=V$, and $\kappa'''(1)=T$.
The supremum in the definition of $\Lambda(a)$ is achieved by $s$ that satisfies $a = \kappa'(s)$
and thus $s \to 1$ as $a \to D$. Moreover
\[ \frac{da}{ds} = \kappa''(s) \]
and $\Lambda(a) = s \kappa'(s) - \kappa(s)$.
Differentiating with respect to $s$, we obtain
\[ \frac{d\Lambda(a)}{ds} = s \kappa''(s) = s \frac{da}{ds} . \]
This implies 
\[ \Lambda'(a) \triangleq \frac{d\Lambda(a)}{da}
	= s \quad \Rightarrow \quad \Lambda'(D) = 1 . \]
Differentiating $\Lambda(a)$ a second time, we have
\[ \Lambda''(a) \triangleq \frac{d\Lambda'(a)}{da}
	= \frac{ds}{da} = \frac{1}{\kappa''(s)}
	\quad  \Rightarrow \quad \Lambda''(D) = \frac{1}{V} . \]
Differentiating a third time, we obtain
\[ \Lambda'''(a) \triangleq \frac{d\Lambda''(a)}{da}
	= \frac{ds}{da} \,\frac{d}{ds} \left( \frac{1}{\kappa''(s)} \right)
	= - \frac{\kappa'''(s)}{(\kappa''(s))^3}
	\quad  \Rightarrow \quad \Lambda'''(D) 
	= - \frac{T}{V^3} \]
which is finite because we have assumed $|T| < \infty$ and $V>0$.
Hence the second-order Taylor expansion of $\Lambda(a)$ is given by
\begin{eqnarray*}
    \Lambda(a)
	& = & \Lambda(D) + (a-D) \Lambda'(D) + \frac{1}{2} (a-D)^2 \Lambda''(D) + O((a-D)^3) \\
	& = & a + \frac{1}{2V} (a-D)^2 + O((a-D)^3)
		\quad \mathrm{as~} a \to D
\end{eqnarray*}
which proves the claim.
\hfill $\Box$

\section{Proof of Proposition~\ref{prop:strongLD}}
\label{app:strongLD}
\setcounter{equation}{0}

For each $i \ge 1$, the random variable $L_i = \log \frac{dP_i(Y_i)}{dQ_i(Y_i)}$ has cgf
\[ \kappa_i(s) = \log \eE_{Q_i} \left( \frac{dP_i}{dQ_i} \right)^s = (s-1) D_s(P_i\|Q_i) \]
under $Q_i$. We have
\[ \kappa_i(1) = 0, \quad \kappa_i'(1) = D_i, \quad \kappa_i''(1) = V_i . \]
Averaging over $i=1,2,\cdots,n$, we obtain
\[ \overline{\kappa}_n(1) = 0, \quad \overline{\kappa}_n'(1) = \overline{D}_n, 
	\quad \overline{\kappa}_n''(1) = \overline{V}_n . \]
Denote by $\overline{\Lambda}_n(\cdot)$ the large-deviations function for $\sum_{i=1}^n L_i$.

Under {\bf (LR1)} {\bf (LR2)} {\bf (LR3)}, the assumptions of Lemma~\ref{lem:HT-Taylor} hold
with $\pP_n = \prod_{i=1}^n P_i$, $\qQ_n = \prod_{i=1}^n Q_i$ playing the roles of $P$ and $Q$.
Evaluating (\ref{eq:HT-Taylor}) at $a=a_n$ of \eqref{eq:an0} and multiplying by $n$, we obtain
\begin{eqnarray}
  n \overline{\Lambda}_n(a_n) 
	& = & n a_n + n \frac{(a_n - \overline{D}_n)^2}{2\overline{V}_n} + O(a_n - \overline{D}_n)^3 \nonumber \\
	& = & n a_n + \frac{(t - c/\sqrt{n\overline{V}_n})^2}{2} + O(n^{-1/2}) \nonumber \\
	& = & n a_n + \frac{t^2}{2} + O(n^{-1/2}) .
\label{eq:nan}
\end{eqnarray}

Denote by $s_n$ the solution to $a_n = \overline{\kappa}_n'(s_n)$ which is
guaranteed to exist by Assumption {\bf (LR3)}.
Since $\overline{\kappa}_n'(1) = \overline{D}_n$, $\overline{\kappa}_n''(1) \ne 0$, and
$\overline{\kappa}_n'(s)$ is continuous at $s=1$, we have $s_n \to 1$ as $a_n \to \overline{D_n}$.	

{\bf (i) Semistrong nonlattice case}: We first show that Chaganty and Sethuraman's conditions 
{\bf (CS1)}, {\bf (CS2)}, {\bf (CS3)} of Sec.~\ref{sec:strongLD} are met. Then \cite[Theorem~3.3]{Chaganty93} 
is applied with $a_n$, $s_n$, $\overline{\kappa}_n$ and $\overline{\Lambda}_n$ respectively 
playing the roles of $m_n$, $\tau_n$, $\psi_n$ and $\gamma_n$ in \cite{Chaganty93}, i.e.,
\begin{equation}
   \qQ_n \left\{ \sum_{i=1}^n L_i \ge n a_n \right\} 
	\sim \frac{e^{-n \overline{\Lambda}_n(a_n)}}{s_n \sqrt{2\pi n \overline{\kappa}_n''(s_n)}} 
	\quad \mathrm{as~} n \to \infty
\label{eq:strongLD-LLR}
\end{equation}
where $L_i = \log \frac{dP_i(Y_i)}{dQ_i(Y_i)}$ for $1 \le i \le n$.
Since $s_n \to 1$ as $n \to \infty$ and $\overline{\kappa}_n''(s_n) \to \overline{\kappa}_n''(1) = \overline{V}_n$,
(\ref{eq:nonlattice}) follows.

Conditions {\bf (CS1)} and {\bf (CS2)} follow directly from {\bf (LR1)} and {\bf (LR2)}. 
Condition {\bf (CS3)} is evaluated using the same method as in \cite[p.~1687]{Chaganty93},
also see the discussion in \cite[pp.~546---548]{Feller71}. Denote by $M_k$ the mgf of $L_k$.
Since $\{L_k\}$ is assumed to be a sequence
of semistrong nonlattice random variables, one of the conditions {\bf (NL1)}, {\bf (NL2)},
or {\bf (NL3)} of Def.~\ref{def:semistrong} applies.
If {\bf (NL1)} applies, then 
\begin{equation}
  \exists \delta_0 > 0 : \quad \sup_{1 \le k \le n}
	\sup_{\delta < |\omega| < \lambda s_n} \left| \frac{M_k(s_n + i\omega)}{M_k(s_n)} \right| 
	\le 1 - a(\delta,\lambda) , \quad \forall 0 < \delta < \delta_0 < \lambda 
\label{eq:moment-ratio}
\end{equation}
for some strictly positive $a(\delta,\lambda)$. 
Hence
\[
   \left| \frac{M_{(n)}(s_n + i\omega)}{M_{(n)}(s_n)} \right|
	= \prod_{k=1}^n \left| \frac{M_k(s_n + i\omega)}{M_k(s_n)} \right| 
	\le [1 - a(\delta,\lambda)]^n 
\]
vanishes exponentially with $n$, and Condition {\bf (CS3)} holds. 
If condition {\bf (NL2)} applies, then
\begin{eqnarray*}
   \sup_{\delta' < |\omega| < \lambda s_n} \left| \frac{M_{(n)}(s_n + i\omega)}{M_{(n)}(s_n)} \right|
	& \le & \sup_{\delta' < |\omega| < \lambda s_n} \prod_{k\in\calI_\delta} \left| \frac{M_k(s_n + i\omega)}{M_k(s_n)} \right| \\
	& \le & [1-a(\delta',\lambda)]^{\lceil \delta n \rceil} = o(1/\sqrt{n}) 
\end{eqnarray*}
and therefore condition {\bf (CS3)} is again satisfied.
Finally, if condition {\bf (NL3)} applies, the same technique is applied 
to the mgf of the sums $L_{i_k} + L_{j_k}, \,k \ge 1$.

To show the inequality ``$\ge n a_n$'' may be replaced with a strict inequality, consider $a_n$ in \eqref{eq:an0} 
where the constant $c$ is replaced by $c+\delta$ for some $\delta > 0$. Then we have
\[ \qQ_n \left\{ \sum_{i=1}^n L_i \ge n a_n + \delta \right\} 
	\le \qQ_n \left\{ \sum_{i=1}^n L_i > n a_n \right\} 
	\le \qQ_n \left\{ \sum_{i=1}^n L_i \ge n a_n \right\} .
\]
By \eqref{eq:an0} and \eqref{eq:nonlattice}, the ratio of the right side to the left side 
is $e^{\delta + o(1)}$. Since $\delta$ is arbitrarily small, we obtain 
$\qQ_n \left\{ \sum_{i=1}^n L_i > n a_n \right\} =  [1+o(1)] \,\qQ_n \left\{ \sum_{i=1}^n L_i \ge n a_n \right\}$.
	
{\bf (ii) Lattice case}: 
If $\sum_{k=1}^n L_k$ is a lattice random variable, so is each component $L_k$. Then
\[ \exists \delta_1 > 0 : \quad
	\sup_{1 \le k \le n} \sup_{\delta < |\omega| < \pi/d_n} \left| \frac{M_k(s_n + i\omega)}{M_k(s_n)} \right| 
	\le 1 - a(\delta) , \quad \forall 0 < \delta < \delta_1
\]
for some strictly positive $a(\delta)$. Therefore
\[ \sup_{\delta < |\omega| < \pi/d_n} \left| \frac{M_{(n)}(s_n + i\omega)}{M_{(n)}(s_n)} \right| 
	\le [1 - a(\delta)]^n , \quad \forall \delta > 0
\]
and {\bf (CS3')} holds. Applying \cite[Theorem~3.5]{Chaganty93} we have, for $n a_n \in \Omega_n$
(also see \eqref{eq:chaganty:lattice})
\begin{equation}
  \qQ_n \left\{ \sum_{i=1}^n L_i \ge n a_n \right\} 
	\sim \frac{s_n d_n}{1-e^{-s_n d_n}} \,\frac{e^{-n \overline{\Lambda}_n(a_n)}}{s_n \sqrt{2\pi n \overline{\kappa}_n''(s_n)}} 
	\quad \mathrm{as~} n \to \infty .
\label{eq:strongLD-LLR-lattice}
\end{equation}
Since $s_n \to 1$ and $\overline{\kappa}_n''(1) = \overline{V}_n$, this proves \eqref{eq:lattice}. 
\hfill $\Box$

\section{Proof of Lemma~\ref{lem:2beta}}
\label{app:2GoodCodes}
\setcounter{equation}{0}

We seek to prove the lower bound \eqref{eq:beta-mix2}, restated here for convenience:
\[ \sum_{P \in \scrP_1 \cap \scrP_n(\calX)} \pi(P) \beta_{1-\epsilon(P),n,P} 
	\ge \hat{\beta}_{1-\epsilon,n} \,[1 + o(1)] ,
	\quad \mathrm{where~} \sum_P \pi(P) \,\epsilon(P) \le \epsilon. 
\]
By the fundamental theorem of linear programming, the minimization problem
\[ \beta_n^* \triangleq \inf_\pi \sum_P \pi(P) \,\beta_{1-\epsilon(P),n,P}
	\quad \mathrm{subj.~to} \quad \sum_P \pi(P) \,\epsilon(P) \le \epsilon 
\]
is achieved by a distribution $\pi$ that has support at two points, say $P_1$ and $P_2$ (which may coincide). 
Let $q_1 = \pi(P_1)$, $\epsilon_1 = \epsilon(P_1)$, and $\epsilon_2 = \epsilon(P_2)$. 
Then
\begin{equation}
   \beta_n^* = \min_{q_1,\epsilon_1,\epsilon_2} \,q_1 \beta_{1-\epsilon_1,n,P_1} + (1-q_1) \beta_{1-\epsilon_2,n,P_2} 
	\quad \mathrm{subj.~to} \quad q_1 \epsilon_1 + (1-q_1) \epsilon_2 = \epsilon .
\label{eq:beta-n*}
\end{equation}

If $\epsilon_1 = \epsilon_2 = \epsilon$, the minimand in \eqref{eq:beta-n*} is equal to $\beta_{1-\epsilon,n}$. 
We show this value cannot be substantially reduced by having $\epsilon_1 \ne \epsilon_2$. The derivations below
are based on Theorem~\ref{thm:NP} which provides the exact asymptotics for expressions such as $\log \beta_{1-\epsilon_1,n,P_1}$.
However convergence is nonuniform over $\epsilon_1 \in (0,1)$, hence a special treatment is needed if $\epsilon_1$ 
approaches either 0 or 1. 
We show below that the minimizing $\epsilon_1$ and $\epsilon_2$ in \eqref{eq:beta-n*}
are equal to $\epsilon$ plus an $o(n^{-1/2})$ term.

Without loss of generality we assume that $q_1 \ge \frac{1}{2}$.
Six possible cases are considered for the minimizing $(q_1,\epsilon_1)$.
The desired lower bound on $\log \beta_n^*$ is derived in each case.
Cases II-V are the ``boundary'' cases for $\epsilon_1$ and $\epsilon_2$.
The main case is Case~VI.
Fix an arbitrarily small $\eta \in (0, \frac{1-\epsilon}{2})$.\\

{\bf Case~I:} $q_1 > 1 - \frac{1}{n}$. Then $\epsilon \ge q_1 \epsilon_1 \ge (1 - \frac{1}{n}) \epsilon_1$ and
\begin{eqnarray*}
  \log \beta_n^* 
    & \stackrel{(a)}{\ge} & \log \beta_{1-\epsilon_1,n,P_1} + \log q_1 \\
	& \stackrel{(b)}{\ge} & \log \beta_{1-\frac{\epsilon}{1-1/n},n,P_1} + \log \left( 1-\frac{1}{n} \right) \\
	& \stackrel{(c)}{\ge} & \log \hat{\beta}_{1-\frac{\epsilon}{1-1/n},n} + o(1) \\
	&  \stackrel{(d)}{=} & - nC + \sqrt{nV_\epsilon} \,\calQ^{-1}\left( \frac{\epsilon}{1 - \frac{1}{n}} \right) 
		- \frac{1}{2} \log n - \overline{A} \left( \frac{\epsilon}{1 - \frac{1}{n}} \right) + o(1) \\
	& = & \log \hat{\beta}_{1-\epsilon,n} + o(1)
\end{eqnarray*}
where inequality (a) follows from \eqref{eq:beta-n*}, (b) from the fact that the function $\beta_{1-\epsilon,n,P}$ 
decreases with $\epsilon$, (c) from \eqref{eq:beta-quasi3}, and (d) from \eqref{eq:beta-epsilon-n}. 

{\bf Case~II:} $\frac{1}{2} \le q_1 \le 1 - \frac{1}{n}$ and $\epsilon_1 < (1-\eta)\epsilon$. Then
\begin{eqnarray*}
  \log \beta_n^* 
    & \stackrel{(a)}{\ge} & \log \beta_{1-\epsilon_1,n,P_1} + \log q_1 \\
	& \stackrel{(b)}{\ge} & \log \beta_{1-(1-\eta)\epsilon,n,P_1} - \log 2 \\
	& \stackrel{(c)}{\ge} & \log \hat{\beta}_{1-(1-\eta)\epsilon,n} - \log 2 + o(1) \\
	& \ge & \log \hat{\beta}_{1-\epsilon,n} + o(1)
\end{eqnarray*}
where inequality (a) follows from \eqref{eq:beta-n*}, (b) from the inequalities $q_1 \ge 1/2$
and $\epsilon_1 < (1-\eta)\epsilon$ and the decreasing property of the function $\beta_{1-\epsilon,n,P}$ with respect to $\epsilon$,
and (c) from from \eqref{eq:beta-epsilon-n}.

{\bf Case~III:} $\frac{1}{2} \le q_1 \le 1 - \frac{1}{n}$ and $\epsilon_2 < (1-\eta)\epsilon$.
Similarly to Case~II, we have
\begin{eqnarray*}
  \log \beta_n^* 
    & \ge & \log \beta_{1-\epsilon_2,n,P_2} + \log (1-q_1) \\
	& \ge & \log \beta_{1-(1-\eta)\epsilon,n,P_2} - \log n \\
	& \ge & \log \hat{\beta}_{1-(1-\eta)\epsilon,n} - \log n + o(1) \\
	& \ge & \log \hat{\beta}_{1-\epsilon,n} + o(1) .
\end{eqnarray*}

{\bf Case~IV:} $\frac{1}{2} \le q_1 \le 1 - \frac{1}{n}$ and $\epsilon_1 > (1+\eta)\epsilon$. Then
\[ \epsilon_2 = \frac{\epsilon - q_1 \epsilon_1}{1-q_1} < \epsilon \frac{1-(1+\eta)q_1}{1-q_1}
	= \epsilon \left( 1 - \frac{\eta q_1}{1-q_1} \right) \le \epsilon \left( 1 - \frac{\eta/2}{1/2} \right) 
    = (1-\eta)\epsilon \]
hence Case~III applies.

{\bf Case~V:} $\frac{1}{2} \le q_1 \le 1 - \frac{1}{n}$ and $\epsilon_2 > 1-\eta$. Then
\[ \epsilon_1 = \frac{\epsilon - (1-q_1) \epsilon_2}{q_1} < \frac{\epsilon - (1-q_1)(1-\eta)}{q_1} \triangleq f(q_1) . \]
The function $f(q_1)$ has derivative given by $f'(q_1) = (1-q_1-\epsilon)/q_1^2$
which is positive (hence $f$ increases) for $q_1 < 1-\epsilon$, and decreases otherwise. 
Also $\epsilon_1 < \max_{1/2 \le q_1 \le 1} f(q_1)$.
If $\epsilon \le \frac{1}{2}$, we have $\epsilon_1 < f(1-\epsilon) = \eta \frac{\epsilon}{1-\epsilon}$
(unconstrained maximum of $f$).
If $\epsilon > \frac{1}{2}$, then $\epsilon_1 < f(\frac{1}{2}) = (2\epsilon-1)+\eta$ (constrained maximum).
In both cases, $\epsilon_1 < (1-\eta)\epsilon$ owing to our condition
$\eta < \frac{1-\epsilon}{2}$, hence Case~II applies.

{\bf Case~VI:} $\frac{1}{2} \le q_1 \le 1 - \frac{1}{n}$ and $(1-\eta)\epsilon \le \epsilon_1 \le (1+\eta)\epsilon$
and $(1-\eta)\epsilon \le \epsilon_2 \le 1-\eta$.
Then $\rho_n \triangleq \frac{q_1}{1-q_1} \in [1, \,n-1]$ and $0 \le \log \rho_n < \log n$.
Let $\delta_n \triangleq \epsilon_1 - \epsilon \in [-\eta\epsilon, \,\eta\epsilon]$, then $\epsilon_1 = \epsilon + \delta_n$ and
\[ \epsilon = \epsilon_1 q_1 + \epsilon_2 (1-q_1) \quad \Rightarrow \quad
	(1-q_1) \epsilon = \delta_n q_1 + \epsilon_2 (1-q_1)  \quad \Rightarrow \quad
	\epsilon_2 = \epsilon - \rho_n \delta_n . 
\]

Since $q_1$ and $1-q_1$ are in $(0,1)$, we may lower bound $\beta_n^*$ as
\begin{equation}
   \beta_n^* \ge \min_{q_1,\epsilon_1,\epsilon_2} \,q_1^{1+\eta} \beta_{1-\epsilon_1,n,P_1} + (1-q_1)^{1+\eta} \beta_{1-\epsilon_2,n,P_2} 
	\quad \mathrm{subj.~to} \quad q_1 \epsilon_1 + (1-q_1) \epsilon_2 = \epsilon .
\label{eq:beta-n*-eta}
\end{equation}
By \eqref{eq:beta-epsilon-n}, there exists $n_0$ such that for all $n > n_0$,
\begin{eqnarray*}
   \beta_{1-\epsilon_1,n,P_1} \ge (1-\eta) \hat{\beta}_{1-\epsilon_1,n} \quad \mathrm{and} \quad 
   \beta_{1-\epsilon_2,n,P_2} \ge (1-\eta) \hat{\beta}_{1-\epsilon_2,n}
\end{eqnarray*}
and therefore the minimand of \eqref{eq:beta-n*-eta} is lower-bounded by
\begin{equation}
   (1-\eta) [q_1^{1+\eta} \hat{\beta}_{1-\epsilon_1,n} + (1-q_1)^{1+\eta} \hat{\beta}_{1-\epsilon_2,n}]
\label{eq:beta-n*-eta-LB}
\end{equation}
which is a strictly convex function of $q_1$. Its minimum is obtained by setting the derivative to zero, hence
\[ 0 = q_1^\eta \hat{\beta}_{1-\epsilon_1,n} - (1-q_1)^\eta \hat{\beta}_{1-\epsilon_2,n} \]
\[ \Rightarrow \quad \eta \log \rho_n + \log \hat{\beta}_{1-\epsilon_1,n} = \hat{\beta}_{1-\epsilon_2,n} . \]
Substituting $\epsilon_1 = \epsilon + \delta_n$ and $\epsilon_2 = \epsilon - \rho_n \delta_n$
and using the expansion \eqref{eq:beta-epsilon-n}, we obtain, for all $n > n_0$,
\begin{equation}
  \sqrt{nV_{\epsilon+\delta_n}} \calQ^{-1}(\epsilon + \delta_n) - \overline{A}(\epsilon + \delta_n) + \eta \log \rho_n
		= \sqrt{nV_{\epsilon-\rho_n\delta_n}} \calQ^{-1}(\epsilon - \rho_n \delta_n) - \overline{A}(\epsilon - \rho_n \delta_n)
\label{eq:min-eq2}
\end{equation}
which implies that both $\delta_n$ and $\rho_n \delta_n$ vanish (hence $\epsilon_1, \epsilon_2 \to\epsilon$) as $n\to\infty$.
Denote by $c > 0$ the derivative of the function $- \calQ^{-1}(\epsilon)$ evaluated at $\epsilon$. 
Then taking the first-order Taylor series expansion of \eqref{eq:min-eq2} around $\rho_n, \delta_n = 0$ 
and dividing by $-1/\sqrt{nV_\epsilon}$ yields the matching condition
\begin{eqnarray}
   c \,\delta_n + O(\delta_n^2) 
	& = & - c \rho_n \delta_n + \frac{\eta \log \rho_n}{\sqrt{nV_\epsilon}} + O \left( \rho_n^2 \delta_n^2 \right) + o(n^{-1/2}) \nonumber \\
   c \,\delta_n & = & \frac{\eta}{\sqrt{nV_\epsilon}} \,\frac{\log \rho_n}{1+\rho_n} + o(n^{-1/2}) .
\label{eq:balance2}
\end{eqnarray}
Denote by $\lambda \approx 0.3$ the maximum of the function $\frac{\log \rho}{1+\rho}$ over $\rho \ge 1$; the maximum is achieved
at $\rho \approx 3.6$ satisfying $\log\rho = 1 + 1/\rho$.
For the optimal $\delta_n,\rho_n$ and $n > n_0$, we have
\begin{eqnarray}
   \log \beta_n^* 
	& \stackrel{(a)}{\ge} & \log(1-\eta) + \log [q_1^{1+\eta} \hat{\beta}_{1-\epsilon_1,n} + (1-q_1)^{1+\eta} \hat{\beta}_{1-\epsilon_2,n}] \nonumber \\
	& \stackrel{(b)}{\ge} & \log(1-\eta) + \log [q_1^{1+\eta} + (1-q_1)^{1+\eta}] + \log \hat{\beta}_{1-\epsilon_1,n} \nonumber \\
	& \stackrel{(c)}{\ge} & \log(1-\eta) - \eta \log 2 + \log \hat{\beta}_{1-\epsilon_1,n} \nonumber \\
	& \stackrel{(d)}{\ge} & \log(1-\eta) - \eta \log 2 + \log \hat{\beta}_{1-\epsilon,n} - \eta \frac{\log \rho_n}{1+\rho_n} + o(1) \nonumber \\
	& \stackrel{(e)}{\ge} & \log(1-\eta) - \eta \log 2 + \log \hat{\beta}_{1-\epsilon,n} - \eta \lambda + o(1)
\label{eq:beta-n*-LB2}
\end{eqnarray}
where (a) follows from \eqref{eq:beta-n*-eta-LB},
(b) holds because $\delta_n \ge 0$, hence $\epsilon_1 \ge \epsilon_2$, hence $\hat{\beta}_{1-\epsilon_1,n} \le \hat{\beta}_{1-\epsilon_2,n}$;
(c) is obtained by replacing the second logarithm by its minimum over $q_1$ (which is achieved at $q_1 = 1/2$);
(d) follows from Taylor expansion of $\log \hat{\beta}_{1-\epsilon_1,n}$ as $\delta_n \to 0$ and \eqref{eq:balance2}; and
(e) from the definition of $\lambda$ above.
Since the asymptotic lower bound \eqref{eq:beta-n*-LB2} holds for arbitrarily small $\eta > 0$, the claim follows.
\hfill $\Box$

\section{Proof of Theorem~\ref{thm:converse-CC3}}
\label{app:converse-CC3}
\setcounter{equation}{0}

There are $J < (n+1)^{|\calX|-1}$ codeword types in $\scrP_1^c$.
For each such type $P_j, \,1 \le j \le J$, denote by $M_j$ the number of codewords of type $P_j$ and
by $\epsilon_j$ the average decoding error
probability conditioned on the codeword having type $P_j$. Hence
\begin{equation}
  M = \sum_{j=1}^J M_j \quad \mathrm{and} \quad \epsilon = \sum_{j=1}^J \epsilon_j \frac{M_j}{M} .
\label{eq:rate-split}
\end{equation}
Let $q_j \triangleq M_j/M$ with is a pmf over $\{1,2,\cdots, J\}$.
Define the index set 
\[ \calJ_+ \triangleq \left\{ j \in \{1,2,\cdots,J\} :~ q_j \ge \exp \left( - \frac{c_1 \,n^{1/4}}{\log^2 n} \right) \right\} \]
which contains at least one element. Moreover
\begin{equation}
   \sum_{j\in\calJ_+} q_j \ge 1 - (J-1) \exp \left( - \frac{c_1 \,n^{1/4}}{\log^2 n} \right) 
	\quad \to 1 \quad \mathrm{as~} n \to \infty
\label{eq:sum-qj}
\end{equation}
and from \eqref{eq:rate-split},
\begin{equation}
  \epsilon = \sum_{j=1}^J \epsilon_j q_j \ge \sum_{j\in\calJ_+} \epsilon_j q_j .
\label{eq:sum-epsilon'j}
\end{equation}
Let
\begin{equation}
  q_j' = \frac{q_j}{\sum_{j\in\calJ_+} q_j} , \quad j \in \calJ_+
\label{eq:q'j}
\end{equation}
which is a pmf over $\calJ_+$. By \eqref{eq:sum-qj}, \eqref{eq:sum-epsilon'j} and \eqref{eq:q'j}, we have
\[ \sum_{j\in\calJ_+} \epsilon_j q_j' 
	\le \frac{\epsilon}{1 - (J-1) \exp \left( - \frac{c_1 \,n^{1/4}}{\log^2 n} \right)}
	\le \epsilon \left[ 1 + 2J \exp \left( - \frac{c_1 \,n^{1/4}}{\log^2 n} \right) \right] . 
\]
Since $q'$ is a pmf over $\calJ_+$, there exists $j\in\calJ_+$ such that
\begin{equation}
  \frac{\epsilon}{2} \le \epsilon_j \le \epsilon \left[ 1 + 2J \exp \left( - \frac{c_1 \,n^{1/4}}{\log^2 n} \right) \right]
\label{eq:epsilon-j-UB}
\end{equation}
and 
\begin{equation}
  q_j = \frac{M_j}{M} \ge \exp \left( - \frac{c_1 \,n^{1/4}}{\log^2 n} \right) .
\label{eq:qj}
\end{equation}
hence $\log M \le \log M_j + \frac{c_1 \,n^{1/4}}{\log^2 n}$.

By Theorem~\ref{thm:converse-CC}(iii) with $\delta_n = \frac{1}{n^{1/4} \log n}$, we have
\begin{eqnarray}
  \log M_j & \le & \zeta_n(P_j;W) + \frac{1}{2} \log n + o(1) \nonumber \\
	& \stackrel{(a)}{\le} & nC - \sqrt{nV_\epsilon} \,\calQ^{-1}(\epsilon_j) 
		- \frac{c_1 \,n^{1/4}}{\log n} [1+o(1)] , \quad 1 \le j \le J \nonumber \\
	& \stackrel{(b)}{\le} & nC - \sqrt{nV_\epsilon} \,\calQ^{-1}\left( \epsilon \left[ 1 
		+ 2J \exp \left( - \frac{c_1 \,n^{1/4}}{\log^2 n} \right) \right] \right) - \frac{c_1 \,n^{1/4}}{\log n} [1+o(1)] \nonumber \\
	& \stackrel{(c)}{=} & nC - \sqrt{nV_\epsilon} \,\calQ^{-1}(\epsilon) - \frac{c_1 \,n^{1/4}}{\log n} [1+o(1)] 
\label{eq:logMj-P2-UB}
\end{eqnarray}
where (a) follows from \eqref{eq:zetan-UB2} and the fact that $P_j \notin \scrP_1$,
(b) from (\ref{eq:epsilon-j-UB}) and the fact that 
the function $-\calQ^{-1}(\epsilon)$ is increasing, and
(c) because the derivative of this function at $\epsilon$ is bounded.
Thus
\begin{eqnarray*}
   \log M & \le & \log M_j + \frac{c_1 \,n^{1/4}}{\log^2 n} \\
	& \le & nC - \sqrt{nV_\epsilon} \calQ^{-1}(\epsilon) - \frac{c_1 \,n^{1/4}}{\log^2 n} \,[1+o(1)]
\end{eqnarray*}
which proves the claim.
\hfill $\Box$

\section{Proof of Theorem~\ref{thm:converse}}
\label{app:converse-avg}
\setcounter{equation}{0}

The method of proof parallels that of Lemma~\ref{lem:2beta}.
Under the average error probability criterion, any $(M,\epsilon)$ code on $\calX^n$ 
is the union of a ``good'' $(M_1,\epsilon_1)$ subcode with codewords in ${\sf F}_1$ 
and a ``bad'' $(M_2,\epsilon_2)$ subcode  with codewords in $\calX^n \setminus {\sf F}_1$ where
\begin{equation}
   M = M_1 + M_2 \quad \mathrm{and} \quad \epsilon = \epsilon_1 \frac{M_1}{M_1 + M_2} + \epsilon_2 \frac{M_2}{M_1 + M_2} .
\label{eq:Meps-2subcodes}
\end{equation}
By Theorems~\ref{thm:converse-CC2} and \ref{thm:converse-CC3} respectively we have
\begin{eqnarray}
   \log M_1 & \le & \log \overline{M}_1(n,\epsilon_1) \triangleq nC - \sqrt{nV_\epsilon} \calQ^{-1}(\epsilon_1) 
		+ \frac{1}{2} \log n + \overline{A}(\epsilon_1) + o(1) \label{eq:logM1-UB} \\
   \log M_2 & \le & \log \overline{M}_2(n,\epsilon_2) \triangleq nC - \sqrt{nV_\epsilon} \calQ^{-1}(\epsilon_2) 
		- \frac{c_1 \,n^{1/4}}{\log  n} [1+o(1)] . \label{eq:logM2-UB}
\end{eqnarray}
Let $q_1 = \frac{M_1}{M_1+M_2} \in (0,1)$, hence $\epsilon = q_1 \epsilon_1 + q_2 \epsilon_2$. 
Using the identity $M = \frac{M_1}{q_1} = \frac{M_2}{1-q_1}$ and applying
the upper bounds (\ref{eq:logM1-UB}) and (\ref{eq:logM2-UB}), we obtain
\begin{eqnarray}
   \log M & = & \min \left\{ \log M_1 - \log q_1, \,\log M_2 - \log (1-q_1) \right\} \nonumber \\
   & \le & \min \left\{ \log \overline{M}_1(n,\epsilon_1) - \log q_1, 
		\,\log \overline{M}_2(n,\epsilon_2) - \log (1-q_1) \right\} \nonumber \\
   & \le & \sup_{0 < q_1 < 1} \min \left\{ \log \overline{M}_1(n,\epsilon_1) 
		- \log q_1, \,\log \overline{M}_2(n,\epsilon_2) - \log (1-q_1) \right\} .
\label{eq:logM-UB-general}
\end{eqnarray}
Since the first argument of the min function decreases continuously from $+\infty$ to a finite value and the second argument
increases continuously from a finite value to $+\infty$ as $q_1$ ranges from 0 to 1, the supremum over $q_1$ is achieved 
when the two arguments are equal:
\begin{equation}
   \log \overline{M}_1(n,\epsilon_1) - \log q_1 = \log \overline{M}_2(n,\epsilon_2) - \log (1-q_1) .
\label{eq:logM-EqualUBs}
\end{equation}
There are three possible cases for the maximizing $q_1$. An upper bound on $\log M$ is derived for each case.\\

{\bf Case~I:} $q_1 > 1 - \frac{1}{n}$. Then $\epsilon \ge q_1 \epsilon_1 \ge (1 - \frac{1}{n}) \epsilon_1$
and (\ref{eq:logM-EqualUBs}) yields
\begin{eqnarray*}
  \log M & \stackrel{(a)}{\le} & \log \overline{M}_1 \left( n, \frac{\epsilon}{1 - \frac{1}{n}} \right) 
		- \log \left( 1-\frac{1}{n} \right) \\
	&  \stackrel{(b)}{=} &  nC - \sqrt{nV_\epsilon} \,\calQ^{-1}\left( \frac{\epsilon}{1 - \frac{1}{n}} \right) 
		+ \frac{1}{2} \log n + \overline{A} \left( \frac{\epsilon}{1 - \frac{1}{n}} \right) + o(1) \\
	& = & \log \overline{M}_1(n,\epsilon) + o(1)
\end{eqnarray*}
where inequality (a) follows from the fact that the function $\overline{M}_1(n,\epsilon)$ 
increases with $\epsilon$, and (b) from (\ref{eq:logM1-UB}).

{\bf Case~II:} $q_1 < \frac{1}{n}$. Then similarly to Case~I,
$\epsilon \ge (1 - \frac{1}{n}) \epsilon_2$. 
Using (\ref{eq:logM-EqualUBs}) and (\ref{eq:logM2-UB}), we obtain
\begin{eqnarray*}
  \log M  & \le & \log \overline{M}_2 \left( n, \frac{\epsilon}{1 - \frac{1}{n}} \right) 
		- \log \left( 1-\frac{1}{n} \right) \\
	& = & nC - \sqrt{nV_\epsilon} \,\calQ^{-1}\left( \frac{\epsilon}{1 - \frac{1}{n}} \right)
		- \frac{c_1 \,n^{1/4}}{\log  n} [1+o(1)] \\
	& = & nC - \sqrt{nV_\epsilon} \,\calQ^{-1}(\epsilon) - \frac{c_1 \,n^{1/4}}{\log  n} [1+o(1)] .
\end{eqnarray*}

{\bf Case~III:} $\frac{1}{n} \le q_1 \le 1 - \frac{1}{n}$. 
Then $\rho_n \triangleq \frac{q_1}{1-q_1} \in [\frac{1}{n-1}, \,n-1]$ and $|\log \rho_n| < \log n$.
Let $\delta_n \triangleq \epsilon_1 - \epsilon \in [-\epsilon, \,1-\epsilon]$, then (\ref{eq:Meps-2subcodes})
yields 
\[ \epsilon = \epsilon_1 q_1 + \epsilon_2 (1-q_1) \quad \Rightarrow \quad
	(1-q_1) \epsilon = \delta_n q_1 + \epsilon_2 (1-q_1)  \quad \Rightarrow \quad
	\epsilon_2 = \epsilon - \rho_n \delta_n . 
\]
With this notation, (\ref{eq:logM-EqualUBs}) becomes
\begin{eqnarray}
   \log \overline{M}_1(n,\epsilon + \delta_n) - \log \rho_n & = & \log \overline{M}_2(n,\epsilon - \rho_n \delta_n) \nonumber \\
	nC - \sqrt{nV_\epsilon} \calQ^{-1}(\epsilon + \delta_n) + \frac{1}{2} \log n + \overline{A}(\epsilon + \delta_n) - \log \rho_n
		& = &  nC - \sqrt{nV_\epsilon} \calQ^{-1}(\epsilon - \rho_n \delta_n) - \frac{c_1 \,n^{1/4}}{\log  n} [1+o(1)] \nonumber \\
	- \sqrt{nV_\epsilon} \calQ^{-1}(\epsilon + \delta_n) 
		& = & - \sqrt{nV_\epsilon} \calQ^{-1}(\epsilon - \rho_n \delta_n) - \frac{c_1 \,n^{1/4}}{\log  n} [1+o(1)] \nonumber \\
\label{eq:min-eq}
\end{eqnarray}
which implies that $\lim_{n\to\infty} \delta_n = \lim_{n\to\infty} \rho_n \delta_n = 0$.
Denote by $c' > 0$ the derivative of the function $- \calQ^{-1}(\epsilon)$ evaluated at $\epsilon$. 
Then first-order Taylor series expansion of (\ref{eq:min-eq}) around $\delta_n = 0$ yields the matching condition
\begin{eqnarray}
   c' \,\delta_n + O(\delta_n^2) 
	& = & - c' \rho_n \delta_n - \frac{c/\sqrt{V}}{n^{1/4} \log n} [1+o(1)] + O \left( \rho_n^2 \delta_n^2 \right) \nonumber \\
   (1+\rho_n) \delta_n & = & - \frac{c}{c' \sqrt{V} n^{1/4} \log n} [1+o(1)] + O(\delta_n^2) + O \left( \rho_n^2 \delta_n^2 \right) .
   \label{eq:balance}
\end{eqnarray}
The only possible asymptotic solution is
\begin{equation}
   \delta_n = - \frac{c}{(1+\rho_n) c' \sqrt{V} \log^2 n} \,[1+o(1)]
\label{eq:deltan-balance}
\end{equation}
which satisfies \eqref{eq:balance} as well as the conditions $\lim_{n\to\infty} \delta_n = 0$ 
and $\lim_{n\to\infty} \rho_n \delta_n = 0$ for all $\rho_n \in [\frac{1}{n-1}, \,n-1]$. 
Moreover for any such $\rho_n$, \eqref{eq:deltan-balance} yields
\begin{eqnarray}
   c' \sqrt{nV_\epsilon} \delta_n - \log q_1
	& = & - \frac{c n^{1/4}}{(1+\rho_n) \log n} \,[1+o(1)]  - \log q_1 \nonumber \\
	& = & \frac{(q_1 - 1) n^{1/4}}{2\log n} \,[1+o(1)]  - \log q_1 \nonumber \\
	& \le & - \frac{c}{n^{3/4} \log n} \,[1+o(1)]
\label{eq:delta-UB}
\end{eqnarray}
with asymptotic equality when $q_1 = 1 - \frac{1}{n}$, or equivalently, $\rho_n = n-1$.
Thus
\begin{eqnarray*}
   \log M & \stackrel{(a)}{\le} & \log \overline{M}_1(n,\epsilon + \delta_n) - \log q_1 \\
   & \stackrel{(b)}{\le} & nC - \sqrt{nV_\epsilon} \calQ^{-1}(\epsilon + \delta_n) + \frac{1}{2} \log n 
		+ \overline{A}(\epsilon+\delta_n) - \log q_1 + o(1) \\
   & \stackrel{(c)}{=} & nC - \sqrt{nV_\epsilon} [\calQ^{-1}(\epsilon) - c' \delta_n + O(\delta_n^2)] + \frac{1}{2} \log n 
		+ \overline{A}(\epsilon) - \log q_1 + o(1) \\
   & \stackrel{(d)}{\le} & nC - \sqrt{nV_\epsilon} \calQ^{-1}(\epsilon) + \frac{1}{2} \log n + \overline{A}(\epsilon) + o(1) \\
	& = & \log \overline{M}_1(n,\epsilon) + o(1)
\end{eqnarray*}
where (a) follows from (\ref{eq:logM-EqualUBs}), (b) from (\ref{eq:logM1-UB}), 
(c) from the fact that $\delta_n \to 0$, and (d) from (\ref{eq:delta-UB}).

In summary, the upper bound $\log M \le \log \overline{M}_1(n,\epsilon) + o(1)$ applies in all three cases.
This concludes the proof.
\hfill $\Box$

\section{Proof of Theorem~\ref{thm:BSC}}
\label{app:BSC}
\setcounter{equation}{0}

{\bf (i) Converse.}
By symmetry of the BSC there is no loss of generality in assuming that $0 < \lambda < \frac{1}{2}$, 
in which case the lattice span $d = \log \frac{1-\lambda}{\lambda}$. Let
\[ na_n \triangleq nC - \sqrt{nV} \,t_\epsilon - \frac{1}{6} S \sqrt{V} (t_\epsilon^2 - 1) . \]
Fix $Q=Q^*$ and any binary sequence $\bx$. The function $\beta_{1-\epsilon}(W^n(\cdot|\bx), Q^n)$
is independent of $\bx$ (as noted in \cite[Appendix~K]{Polyanskiy10}), and its exact asymptotics are obtained 
from Theorem~\ref{thm:NP}(ii):
\begin{equation}
   \beta_{1-\epsilon}(W^n(\cdot|\bx), Q^n) = \frac{d \,e^{d/2}}{1-e^{-d}}
	\,\frac{\exp\{-n a_n - \frac{1}{2} t_\epsilon^2 + o(1) \} }{\sqrt{2\pi Vn}}
\label{eq:beta-NP-BSC}
\end{equation}
for all $n$ such that $na_n + \frac{d}{2} \in \Omega$. 
The function $\beta_{1-\epsilon}$ is piecewise linear in $\epsilon$, with breaks at the points
satisfying $na_n + \frac{d}{2} \in \Omega$. \footnote{
	The expression \eqref{eq:beta-NP-BSC} could be derived independently of Theorem~\ref{thm:NP}(ii) by noting
	that the random count $Z_n \triangleq \sum_{i=1}^n \mathds1\{Y_i=x_i\}$ is a sufficient statistic for the test; 
	$Z_n$ is a binomial random variable with $n$ trials and parameter $1-\lambda$ and $\frac{1}{2}$ under
	$W^n(\cdot|\bx)$ and $Q^n$, respectively. One can then derive \eqref{eq:beta-NP-BSC}
	using the exact asymptotics \eqref{eq:binomial-exact} for the tail of the binomial distribution, together with 
	the asymptotic expression $\Pr\{\mathrm{Bi}(n,p) = [np + \sqrt{np(1-p)} t]\} \sim \frac{\phi(t)}{\sqrt{np(1-p)}}$.
}
By convexity of the function $\exp\{\cdot\}$, the equality in \eqref{eq:beta-NP-BSC} is replaced with $\ge$ 
for $na_n + \frac{d}{2} \notin \Omega$.
Applying Theorems \ref{thm:poly-converse-max} and \ref{thm:poly-converse-avg}, we obtain
\[ \log M^*(n,\epsilon) \le n a_n + \frac{1}{2} t_\epsilon^2 + \frac{1}{2} \log (2\pi Vn) - \log \frac{d \,e^{d/2}}{1-e^{-d}} + o(1) ,
	\quad na_n + \frac{d}{2} \in \Omega . 
\]

{\bf  (ii) Achievability.}
We use iid random codes drawn from $Q^*$ and analyse the error probability starting from the expression \eqref{eq:UnionBound}.
We select the threshold $z_n^*$ to be a lattice point. Recall $k = \max\{1, [\frac{1}{d}]\}$ and that
the normalized lattice random variable $T_n$ has span $\frac{d}{\sqrt{nV}}$.
We choose $t_n^*$ to be the largest $t\in\Omega$ such that $t \le \hat{t}_{\epsilon,n} + \frac{kd}{\sqrt{nV}}$. 
Hence $z_n^* = nC - \sqrt{nV} t_n^*$ and
\[ P_{Z_{1,n}} \{Z_{1,n} < z_n^*\} = 1 - F_{T_n}(t_n^*) = 1 - F_{T_n} \left( \hat{t}_{\epsilon,n} 
   + \frac{kd}{\sqrt{nV}} \right) = \epsilon - \frac{kd}{\sqrt{nV}} \phi(t_\epsilon) + o(n^{-1/2}) .
\]
Moreover $t_n^* = \hat{t}_{\epsilon,n} + \frac{(k-\gamma_n)d}{\sqrt{nV}}$ where 
$\gamma_n$ is a sawtooth function of $\hat{t}_{\epsilon,n}$ that oscillates between 0 and 1.

Next, the random variables $L = \log \frac{W(Y|X)}{Q^*(Y)}$ and $L' = \log \frac{W(Y|X')}{Q^*(Y)}$
are iid under the tilted distribution $\tilde{P}_{L'L}$ of \eqref{eq:Ptilde-L'L}. Hence
the normalized random variables $W_n$ and $W_n'$ of \eqref{eq:Wn'Wn} are independent lattice random variables
with zero offset and span $\frac{d}{\sqrt{nV}}$. 
Denote their marginal pmf by $p_n$, and note that $p_n(0) \sim \frac{d}{\sqrt{nV}} \phi(t_\epsilon)$.
From \eqref{eq:PL'L-0}, we obtain
\begin{eqnarray*}
  \tilde{P}_{L'L}^n\{ 0 \le W_n \le W_n'\}
  & = & e^{- z_n^*} \sum_{w \ge 0} p_n(w) \sum_{w' \ge w} p_n(w') e^{-\sqrt{nV} w'} \\
  & \sim & e^{- z_n^*} \sum_{w \ge 0} p_n(w)^2 e^{-\sqrt{nV} w} \sum_{k'\in\nN} e^{-dk'} \\
  & = & e^{- z_n^*} \sum_{w \ge 0} \frac{p_n(w)^2}{1-e^{-d}} e^{-\sqrt{nV} w} \\
  & \sim & e^{- z_n^*} \frac{p_n(0)^2}{(1-e^{-d})^2} 
	\sim e^{- z_n^*} \left( \frac{d}{1-e^{-d}} \right)^2 \frac{\phi(t_\epsilon)^2}{nV}
\end{eqnarray*} 
which is the same as \eqref{eq:PL'L-1}, multiplied by the constant $\left( \frac{d}{1-e^{-d}} \right)^2 \ge 1$.
Hence \eqref{eq:UnionBound} yields
\[ \Pr\{\mathrm{Error}\} = \epsilon - \frac{\phi(t_\epsilon)}{\sqrt{nV}} \left[ kd
	+ e^{nR_n - z_n^*} \left( \frac{d}{1-e^{-d}} \right)^2 \frac{\phi(t_\epsilon)}{\sqrt{nV}} \right] + o(n^{-1/2}) .
\]
The $O(n^{-1/2})$ term vanishes if we use the same $nR_n$ as in the nonlattice case (with $\rho=0$)
minus the quantity $2 \log \frac{d}{1-e^{-d}} + (k-\gamma_n) d - \log (kd) - 1 \ge -\gamma_n d$.
(Since $x - \log x - 1 \ge 0 \,\forall x \in \rR$, this motivated our choice of the integer $k$).
Since $\gamma_n \ge 0$, we can relax this quantity to the constant $2 \log \frac{d}{1-e^{-d}} + kd - \log (kd) - 1$,
hence the claim.
\hfill $\Box$

\end{document}